\documentclass[%
 reprint,
superscriptaddress,
 amsmath,amssymb,
 aps,
]{revtex4-2}

\usepackage{graphicx}
\usepackage{dcolumn}
\usepackage{bm}
\usepackage{float}
\usepackage{tabularx}
\usepackage{setspace}
\usepackage{outlines}
\usepackage{color}
\usepackage{wrapfig}

\usepackage[most]{tcolorbox}
\usepackage{varwidth}
\usepackage{capt-of}
\tcbuselibrary{listings}

\tcbset{enhanced,colback=white!5!white,colframe=white!25!black,fonttitle=\bfseries}

\usepackage{xr}
\makeatletter
\newcommand*{\addFileDependency}[1]{
  \typeout{(#1)}
  \@addtofilelist{#1}
  \IfFileExists{#1}{}{\typeout{No file #1.}}
}
\makeatother
\newcommand*{\myexternaldocument}[1]{%
    \externaldocument{#1}%
    \addFileDependency{#1.tex}%
    \addFileDependency{#1.aux}%
}
\myexternaldocument{SI}

\newcommand{\bp}{\boldsymbol{\pi}}
\newcommand{\bps}{\boldsymbol{\psi}}
\newcommand{\bch}{\boldsymbol{\chi}}
\newcommand{\bt}{\boldsymbol{\theta}}

\usepackage{appendix}

\begin{document}

\title{Limits on the computational expressivity of non-equilibrium biophysical processes}

\author{Carlos Floyd}
\email{csfloyd@uchicago.edu}
\affiliation{The Chicago Center for Theoretical Chemistry, The University of Chicago, Chicago, Illinois 60637, USA
}
\affiliation{The James Franck Institute, The University of Chicago, Chicago, Illinois 60637, USA
}
\author{Aaron R.\ Dinner}
\affiliation{The Chicago Center for Theoretical Chemistry, The University of Chicago, Chicago, Illinois 60637, USA
}
\affiliation{The James Franck Institute, The University of Chicago, Chicago, Illinois 60637, USA
}
\affiliation{Department of Chemistry, The University of Chicago, Chicago, Illinois 60637, USA
}
\author{Arvind Murugan}
\affiliation{The James Franck Institute, The University of Chicago, Chicago, Illinois 60637, USA
}
\affiliation{Department of Physics, The University of Chicago, Chicago, Illinois 60637, USA
}
\author{Suriyanarayanan Vaikuntanathan}
\email{svaikunt@uchicago.edu}
\affiliation{The Chicago Center for Theoretical Chemistry, The University of Chicago, Chicago, Illinois 60637, USA
}
\affiliation{The James Franck Institute, The University of Chicago, Chicago, Illinois 60637, USA
}
\affiliation{Department of Chemistry, The University of Chicago, Chicago, Illinois 60637, USA
}

\date{\today}
\begin{abstract}
Many biological decision-making processes can be viewed as performing a classification task over a set of inputs, using various chemical and physical processes as ``biological hardware." In this context, it is important to understand the inherent limitations on the computational expressivity of classification functions instantiated in biophysical media. Here, we model biochemical networks as Markov jump processes and train them to perform classification tasks, allowing us to investigate their computational expressivity. We reveal several unanticipated limitations on the input-output functions of these systems, which we further show can be lifted using biochemical mechanisms like promiscuous binding.  We analyze the flexibility and sharpness of decision boundaries as well as the classification capacity of these networks. Additionally, we identify distinctive signatures of networks trained for classification, including the emergence of correlated subsets of spanning trees and a creased ``energy landscape" with multiple basins. Our findings have implications for understanding and designing physical computing systems in both biological and synthetic chemical settings.
\end{abstract}

\maketitle


\begin{twocolumngrid}
\section{Introduction}
The complexity of biochemical interaction networks can be tamed by identifying network motifs and determining the computations (i.e., information processing functions) that they perform \cite{bray1995protein, bray2009wetware, alon2019introduction, avanzini2023circuit}. 
One of the most prevalent categories of such computations is a form of classification \cite{murugan2015multifarious, evans2024pattern, antebi2017combinatorial, klumpe2023computational, parres2023principles}, in which a space of inputs leads to a set of discrete states.  For example, viewing kinase activity as an input, the Goldbeter-Koshland push-pull circuit \cite{goldbeter1981amplified} implements a sigmoidal transition between binary states (Figure \ref{MarkovDefinition}A,B). Phase separation in the cell can similarly lead to sharp boundaries in the space of molecular concentrations \cite{yoo2019cellular}. Additionally, the ``glycan code'' can be viewed as a much richer encoding of cell state, with enzyme activities in the Golgi apparatus acting as inputs (Figure \ref{MarkovDefinition}C,D) \cite{varki2015essentials, yadav2022glycan, jaiman2020golgi}.  These biochemical systems draw decision boundaries through their input spaces, demarcating them into regions which map to discrete states (classes).  Training of these systems presumably occurs over evolutionary time to yield sets of kinetic rates and chemical conditions that allow them to perform their computational tasks precisely.

Previous works have studied aspects of computation in physical networks \cite{stern2021supervised, anisetti2023learning, dillavou2022demonstration},  specific chemical model systems \cite{evans2024pattern, hjelmfelt1991chemical,magnasco1997chemical,chen2014deterministic,chan2022computational, lakin2023design}, and notably in promiscuously interacting molecular networks at equilibrium \cite{murugan2015multifarious, evans2024pattern, poole2017chemical, zhong2017associative, su2022ligand, parres2023principles}.   Although such studies have illustrated an analogy between neural networks and biochemical networks, it is currently not clear what constraints on the amount of information that can be encoded (i.e., the expressivity) is introduced through the use of molecular activities as representations.  To our knowledge, a general investigation of classification ability of non-equilibrium biological processes has not been carried out.  

In this work, we use tools developed to describe far-from-equilibrium systems to investigate this central and important question. Our main results reveal strong and surprising limits on the ability of non-equilibrium biological systems to perform classification tasks. These constraints are derived from a new class of non-equilibrium response limitations reported by some of us in Ref.~\citenum{floyd2024learning}. Importantly, we also show how these constraints can be systematically lifted using biochemical mechanisms like promiscuous binding. Together, our work lays down fundamental design principles that predict the minimal necessary ingredients needed for a biological system to perform complex computational tasks. 

\begin{figure*}[ht!]
\begin{center}
\includegraphics[width=\textwidth]{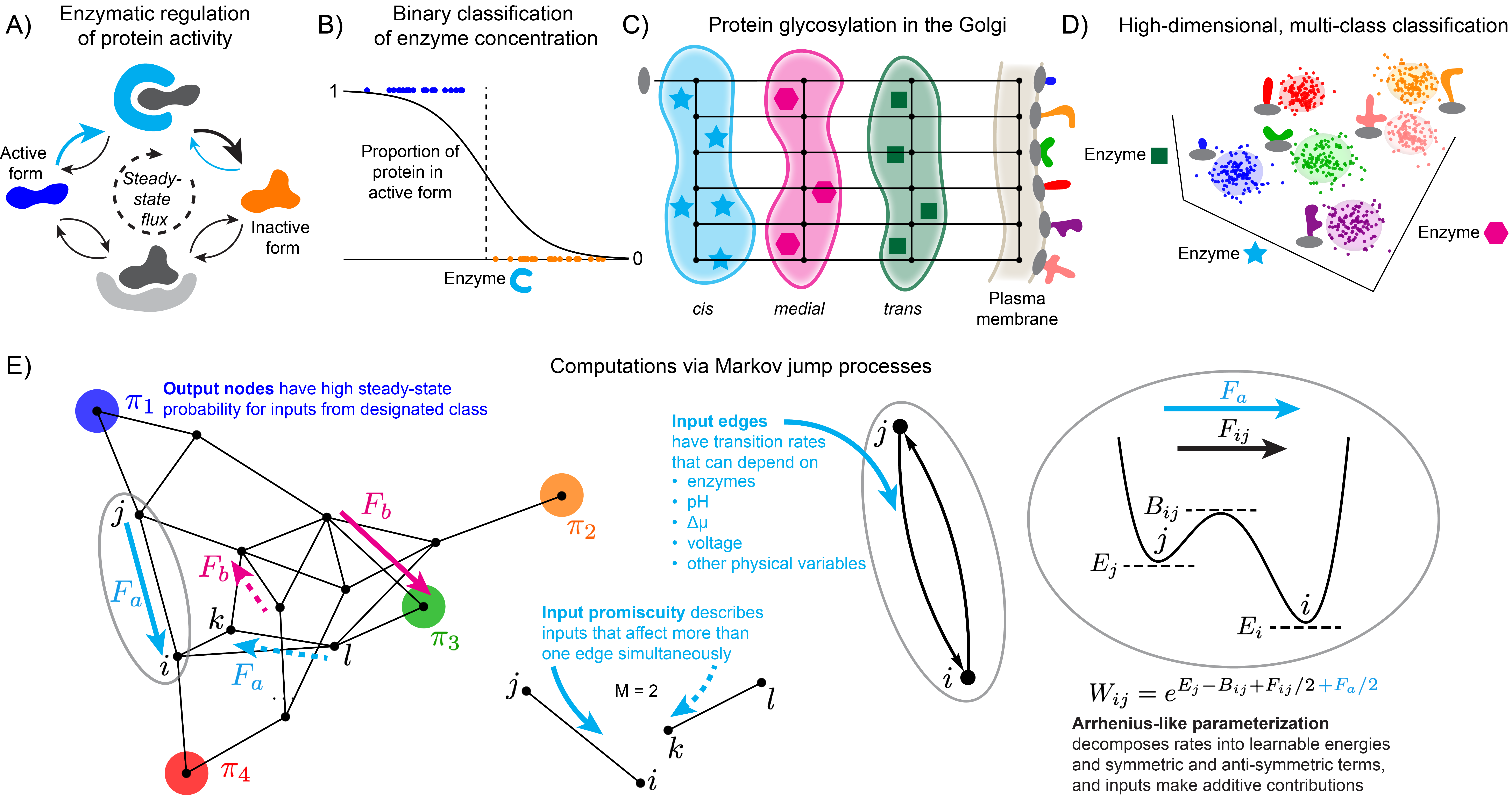}
\caption{Classification tasks performed by biochemical networks.  A)  The push-pull circuit of enzyme activation.  The input here is the activity of activating enzyme, shown in cyan, which affects the colored transition rates in the corresponding Markov network.  B)  Schematic graph of the binary (active vs.\ inactive) classification task, which computes a soft threshold on the activity of activating enzyme.  Colored points represent desired outputs, which is approximated by the learned function shown in black.  C)  Schematized representation of the process of protein glycosylation in the Golgi apparatus, adapted from the model in Ref.\ \citenum{yadav2022glycan}.  Proteins shown as gray ellipses traverse through many cisternae, and the state of the cell dictates the set of glycosyltransferase enzymes found in each cisternae and, in turn, the sugars attached to the proteins.  A decorated protein ends up in one of many distinct glycan forms on the plasma membrane, where it serves as an encoding of the cell state.  D)  Schematic graph of how protein glycosylation yields many output states which cluster based on the set of enzymes in the Golgi cisternae.  The colors of data points represent the output glycan identities at a given point in enzyme space, and the colored ellipsoids represent decision boundaries which approximately achieve this desired classification.  E) Drawing of a random Markov graph with $15$ nodes and $25$ edges.  The output nodes are labeled, and the input forces (with positive orientation) are drawn labeled with arrows.  In classification tasks using this network the solid arrows are always used as inputs, and the dashed arrows are used when $M=2$.  Input edge driving, input promiscuity, and Arrhenius-like parameterization of the edge rates are illustrated. }
\label{MarkovDefinition}
\end{center}
\end{figure*}

\section{Limit on expressivity of binary classification from non-equilibrium thermodynamics}

Cells are frequently required to make binary decisions that require integrating from many different input signals.  Examples include decisions made in processes such as chemotaxis, transcription regulation in response to heat shock, quorum sensing, and many others~\cite{bray1995protein,bray2009wetware}.  These decisions are made using networks of biochemical components based on complex combinations of input signals from the environment.  Can these biochemical networks compute arbitrarily complicated functions of their input signals, or, if not, what ingredients are needed to allow for more complicated decision making?  

To address this question, we work with a general mesoscale Markov state characterization of biological processes (Figure \ref{MarkovDefinition}E). Nodes in the Markov network are coarse representations of the state of the system. Edges encode rates of transitions between the states and can be functions of, for example, temperature, pH, enzyme activities, and chemical potential gradients.  This class of physical models is commonly used to represent kinetic schemes of chemical reaction networks \cite{owen2020universal, owen2023size, fernandes2023topologically, aslyamov2024nonequilibrium, harunari2024mutual, mahdavi2023flexibility, liang2024thermodynamic, arunachalam2023thermodynamic}.  We model inputs to the system as modulating the rates along designated edges of the Markov state network. The output is encoded in the steady-state properties of the network, specifically the occupancy of a few designated output nodes.  Our main results rely on non-equilibrium thermodynamic descriptions of the steady state and its response to perturbations, and we obtain several general limits on how effectively the Markov state networks can classify inputs and how sharp the decision boundaries drawn by this physical system can be \cite{schnakenberg1976network, owen2020universal}.  We describe this effectiveness with the term expressivity, referring to the notion in machine learning of a model's ability to account for and represent complex features in a dataset \cite{hart2000pattern}.

A Markov jump process can be represented by a graph with $N_\text{n}$ nodes and $N_\text{e}$ edges and a probability vector $\mathbf{p}(t)$ over this set of nodes.  The rate of jumping from node $j$ to $i$ is denoted $W_{ij} = e^{E_j - B_{ij} + F_{ij}/2 + F_a/2}$, where $E_j$, $B_{ij} = B_{ji}$, and the non-equilibrium forces $F_{ij} = - F_{ji}$ are learnable parameters (Figure \ref{MarkovDefinition}E).  We add an input $F_a$ to the value of $F_{ij}$ if edge $ij$ has been assigned as an input edge.  We represent the input variables as a $D$-dimensional vector $\mathbf{F}$, and we represent the $N_\text{n} + 2 N_\text{e}$ learnable parameters $\{E_j\}_{j=1}^{N_\text{n}}\cup \{B_{ij},F_{ij}\}_{ij \in \mathcal{E}}$, with $\mathcal{E}$ the set of edges, as a vector $\bt$ (see the Methods for physical interpretation of these parameters).  Under the master equation dynamics $\dot{\mathbf{p}}(t) = \mathbf{W}(\mathbf{F};\bt)\mathbf{p}(t)$, we view the steady state $\bp(\mathbf{F};\bt) \equiv \lim_{t \rightarrow \infty} \mathbf{p}(t)$ as performing a parameterized computation on the inputs $\mathbf{F}$.  Specifically, we use a one-hot encoding in which the values $\pi_\rho$ at designated output nodes should be near $1$ when inputs $\mathbf{F}^\rho$ from the corresponding class $\rho$ are presented.  

\end{twocolumngrid}
\begin{onecolumngrid}
\begin{tcolorbox}[title=Box 1: Computational expressivity from the matrix-tree theorem,  halign title=flush center, width=\textwidth, fontupper=\small]
    \begin{center}
    \includegraphics[width=0.95\textwidth]{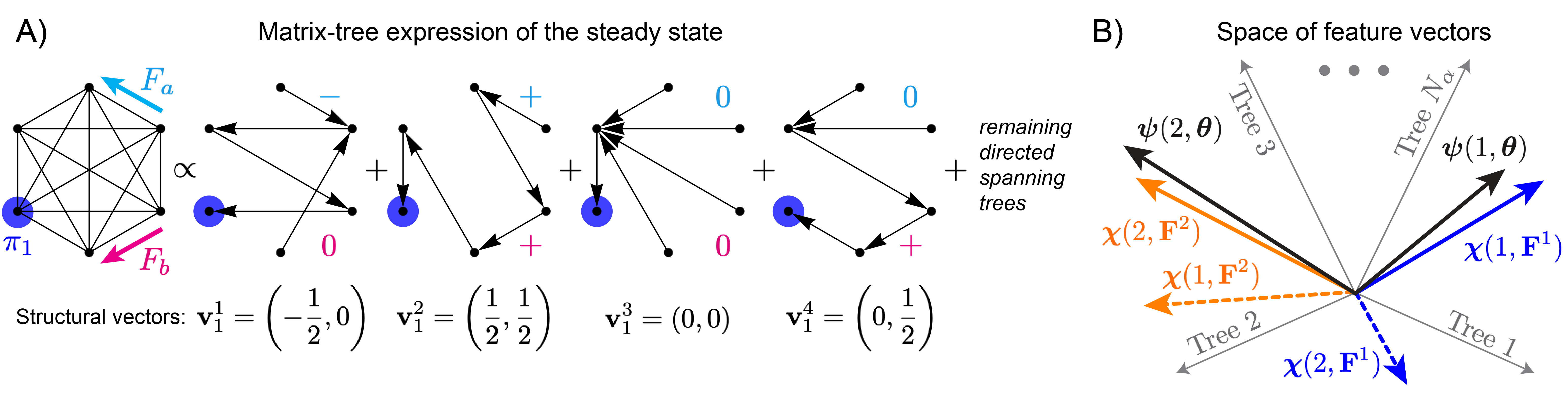}
    \captionof{figure}{The matrix-tree theorem. A)  Computing the steady-state occupancy $\pi_1$ by summing directed tree weights over spanning trees.  In each directed tree, the input forces make a positive, negative, or zero contribution to the tree weight.  The structural vectors $\mathbf{v}^\alpha_1$ are shown below each tree.  B)  Schematic illustration of the high-dimensional space of feature vectors $\bps(i;\bt)$ and $\bch(i,\mathbf{F})$.  The depicted arrangement of vectors could solve a binary classification problem.  }
    \label{MatrixTree}
    \end{center}
    
    \begin{minipage}[t]{0.48\textwidth} 
    
    \ \ \ \ \ To study expressivity analytically, we use the matrix-tree theorem expression for the steady-state probability $\bp(\mathbf{F};\bt)$ \cite{schnakenberg1976network, hill2012free}:
    \begin{equation}
        \pi_i(\mathbf{F};\bt) = \frac{\sum_{T^\alpha\in\mathcal{T}} w(T^\alpha_i, \mathbf{F};\bt)}{\sum_k \sum_{T^\alpha\in\mathcal{T}} w(T^\alpha_k, \mathbf{F};\bt)} \label{eqmtt0} \\
    \end{equation}
    Here, $\mathcal{T}$ represents the set of $N_\alpha$ spanning trees, $T_i^\alpha$ represents the $\alpha^\text{th}$ spanning tree whose edges have been directed to point toward node $i$ as a root, and the directed tree weight $w(T_i^\alpha)$ represents the product of all rate matrix elements $W_{lm}$ corresponding to the directed edges $l\leftarrow m$ in $T_i^\alpha$ (Figure \ref{MatrixTree}A).  This formula thus constructs the steady state for node $i$ by summing over all possible kinetic pathways into node $i$ and then normalizing with respect to all nodes.
    
    \ \ \ \ \ In the Supplementary Material we show how to rewrite Equation \ref{eqmtt0} as 
    \begin{equation}
        \pi_i(\mathbf{F};\bt) =\frac{\bps(i;\bt)*\bch(i,\mathbf{F})}{\sum_k \bps(k;\bt)*\bch(k,\mathbf{F})}. \label{eqmtt}
    \end{equation}
    We interpret $\bps(i;\bt)$ as a learnable feature vector with elements $\psi_\alpha(i;\bt) = e^{\mathbf{u}_i^\alpha \cdot \bt} > 0$ corresponding to the trees $T_i^\alpha$; similarly,  $\bch(i,\mathbf{F})$ is an input feature vector with elements $\chi_\alpha(i,\mathbf{F})  = e^{\mathbf{v}_i^\alpha \cdot \mathbf{F}} >0$.  The operation $*$ is a dot product over trees. The structural vectors $\mathbf{u}_i^\alpha \in \mathbb{R}^{N_\text{n} + 2N_\text{e}}$ encode the topology, i.e., which elements of $\bt$ enter exponentially into the tree weights $w(T_i^\alpha)$ and their signs; $\mathbf{v}_i^\alpha \in \mathbb{R}^{D}$  records similar information for $\mathbf{F}$.  The learnable feature vector $\bps(i;\bt)$ is therefore a non-linear encoding of the parameters $\bt$, while the input feature vector $\bch(i,\mathbf{F})$  is a non-linear encoding of the input force $\mathbf{F}$.  The goal of
    \end{minipage}
    \hfill 
    \begin{minipage}[t]{0.48\textwidth} 
    training is to adjust $\bt$ so that when $\mathbf{F}^\rho$ is drawn from the class assigned to node $\rho$, $\bps(\rho;\bt)$ has a larger overlap (dot product) with $\bch(\rho,\mathbf{F}^\rho)$ than any other $\bps(\rho';\bt)$ has with $\bch(\rho',\mathbf{F}^\rho)$ for $\rho'\neq \rho$ (Figure \ref{MatrixTree}B).

    \ \ \ \ \ An equivalent formulation of Equation \ref{eqmtt} shows that steady states of Markov jump processes implement a rational polynomial function of exponentiated input variables.  Defining $y_a \equiv e^{F_a/2} > 0$, we rewrite the matrix-tree expression for $\pi_i$ as 
    \begin{equation}
    \pi_i(\mathbf{F};\bt)=\frac{\sum_{\mu}\zeta^i_\mu(\bt)y^\mu(\mathbf{F})}{\sum_{\mu}\bar{\zeta}_\mu(\bt)y^\mu(\mathbf{F})} \label{eqmttzeta}.
    \end{equation}
    We use multi-index notation for $\mu=\{\mu_a\}_{a \in \mathcal{A}}$ where $\mathcal{A}$ is the set of $D$ input labels, and we use the notation $y^\mu \equiv \prod_{a\in\mathcal{A}}y_a^{\mu_a}$. Each $\mu_a$ runs over the values $\{-M, -(M-1), \ldots, M-1, M\}$, where the input promiscuity $M$ is the number of edges affected per input.  The coefficients in the denominator are defined as $\bar{\zeta}_{\mu}(\bt) \equiv \sum_{k=1}^{N_\text{n}}\zeta^k_{\mu}(\bt)$. The $(2M+1)^D$ monomials $y^\mu(\mathbf{F})$ in Equation \ref{eqmttzeta} combinatorically depend on the different mixtures $\mu$ of input driving, representing a net total $\mu_a$ of signed contributions from the input force $F_a$, $\mu_b$ such contributions for $F_b$, and so on for each input.  The coefficients $\zeta^i_\mu(\bt)$, which are functions of the parameters $\bt$, are the sums of directed tree weights over all trees rooted at node $i$ which have the corresponding mixture $\mu$ of signed input contributions.  The monomial coefficients $\zeta^i_\mu(\bt)$ thus represent learnable amplitudes of each polynomial ``basis function'' $y^\mu(\mathbf{F})$.  Classification will be successful if, for $\mathbf{F}^\rho$ drawn from class $\rho$, the coefficients $\zeta^\rho_\mu(\bt)$ and monomials $y^\mu(\mathbf{F}^\rho)$ are large for the same $\mu$.

\end{minipage}
\end{tcolorbox}
\end{onecolumngrid}
\begin{twocolumngrid}

\begin{figure*}[ht!]
\begin{center}
\includegraphics[width=\textwidth]{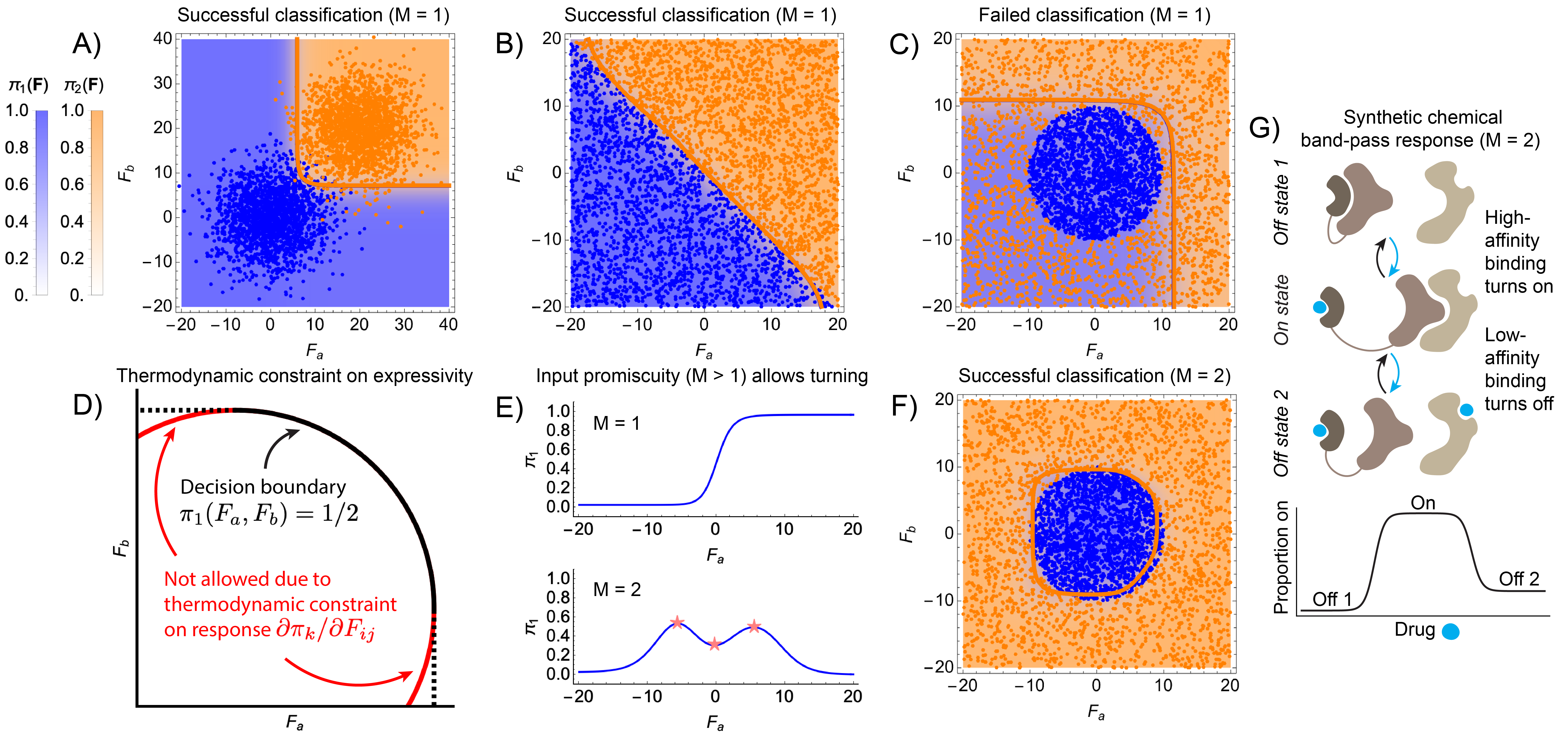}
\caption{Overcoming inflexible decision boundaries by increasing the input promiscuity hyperparameter $M$.  A)  Plot of the learned classification functions $\pi_1(\mathbf{F})$ and $\pi_2(\mathbf{F})$ shown as colored density plots over the input force space.  On top of this, scatter plots show the dataset, colored by assigned class, which was used to train the network.   Solid lines show the contour $\pi_1(\mathbf{F}) = 1/2$ in blue and $\pi_2(\mathbf{F}) =1/2$ in orange; note that these are overlapping.  The network shown in Figure \ref{MarkovDefinition}E is used for all classification tasks in this figure.  B,C)  Same as panel A, but for different classification tasks.  D)  Schematic illustration of the monotonicity constraint.  E)  Plots illustrating that increasing $M$ from 1 to 2 allows for non-monotonic dependence of a steady-state occupation on an input driving force.  F)  Same as panel C, but for the network in Figure \ref{MarkovDefinition}E which also includes driving along the dashed arrows ($M=2$).  G)  Schematic illustration of a recently designed synthetic chemical band-pass system using promiscuous input binding \cite{shui2024protein}.  A drug binds through a high-affinity pathway to activate a protein and through a second, low-affinity pathway to deactivate the protein, leading to a non-monotonic dependence of activation on the drug. }
\label{BinaryClass}
\end{center}
\end{figure*}

In Box 1 we describe an analytical formula for $\bp(\mathbf{F};\bt)$ based on the matrix-tree theorem \cite{schnakenberg1976network, hill2012free}.  We show how it can be formulated in several equivalent ways: as a linear attention-like function with non-linear learnable feature vectors $\boldsymbol{\psi}(i;\bt)$ and input feature vectors $\boldsymbol{\chi}(i,\mathbf{F})$, and as a rational polynomial with learnable scalar coefficients $\zeta^i_\mu(\bt)$.  In the Methods we also describe how structural incompatibility, which we define as a mismatch between the choice of assigned input edges and output nodes, prevents solving certain classification problems.  Reassigning the input edges and output nodes in the same graph can circumvent this issue, and in this paper we only consider structurally compatible problems in order to focus on other issues of expressivity.

As a first illustration of how physical constraints can limit expressivity of such systems, we train the network shown in Figure \ref{MarkovDefinition}E to perform a series of binary classification tasks (Figure \ref{BinaryClass}A-C).  In each case, we assign node $1$ to blue points and node $2$ to orange points, and we train the network, as described in the Methods section, to obtain a set of learned parameters $\bt^*$.  We indicate the results by drawing contours at $\pi_1(F_a,F_b;\bt^*) = 1/2$ and $\pi_2(F_a,F_b;\bt^*) = 1/2$.  In the Supplementary Material we show examples of the learned parameters in trained networks.  The network successfully classifies the points in Figures \ref{BinaryClass}A,B but not \ref{BinaryClass}C.  The failure in Figure \ref{BinaryClass}C is not a limitation of the training protocol. Rather, it emerges from a fundamental constraint on the response of non-equilibrium steady states as the forces $F_{a}$ and $F_b$ are tuned. 

Specifically, for any choice of $i,j,$ and $k$, and with other parameters held fixed, the derivative $\partial \pi_k / \partial F_{ij} $ has a fixed sign across the entire range of $F_{ij}$ \cite{floyd2024learning}; in other words $\pi_k(F_{ij})$ is a strictly monotonic function.  Thus, for fixed $F_b$, $\pi_1(F_a, F_b; \bt^*)$ must be a monotonic function of $F_a$ which implies that it can take the value $1/2$ at most once along any line drawn parallel to $F_b=0$ (Figure \ref{BinaryClass}D).  By symmetry, the function $\pi_1(F_a, F_b; \bt^*)$ must also be a single-valued function of $F_b$ along any line parallel to $F_a= 0$.  We refer to this limitation on the flexibility of the decision boundary as the monotonicity constraint, which implies that the learnable decision boundaries are not invariant to a rotation of the input space.  This corresponds to a specific failure mode of computations by non-equilibrium biophysical systems modeled as Markov jump processes. 

\section{Improving expressivity by increasing input promiscuity}
Biologically, $F_a$ can be interpreted, for example, as the logarithm of the chemostatted activity of an enzyme, which scales the pseudo first-order kinetic rate across an edge (see the Methods). In biochemical kinetics it is common for some species to be involved in multiple reactions simultaneously, making it plausible for $F_a$ to drive multiple edges \cite{owen2023size}.  We find that allowing for such promiscuous interactions of the input improves classification expressivity, and one way this happens is by lifting the monotonicity constraint.  We assume for simplicity that each of the $D$ input variables $\{F_a\}_{a \in \mathcal{A}}$, where $\mathcal{A}$ is the set of input labels, affects the same number of edges.  We denote this number of edges by the hyperparameter $M$.  Setting $M>1$ lifts the monotonicity constraint because the condition for $\pi_k(F_{ij})$ to be a monotonic function is that all other edge parameters are held fixed; with $M>1$ this is no longer true since several edge parameters change simultaneously as an input is varied.  

To better understand the gain in the decision boundary's flexibility allowed by setting $M>1$, in the Supplementary Material we analyze the steady-state representation in the rational polynomial form of the matrix-tree expression, Equation \ref{eqmttzeta}.  Considering the case $D=1$ and identifying turning points as roots of $\partial \pi_i / \partial F_a$, we show that the maximum number $R$ of such roots obeys
\begin{equation}
R = \ 
\begin{cases} 
      0 & M = 1 \\
      2M-1 & M > 1, \\
   \end{cases}
\end{equation}
which is a direct measure of the classifier's expressivity;
see Figure \ref{BinaryClass}E for an illustration.  Thus, once $M>1$, $\pi_i$ is no longer subject to the monotonicity constraint and behaves like a polynomial of degree up to $2M+1$ (but we show in the next section that additional constraints exist on the polynomial coefficients $\zeta^i_\mu(\bt)$).  In other words, input promiscuity allows the non-equilibrium biological process to be more expressive and draw out decision boundaries that can classify more complex data structures.  Indeed, returning to the previously failed classification with $M=1$ (Figure \ref{BinaryClass}C), we see that setting $M=2$ allows the same network to now learn a decision boundary which successfully encloses the data assigned to class 1 (Figure \ref{BinaryClass}F).  This implies that classifying a finite band of input signal levels (like a band-pass filter) requires setting $M>1$ along the corresponding input dimension.  A recent development in synthetic biology has in fact shown that binding drug binding to receptor molecules via two distinct binding pathways can be used to design band-pass like responses to the drug (Figure \ref{BinaryClass}G) \cite{shui2024protein}.

To quantify the binary classification ability for arbitrary $M$ and $D$, we consider the classic measure called the Vapnik-Chervonenkis (VC) dimension \cite{vapnik2013nature}.  This represents the largest number $N_{VC}$ of points which, for at least one fixed configuration of the points in the input space, a set of classifiers can correctly classify for any of the $2^{N_{VC}}$ assignments of binary labels to the points.  A theorem by Dudley \cite{dudley1984course, ben1998localization} states that if a classifier $h(\mathbf{F}) \lessgtr 0 $ belongs to a vector space $\mathcal{H}$ of real scalar-valued functions, then the VC dimension of the set of all classifiers in $\mathcal{H}$ is equal to the dimension of $\mathcal{H}$.  Given the representation of the contour $\pi_i(\mathbf{F};\bt) = 1/2$ in the rational polynomial form of the matrix-tree expression, Equation \ref{eqmttzeta}, we see that its vector space is spanned by the $(2M+1)^D$ coefficients $\zeta^i_{\mu}(\bt) - \bar{\zeta}_{\mu}(\bt)/2$. We thus estimate the VC dimension of this classifier as 
\begin{equation}
N_{VC} \leq (2M+1)^D.
\end{equation}
This should be viewed as an upper bound in two senses.  First, for $M=1$, the monotonicity constraint imposes that $N_{VC}$ is strictly less than $(2M+1)^D = 3^D$.  Second, even for $M>1$ the $N_\text{n}(2M+1)^D$ coefficients $\zeta^i_\mu(\bt)$ are not all independent degrees of freedom, as we illustrate in the next section. 

These findings suggest that enzyme promiscuity $M$ significantly increases the complexity (measured by VC dimension) of the classification tasks a biochemical circuit can perform, scaling roughly as $\sim M^D$.  Input promiscuity is a known feature of many biochemical networks: for example, many transcription factors \cite{alon2019introduction} as well as glycosyltransferases in the Golgi apparatus \cite{varki2015essentials, yadav2022glycan, jaiman2020golgi} are known to act on several targets.   Other input variables, such as temperature, voltage, or chemical potential gradients, may also affect multiple edge rates simultaneously.  Our work thus shows how input promiscuity could enable biological processes to perform more expressive computations.

\section{Storing more classes by increasing input promiscuity}

We now generalize the binary classification task and ask how many different classes can be stored as a function of the hyperparameters $M$ and $D$. Classifying many different classes is crucial in biology. For example, deciphering the glycan code, which specifies one of several hundred different cell states, or recognizing previously encountered antigens during an immune response both require making single decisions among large numbers of possibilities \cite{varki2015essentials, mayer2015well, mayer2019well}.  How these biochemical systems achieve these complex classification tasks (e.g., through microscopic sensing events like estimating antigen binding affinities) remains an important and open question. 

\begin{figure*}[ht!]
\begin{center}
\includegraphics[width=\textwidth]{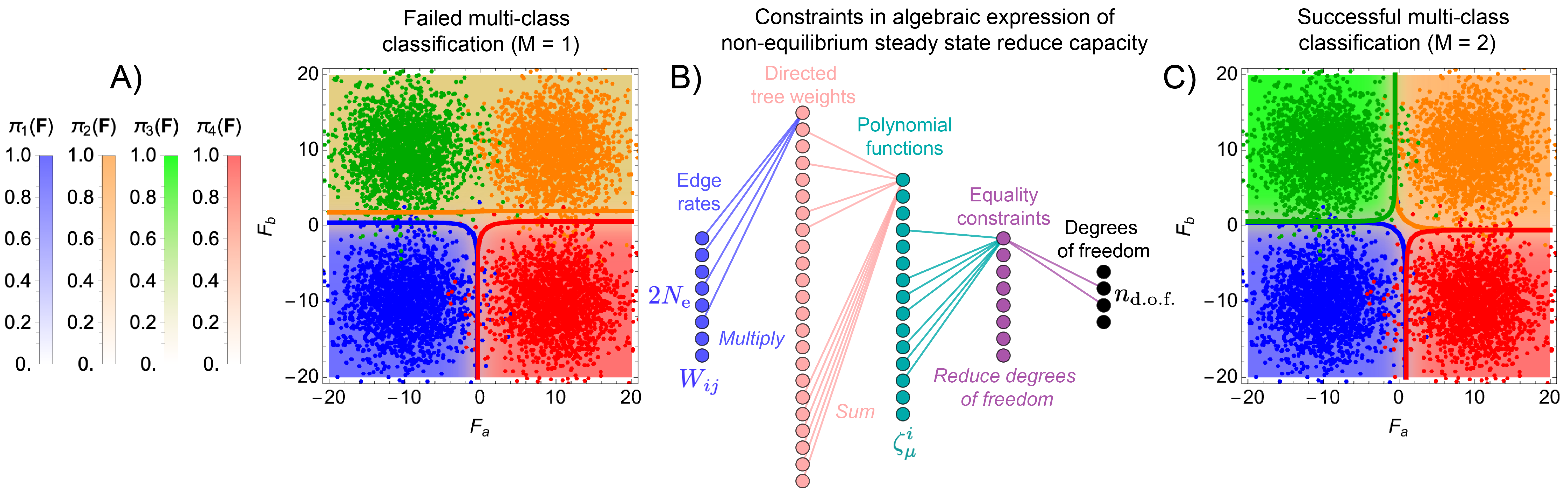}
\caption{Input promiscuity allows overcoming reduced degrees of freedom and increases multi-class capacity. 
 A)  Plot of the learned classification functions $\pi_1(\mathbf{F})$, $\pi_2(\mathbf{F})$, $\pi_3(\mathbf{F})$, and $\pi_4(\mathbf{F})$ for the network shown in Figure \ref{MarkovDefinition}E, with only the solid arrows used for inputs ($M=1$).  B)  Schematic illustration of how the learnable parameters $W_{ij}$ (the edge rates on the left) are first transformed through multiplication along directed trees into directed tree weights, then summed together to yield the polynomial functions $\zeta^i_\mu(\bt)$ which appear in Equation \ref{eqmttzeta}.  Although each function $\zeta^i_\mu(\bt)$ is uniquely defined, there exist equality constraints owing to the physics of Markov networks which reduce the effective number of degrees of freedom below the number needed to solve the four-class classification task in panel A.  C)  By including driving along the dashed arrows in Figure \ref{MarkovDefinition}E (setting $M=2$), there are sufficiently many degrees of freedom to solve the four-class classification task.  }
\label{MultiClassFrustration}
\end{center}
\end{figure*}

For $M=1$, a simple consideration suggests that the largest number of classes which can be distinguished is $2^D$.  This is because, due to the monotonicity constraint (Fig.~\ref{BinaryClass}D), each decision boundary needs to distinguish at least a half-line for $D=1$, a quadrant for $D=2$, an octant for $D=3$, and so on.  In the best case problem definition each data cloud will occupy a different one of these $2^D$ regions so that they may all be distinguished.  In Figure \ref{MultiClassFrustration}A we attempt to distinguish four classes by placing them in the four quadrants of the $(F_a, F_b)$ plane, but we find, contrary to expectation, that no network can separate them all with $M=1$.  

This limitation on expressivity results again from constraints implicit in the parametric form of the non-equilibrium steady state. Specifically, the $N_\text{n}(2M+1)^D$ quantities $\{\zeta^i_\mu(\bt)\}_{i=1}^{N_\text{n}}$ in the rational polynomial form of the matrix-tree expression, Equation \ref{eqmttzeta}, are complicated polynomial functions of the edge rates $\{W_{ij}\}_{ij \in \mathcal{E}}$, and there exist equality constraints which relate these functions to each other (Figure \ref{MultiClassFrustration}B) \cite{floyd2024learning}.  These constraints reduce the number of degrees of freedom below the number of conditions which must be satisfied to put four classes in four separate quadrants.

In the Supplementary Material we count constraints to show that the maximum number of degrees of freedom $n_\text{d.o.f.}$ among the $3N_\text{n}$ functions $\{\zeta^i_\mu\}_{i=1}^{N_\text{n}}$ for $M=1, \ D=1$ is
\begin{equation}
    n_\text{d.o.f.} = 2 N_\text{n}. \label{eqndof2}
\end{equation}
We also show that the maximum number of degrees of freedom among the $9N_\text{n}$ functions $\{\zeta^i_\mu(\bt)\}_{i=1}^{N_\text{n}}$ for $M=1, \ D=2$ is
\begin{equation}
    n_\text{d.o.f.} = \text{min}(2 N_\text{e}, 3 N_\text{n}). \label{eqndof3}
\end{equation}
The minimum is taken because the number of degrees of freedom cannot exceed the number of edge rates, and for sparse graphs it may be the case that $2N_\text{e} < 3 N_\text{n}$.


To see how this reduced $n_\text{d.o.f.}$ prevents solving certain classification problems, in the Supplementary Material we consider inequality conditions among the functions $\{\zeta^i_\mu(\bt)\}_{i=1}^{N_\text{n}}$ which must be satisfied if the steady-state probabilities of the output nodes are to be the largest among all nodes in their respective regions of the input space.  For $M=1, \ D=1$, the number of such conditions scales as $2N_\text{n}$, and $n_\text{d.o.f.}$ also scales as $2N_\text{n}$.  For $M=1, \ D=2$, the number of such conditions scales as $4N_\text{n}$, while the maximum $n_\text{d.o.f.}$ scales as $3N_\text{n}$.  We thus expect that two-class classification is possible for $M=1, \ D=1$, but four-class classification is impossible for $M=1, \ D=2$.  In the Supplementary Material we support this basic argument with a more detailed analysis.  The reduced number of degrees of freedom from equality constraints among the coefficients of the classifier function represents an additional failure mode of classification in Markov networks. 
 
As in the case of binary classification, we find that setting $M>1$ allows for greater expressivity.  With $M=2$ there are enough functions $\zeta^i_\mu(\bt)$ that, even with equality constraints, $n_\text{d.o.f.}$ tend to saturate the limit of $2 N_\text{e}$.  This provides enough degrees of freedom to successfully separate the four classes (Figure \ref{MultiClassFrustration}C).  This result suggests that input promiscuity may be a key component for the remarkable feats of multi-class classification used in biological processes like adaptive immunity or deciphering the glycan code. 

\section{Expressivity requires non-equilibrium driving}
A hallmark of biophysical processes is that they are sustained far from thermodynamic equilibrium through continual consumption of chemical free energy.  We now explain how this feature is a necessary ingredient for some of the aforementioned computational abilities.  In the absence of any non-equilibrium driving, either through the learned parameters $F_{ij}$ or the input variables $F_a$, the steady-state distribution is a Boltzmann form $\pi_i \propto e^{-E_i}$ and does not depend on the $B_{ij}$ parameters.  Beating this restrictive functional form and achieving non-trivial classification expressivity thus requires non-equilibrium driving.  In the Supplemental Material, we use the linear attention-like form of the matrix-tree expression, Equation \ref{eqmtt}, to show how the non-equilibrium parameters $F_{ij}$ allow for the greatest flexibility in positioning the learnable feature vectors $\bps(i;\bt)$, thereby enabling expressive computations. 

To demonstrate this numerically, we measure how classification accuracy depends on the amount of allowed non-equilibrium driving.  To do this we consider an input modality in which input variables $B_a$ present additive contributions to the $B_{ij}$ parameters along input edges rather than the $F_{ij}$ parameters along those edges.  In this way, the only non-equilibrium driving in the system comes from the learned $F_{ij}$ parameters which we then constrain in magnitude.  We train the network in Figure \ref{MarkovDefinition}E for the classification task shown in Figure \ref{BinaryClass}A for several values of $F_\text{max}$, which we impose during training as a ceiling on the absolute value of any learned $F_{ij}$ parameter.  In Figure \ref{NonEqDriving} we plot the classification accuracy of the trained networks, showing a continuous increase in performance as a function of $F_\text{max}$.  This implies that under the linear dynamics of Markov jump processes it is necessary to break detailed balance to perform non-trivial computations, but we note that if the chemical dynamics are non-linear then computations can be expressive even in equilibrium conditions \cite{parres2023principles, klumpe2023computational}.  

An additional perspective on non-equilibrium Markov networks trained for classification can be gained from recent work exploring the analogies between transformers, which implement softmax-based filters over key, query, and value vectors, and models of dense Hopfield networks, in which relaxational dynamics in the landscape of a softmax-like energy function allows for storage of a far greater number of associative memories compared with the usual quadratic Hopfield energy functions \cite{ramsauer2020hopfield, lucibello2024exponential}. The linear attention-like form of the matrix-tree expression, Equation \ref{eqmtt}, motivates consideration of the corresponding ``energy'' function that describes Markov steady states, which we identify in analogy with Ref.\ \citenum{lucibello2024exponential} as $\mathcal{F} \equiv - \ln \sum_{k} \bps(k;\bt) * \bch(k;\mathbf{F})$, where $\bps(k;\bt)$ and $\bch(k;\mathbf{F})$ are the non-linear feature vectors in the attention function.  Plotting this function over the input space $(F_a, F_b)$ in trained and untrained graphs reveals a landscape characterized by flat, sloping basins delimited by ``creases'' of high curvature.  In trained graphs some of these creases co-localize with the learned decision boundaries, separating the input space into regions in which different subsets of trees dominate the contribution to $\nabla_\mathbf{F} \mathcal{F}$.  Some of these creases also represent topological features of the graph which cannot be removed by training, such as mismatched limits (e.g., $F_a \rightarrow \infty$ followed by $F_b \rightarrow \infty$ or vice versa) at the corners of the input domain.  In the Supplementary Material we expand on this analogy between the steady states of Markov processes and dense associative memories, but we leave a full treatment of this connection to future work.  

We also show in the Supplementary Material that in networks trained for classification, the trees with contribute to the steady states of the output nodes are more localized than in untrained networks.  We also find that the subset of trees dominating the steady state are clustered together, and these clusters correspond to the class that the input driving is drawn from.  This suggests that the steady states of evolved reaction networks are far from uniform and contain relatively few subsets of pathways which dominate the network flow. 

\begin{figure}[ht!]
\begin{center}
\includegraphics[width=0.8 \columnwidth]{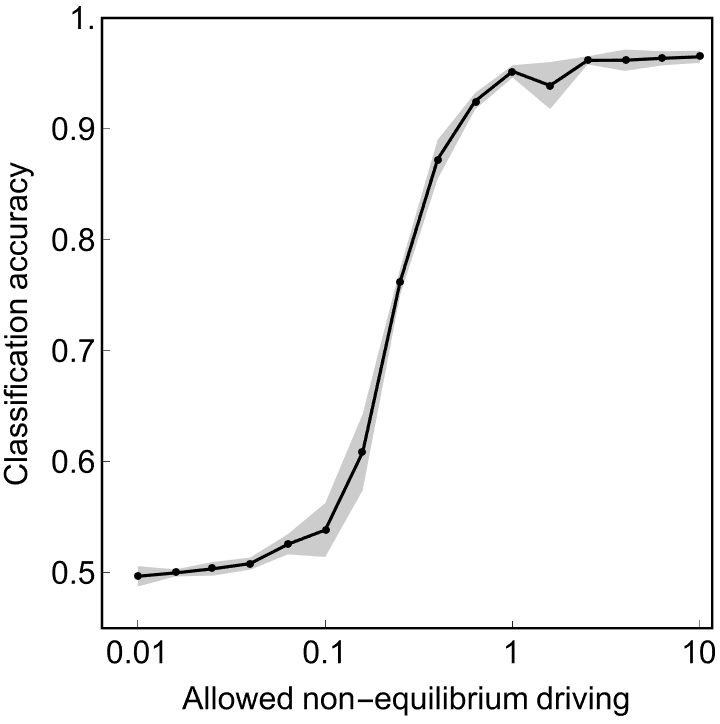}
\caption{Accurate classification requires non-equilibrium driving.  This plot corresponds to the classification task in Figure \ref{BinaryClass}A using $B_a$ and $B_b$ as inputs instead of $F_a$ and $F_b$.  The classification accuracy, defined as the average of $\pi_\rho(\mathbf{F}^\rho;\bt)$ over $10^3$ randomly drawn samples of $\mathbf{F}^\rho$ from classes $\rho = 1,2$, is shown as a function of $F_\text{max}$, the maximum absolute value of $F_{ij}$ that is allowed on any edge.  Five training trials for each value of $F_\text{max}$ were performed, and the gray area illustrates the standard deviation of accuracy over these trials.}
\label{NonEqDriving}
\end{center}
\end{figure}

\begin{figure*}[ht!]
\begin{center}
\includegraphics[width=0.9\textwidth]{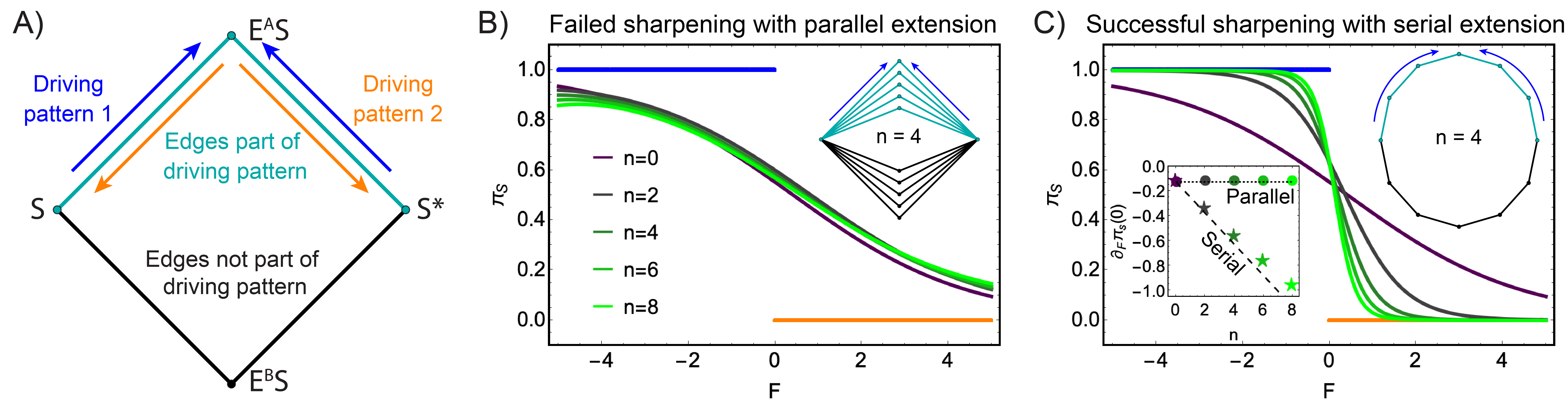}
\caption{Sharpening decision boundaries for a biochemical motif. A) Labeled illustration of the four-species push-pull network, with driving patterns drawn as blue and orange arrows.  B,C) Plots of the trained $\pi_S(F;\bt)$ curves for increasing nodes in the parallel (B) and serial (C) extension.  The inset shows the slopes $\partial_F \pi_S(F=0)$ for the serial and parallel extensions as a function of $n$.  The dashed and dotted lines show the bound obtained from $M_R^\mathcal{O}$ for the enumerated trees in serial and parallel extensions. 
}
\label{Sharpness}
\end{center}
\end{figure*}

\section{Network topology dictates sharpness of the decision boundary}
The flexibility of the decision boundary may be distinguished from another important feature relevant for biological discrimination, which is the boundary's sharpness.  We can quantify this as the norm of the gradient $\nabla_\mathbf{F} \pi_i(\mathbf{F};\bt)$ evaluated at a location separating the discrimination regimes.  Sharpness and the related measure of steady-state amplification are topics which have received much research focus, particularly in models of cooperative binding, cellular sensing, ultrasensitive covalent modifications, and kinetic proofreading \cite{goldbeter1981amplified, murugan2012speed, murugan2014discriminatory, owen2023size, owen2020universal, fernandes2023topologically,  owen2023thermodynamic, lammers2023competing, aslyamov2024nonequilibrium, arunachalam2023thermodynamic, liang2024thermodynamic}.  A consensus among these works is that greater sharpness requires greater expenditure of chemical free energy; this idea is often expressed in the form of inequalities reminiscent of the thermodynamic uncertainty relations \cite{barato2015thermodynamic, gingrich2016dissipation, horowitz2020thermodynamic}.  Here, we extend this line of research by explicitly framing it within the context of a computational classification problem. We demonstrate systematic methods to sharpen the decision boundary by increasing the number of input-driven transitions that occur serially along a reaction pathway (multiple forms of an intermediate molecule), rather than through parallel reaction pathways (multiple different intermediate molecules).  These serially driven transitions tend to yield directed spanning trees with greater net input driving.  Serially driven edges occur biologically, for instance, in common models of how ligands drive cooperative binding reactions and how ATP or GTP drives multi-step kinetic proofreading \cite{alon2019introduction, owen2020universal}.  Parallely driven edges occur in some types of general enzymatic schemes \cite{owen2020universal, owen2023thermodynamic, gunawardena2012linear}, and both kinds of driven edges occur in common models of the flagellar motor \cite{owen2023size}.  Additionally, we show that it is important to consider not only the extremes of the trees' net input driving but also the multiplicity of trees with intermediate net input driving, as a large number of such intermediate trees can reduce sharpness.  

For simplicity we study sharpness with $D=1$ but note that the argument extends straightforwardly to $D>1$, because to compute the gradient norm $\nabla_\mathbf{F} \pi_i(\mathbf{F};\bt)$ we merely sum the one-dimensional terms $(\partial \pi_i / \partial F_a)^2$ for $a\in \mathcal{A}$.  We first consider a classic biological motif, the Goldbeter-Koshland push-pull network (Figure \ref{Sharpness}A), in which a substrate is shuffled between a non-phosphorylated ($S$) 
and phosphorylated ($S^*$) form by competing kinase ($E^A$) and phosphatase ($E^B$) enzymes.  Our input is the chemostatted activity of kinase, which affects the transition rates $E^AS \leftarrow S$ and $E^AS \leftarrow S^*$ by the same factor $F$.  

We train this network to classify inputs $F\in(-5,0)$ with high $\pi_S$ probability and inputs with $F\in(0,5)$ with high $\pi_{S^*}$ probability, with results shown in Figure \ref{Sharpness}B.  The learned $\pi_S(F;\bt)$ curve has the right qualitative features but is not very sharp.  To sharpen this transition, we consider systematically adding driven edges (increasing $M$) in one of two ways, either in parallel (Figure \ref{Sharpness}B) or in serial (Figure \ref{Sharpness}C) with the original driven edges.  Training each of these networks with increasing numbers $n$ of additional pairs of driven edges (also adding undriven edges on the bottom for symmetry), we see that adding edges in parallel fails to sharpen the transition, while adding edges in serial succeeds.  

To explain this difference, in the Supplementary Material we use the $1$D rational polynomial form of the matrix-tree expression, Equation \ref{eqmtt}, to maximize $\partial \pi_S / \partial F$ with respect to the learnable coefficients $\zeta^S_m$ and $\bar{\zeta}_m$ (treated for now as free and independent).  The multi-index $\mu$ used in Equation \ref{eqmttzeta} simplifies here to the single index $m$.  We show that
\begin{equation}
    \text{max}_{\{\zeta^S_m\},\{\bar{\zeta}_m\}}\left |\frac{\partial \pi_S(F;\bt)}{\partial F}\right| =\frac{M_R}{8} \label{eqsharpbound}
\end{equation}
where $M_R = M_\text{max} - M_\text{min}$ is the range in exponential powers of $e^{F/2}$ among all directed spanning trees.  A tighter approximation to the bound can be obtained by replacing $M_R$ with $M_R^\mathcal{O} \leq M_R$, which is the range in exponential powers among only the directed spanning trees rooted on the output nodes.  We further explain in the Supplementary Material that the directed spanning trees of the parallelly extended push-pull networks prevent $M^\mathcal{O}_R$ from scaling with $n$, whereas the spanning trees for serially extended networks allow $M^\mathcal{O}_R \sim n$, which enables increasingly sharp transitions as more edges are added. 

That the sharpness is bounded by the maximal net number of driven edges encountered among directed spanning trees is resonant with (though technically distinct from) recent results in Refs.\ \citenum{owen2020universal, owen2023size, aslyamov2024nonequilibrium, arunachalam2023thermodynamic, liang2024thermodynamic}.  Saturating the bound requires that  trees with the greatest net contributions from the input driving ($m=M_\text{min}$ or $m=M_\text{max}$) have the largest coefficients $\zeta^S_m$.  In the Supplementary Material we illustrate that this bound fails to be tight in ladder networks with a high multiplicity of spanning trees that have intermediate contributions from the input driving ($M_\text{min} < m < M_\text{max}$).  This prevented sharpening can be seen as an alternate repercussion of having reduced degrees of freedom: as discussed in the previous section the functions $\zeta_m^S(\bt)$ in the classifier cannot be adjusted freely because they are connected to each other in complex ways via overlapping spanning trees (Figure \ref{MultiClassFrustration}B).  This implies that highly inter-connected biochemical networks, with many overlapping reaction pathways, may be unable to exhibit sharp responses.

\section{Discussion}
We have explored the computational expressivity of classifiers implemented in trained non-equilibrium biochemical networks, which we model as Markov jump processes.  An analytical solution for the steady states of these systems can be written in several equivalent ways, highlighting complementary interpretations of the classifier as computing a linear softmax operation using learnable, nonlinear feature vectors (Equation \ref{eqmtt}), and as computing a rational polynomial function with learnable scalar coefficients (Equation \ref{eqmttzeta}).  The feature vectors and coefficients are themselves complicated functions of the tree weights of the physical network, and because of this dependency they are significantly constrained relative to abstract parametric classifiers having the same functional form as the matrix-tree expression.  We identified several limitations to expressivity, including monotonic responses $\pi_k(F_{ij})$ and a reduction in degrees of freedom of the classifier function. We further showed that increasing input promiscuity (setting $M>1$) helps mitigate these limitations, by creating additional turning points of $\pi_k(F_{ij})$ and allowing the number of degrees of freedom in the graph to saturate at $2 N_\text{e}$. With even modest input promiscuity, chemical reaction-based classifiers prove to be capable of solving difficult classification tasks, demonstrating non-linear information processing performance reminiscent of neural networks \cite{bray1995protein, bray2009wetware}.  

Key biological implications follow from the sensitive dependence of computational expressivity on the input promiscuity hyperparameter $M$, which we define as the number of edges driven by a single input variable.  Biologically, $M>1$ occurs when a single input variable, such as an enzyme activity, temperature, or chemical potential gradient, simultaneously affects more than one chemical transition.  Input promiscuity in a real biochemical network may at first glance seem counterproductive, by decreasing the networks' modularity \cite{alon2019introduction}, but our results show that it serves to significantly expand networks' computational capabilities.  In the context of cooperative binding, there is also a relationship between $M$ and the Hill coefficient, which determines the sharpness of switch-like input responses \cite{owen2023size}.  We hope to connect our general findings to specific biochemical systems in the future.  

Generalizing the results of this paper beyond the particular chemical dynamics and definition of classification which we have adopted is another avenue for future work.  Although the (pseudo) first-order reaction networks modeled by Markov jump processes have less rich dynamics than non-linear chemical kinetics, there are still many biochemical systems to which the matrix-tree theorem which underlies our results can be applied \cite{owen2020universal, owen2023size}.  For example, approximations based on time-scale separations can in some cases be used to create an effective linear system out of non-linear kinetic schemes \cite{gunawardena2012linear}.  Fully general analyses based on chemical reaction network theory may be feasible in the future using recent theoretical developments \cite{feinberg2019foundations, dal2023geometry, rao2016nonequilibrium, avanzini2023circuit}.  Additionally, we adopted here the common convention used in machine learning of one-hot encoding to specify classification outcomes, but biologically it may be more realistic to specify whole profiles of chemical concentrations as computational outputs.  In previous work \cite{floyd2024learning} we showed analytically that the ratio $(\partial \pi_k / \partial F_{ij} ) / (\partial \pi_{k'} / \partial F_{ij} )$ is independent of $F_{ij}$ for any $k,k'$, which can be shown to imply that the monotonicity constraint holds under any linear mapping $\tilde{\bp} = \mathbf{R} \bp$.  Thus, the expressivity limitations identified in this work should at least hold for output profiles which are arbitrary linear transformations of one-hot encoded outputs.    

A final aspect of the physics not emphasized in this work is that chemical dynamics are inherently stochastic, and fluctuations about the steady-state mean are important, especially when copy numbers are small.  Decision-making under fluctuations has often been treated using the framework of information theory \cite{lammers2023competing, wang2020price, tkavcik2009optimizing, voliotis2014information}.  A general trend from this line of research is that maximizing information flow requires reducing fluctuations, which in turn requires greater expenditure of chemical free energy.  Some forms of chemical computation even harness this stochasticity to generatively model probability distributions \cite{poole2017chemical}.  Combining insights from these works with our results on classification expressivity can help paint a full picture of how biochemical systems use fuzzy logic to make decisions.  

\section{Methods}
 
\subsection{Network inputs}
To present an input to the Markov network we adjust the parameters along designated input edges (Figure \ref{MarkovDefinition}E).  In this paper we mostly present inputs via additive contributions $F_a$ to the $F_{ij}$ parameters along the edges assigned to input $a$, although we also consider presenting additive contributions $B_a$ to the corresponding $B_{ij}$ parameters.  What the edge inputs $F_{ij}$ and $B_{ij}$ represent physically depends on the specific model system which one has in mind, but we next elaborate on their general physical properties.  

A contribution to the anti-symmetric term $F_{ij}$ generally exists due to a broken time-reversal symmetry \cite{raz2016mimicking}, or under coarse-graining, whereby the degrees of freedom of at least two baths with a potential gradient between them are not explicitly modeled in the system dynamics \cite{peliti2021stochastic}.  A pertinent example is the chemical potential difference $\Delta \mu$ between ATP and its hydrolysis products.  The assumption that the concentrations of these species are chemostatted away from equilibrium implies that their concentrations do not enter as model variables, and they instead break detailed balance by appearing as contribution to the parameter $F_{ij} \sim \Delta \mu$ along the coarse-grained transition $i\leftarrow j$ which hydrolyzes ATP and release its products.  Transitions which couple to baths of different temperatures, voltage, or osmotic pressure could also have non-zero $F_{ij}$ parameters.  Additionally, reactions coarse-grained so that a bound state of a chemostatted species can be accessed through multiple pathways will have non-zero values of $F_{ij}$ for all transitions into the bound state.  For example, both the transitions $E^AS \leftarrow S$ and $E^A \leftarrow S^*$ in Figure \ref{Sharpness}A are driven by the same $F_a \sim \ln [E]$, where the enzyme activity $[E]$ is assumed to be held fixed during the system dynamics and controlled as an input variable.  

Contributions to $B_{ij}$ represent symmetric changes to the transition rates between two states.  We give two specific biological examples but note that many others are possible.  First, we consider a coarse-graining scheme in which the enzymatic reaction 
    $E+S \rightleftharpoons ES \rightleftharpoons E+S^*$
is assumed to be very fast relative to the other dynamics involving $S$ and $S^*$, and in which the enzyme activity is fixed.  It is then possible to show that an effective reaction $S \rightleftharpoons S^*$ has the first order rate constants
\begin{equation}
    k_{S\rightarrow S^*} = [E] \frac{k_{E+S \rightarrow ES}k_{ES \rightarrow E+S^*}}{k_{ES \rightarrow E+S^*} + k_{ES \rightarrow E+S}}
\end{equation}
and 
\begin{equation}
    k_{S^*\rightarrow S} = [E] \frac{k_{ES \rightarrow E+S}k_{E+S^* \rightarrow ES}}{k_{ES \rightarrow E+S^*} + k_{ES \rightarrow E+S}}.
\end{equation}
The ratio $k_{S\rightarrow S^*} / k_{S^*\rightarrow S}$ is independent of $[E]$, but $[E]$ symmetrically scales both rates, thus appearing as a contribution $B_{ij} \sim \ln [E]$ in the Markov network model.  We refer to Refs.\ \citenum{gunawardena2012linear, owen2023thermodynamic, nam2022linear} for additional details on how enzymatic reactions may be coarse-grained into Markovian descriptions using the so-called ``linear framework.''  Second, we consider a tension-gated ion channel in which the channel's dynamics of opening and closing are fast, and its probability of being open is a function of membrane tension which we view as an input variable.  We can take the ion concentrations on either side of the membrane as among the coarse-grained model variables, and the transition rate through the channel depends on the probability of it being open \cite{haswell2011mechanosensitive}.  This dependence symmetrically scales both directions of ion flow, so that the tension effectively modulates the $B_{ij}$ parameter along the Markovian transition from ions inside the membrane to those outside the membrane.

\subsection{Training}
We train Markov networks using an approximation to gradient descent, as discussed in Refs.\ \citenum{scellier2017equilibrium, scellier2019equivalence, stern2021supervised}.  This method requires a variational quantity $\mathcal{L}(\mathbf{p};\mathbf{F},\bt)$ which is minimized by the steady-state distribution $\bp(\mathbf{F},\bt) = \text{argmin}_{\mathbf{p}}\mathcal{L}(\mathbf{p};\mathbf{F},\bt)$.  Two considerations lead to equivalent choices of $\mathcal{L}$: one is that $\mathcal{L}$ should be the KL divergence  $\sum_k p_k \ln (p_k / \pi_k)$, which has been shown to act as a Lyapunov function for the evolution of $\mathbf{p}(t)$ and is minimized to zero at steady state \cite{schnakenberg1976network}.  The other consideration is based on the observation that, from Equation \ref{eqmtt}, $\pi_i \propto e^{-\Phi_i(\mathbf{F};\bt)}$ with $\Phi_i = -\ln\left(\sum_{T^\alpha \in \mathcal{T}}w(T_i^\alpha)\right)$ is a Boltzmann-like distribution which must hence maximize a constrained entropy functional.  We show in the Supplementary Material that both of these considerations lead to equivalent update rules which require numerically estimating the vectors $\partial_{\bt} \pi_i$ during training.

\subsection{Structural compatibility}
We find that for a given graph structure, not every classification problem (i.e., an assignment of input edges, output nodes, and sets of input data) can be solved.  For example, we may assign an input $F_{a}$ along edge $i\leftarrow j$ and assign node $j$ to be large when $F_{a} > 0$; this will be very difficult to achieve because all of the spanning trees into node $j$ which involve edge $j\leftarrow i$ will have an exponentially small contribution from the input force, making the input feature vector $\bch(j,\mathbf{F})$ in Equation \ref{eqmtt} small; this cannot be helped no matter how we learn the parameters $\bt$.  If we flipped the sign of the input $F_{a}$ in the assignment, then the problem may become structurally compatible and hence solvable.  We refer to this mismatch between input force assignment and achievable output node assignment as structural incompatibility.  For a fixed set of hyperparameters (i.e., number of nodes, edges, input edges, output nodes, etc.), some problems will be structurally compatible and some will not be.  This issue is thus separate from more intrinsic properties of computational expressivity that depend on hyperparameters like $M$ and $D$, but it implies that we cannot define classification problems completely arbitrarily.  We leave to future work a dedicated study of what determines structural compatibility, which may be posed as a determining a feasible region as in constrained optimization \cite{nocedal1999numerical}.  Throughout this paper, we will bypass this issue by assuming that we have defined structurally compatible problems to focus on other constraints on expressivity, but we note that this represents one limitation on using chemical reaction networks as classifiers.

\section*{Acknowledgements}
We wish to thank Menachem Stern, Martin Falk, Kristina Trifonova, Tarek Tohme, Aleksandra Walczak, and Thierry Mora for helpful discussions.  This work was mainly supported by the National Institute of General Medical Sciences of the NIH under Award No. R35GM147400 by funding to SV and CF. AM was funded by the National Institute of General Medical Sciences of the NIH under Award no. R35GM151211.  We acknowledge support from the National Science Foundation through the Physics Frontier Center for Living Systems (PHY-2317138).  CF acknowledges support from the University of Chicago through a Chicago Center for Theoretical Chemistry Fellowship.   
\end{twocolumngrid}

\clearpage

\singlespacing

\widetext
\begin{center}
	\textbf{\LARGE Supplementary Material}
\end{center}
\setcounter{section}{0}

\begin{onecolumngrid}

\section{Training Markov networks}
\subsection{Training algorithm}
The results in this paper are obtained by training Markov networks using an algorithm based on Refs.\ \citenum{stern2021supervised, stern2023learning, scellier2017equilibrium, scellier2019equivalence} which approximates gradient descent on a loss landscape.  Broadly, this approach involves presenting a labeled example $\mathbf{F}$ to the system, obtaining its ``free'' output $\bp(\bt)$ under its current set of learnable parameters $\bt$,  ``nudging'' the system toward the desired output $\bp'$ for that label, and then updating $\bt$ according the different outputs in the free and nudged configurations.  Throughout this section we suppress the dependence of quantities on $\mathbf{F}$.  The update rule
\begin{equation}
	\Delta \bt = \frac{\partial \mathcal{L}[\bp(\bt);\bt]}{\partial \bt} - \frac{\partial \mathcal{L}[\bp';\bt]}{\partial \bt} \label{eqtraincp1}
\end{equation}
involves a variational quantity $\mathcal{L}[\mathbf{p};\bt]$ which is minimized at the steady state $\lim_{t\rightarrow \infty}\mathbf{p}(t) = \bp$ of the network.  This algorithm has recently been used to train physical systems to act like neural networks \cite{stern2021supervised, stern2023learning}.  Typical physical systems at or near equilibrium minimize a variational quantity, such as an energy function (like the elastic energy in a spring networks) or dissipation function (like the total dissipated power in resistor networks).  A useful feature of these variational quantities is that their derivatives with respect to $\bt$ sometimes decompose into functions which are local to nodes in the network, meaning that Equation \ref{eqtraincp1} can be applied without global information; each node only requires information in from its immediate neighborhood in the network to update its value.  

We find below that while variational quantities $\mathcal{L}$ for the steady states of far-from-equilibrium Markov networks can be defined and used in Equation \ref{eqtraincp1}, their derivatives with respect to $\bt$ do not decompose exactly into local quantities.  As a result, using this approximate gradient descent technique to train Markov networks currently has no inherent advantage over exact gradient descent; both approaches require numerically estimating similar derivatives.  Despite this, we describe the set-up for using this algorithm here because it may be possible in future work to obtain approximations to the derivatives $\partial_{\bt} \mathcal{L}$ in terms of easily computable local quantities, which would allow for a highly scalable training approach conducive to large system sizes.  We note that other training approaches may be possible, such as genetic algorithms \cite{arunachalam2023thermodynamic, parres2023principles} or automatic differentiation \cite{jhaveri2024discovering}.  Rather than compare each approach, our focus in this paper is on the underlying limitations of computation by the trained networks.

\subsection{Variational quantities of Markov steady states}
We describe two variational quantities $\mathcal{L}[\mathbf{p};\bt]$ which are minimized by the steady state $\mathbf{p} = \bp(\bt)$.  The first is the Kullback-Leibler (KL) divergence between $\mathbf{p}$ and $\bp$:
\begin{equation}
	\mathcal{L}^{\text{KL}}[\mathbf{p};\bt] \equiv \sum_k p_k \ln\frac{p_k}{\pi_k(\bt)}.
\end{equation}
It can be shown that $\mathcal{L}^{\text{KL}}$ acts as a Lyapunov function under the dynamics $\dot{\mathbf{p}} = \mathbf{W} \mathbf{p}$, such that $\partial_t \mathcal{L}[\mathbf{p}(t);\bt] \leq 0$ along any approach toward steady state \cite{schnakenberg1976network}.  At steady state, $\mathcal{L}^{\text{KL}}$ is minimized to zero.   

While the variational nature of $\mathcal{L}^{\text{KL}}$ can be established through its connection to a dynamical Lyapnuov functional, it is of interest to consider an alternative route to a variational quantity through analogy to a thermodynamic functional, namely, the Helmholtz free energy.  To make this analogy, we first note that Equation \ref{eqmtt0} of the main text can be expressed in the Boltzmann-like form 
\begin{equation}
	\pi_i(\bt) = \frac{e^{-\Phi_i(\bt)}}{Z(\bt)} \label{eqpiBoltz}
\end{equation}
where 
\begin{equation}
	\Phi_i(\bt) = -\ln \sum_{T^\alpha \in \mathcal{T}} w(T^\alpha_i;\bt)
\end{equation}
is the ``non-equilibrium potential'' and 
\begin{equation}
	Z(\bt) = \sum_{k = 1}^{N_\text{n}} e^{-\Phi_k(\bt)}
\end{equation}
is the ``partition function.''  As the Boltzmann distribution minimizes the Helmholtz free energy, i.e., it maximizes the entropy subject to the constraint of normalization and an average energy, we can write a variational ``free energy'' which Equation \ref{eqpiBoltz} minimizes:
\begin{equation}
	\mathcal{L}^{\text{FE}}[\mathbf{p};\boldsymbol{\theta}] = \sum_k p_k \ln p_k + \sum_k \Phi_k(\boldsymbol{\theta}) p_k + \lambda(\bt)\left(\sum_k p_k -1 \right). \label{eqfreeenergy}
\end{equation}
The Lagrange multiplier $\lambda(\bt)$ enforces the normalization of $p_k$.  The the derivative of $\mathcal{L}^{\text{FE}}$ with respect to $p_k$ is
\begin{eqnarray}
	\frac{\partial \mathcal{L}^{\text{FE}}}{\partial p_k} = \ln p_k + 1 + \lambda(\bt) + \Phi_k(\bt).
\end{eqnarray}
The minimizer $\bp(\bt) = \text{argmin}_\mathbf{p}\mathcal{L}^{\text{FE}}[\mathbf{p};\bt]$ will satisfy $ \frac{\partial \mathcal{L}}{\partial p_k} = 0$, and it can be shown that the solution of this is Equation \ref{eqpiBoltz}, with the interpretation that $\lambda(\bt) = \ln Z(\bt) - 1$.  Inserting this, we can rewrite $\mathcal{L}$ as
\begin{eqnarray}
	\mathcal{L}^{\text{FE}} &=& \sum_k p_k \ln p_k + \sum_k p_k\left(\Phi_k(\bt) - \ln Z(\bt)\right) - \sum_k p_k +1 - \ln Z(\bt) \nonumber \\
	&=& \sum_k p_k \ln p_k + \sum_k p_k\ln\pi_k(\bt) - \sum_k p_k +1 - \ln Z(\bt) \nonumber \\
	&=& \sum_k p_k \ln\frac{p_k}{\pi_k(\bt)} - \sum_k p_k +1 - \ln Z(\bt).
\end{eqnarray}

We see that
\begin{equation}
	\mathcal{L}^{\text{FE}} = \mathcal{L}^{\text{KL}} -\sum_k p_k +1 - \ln Z(\bt),
\end{equation}
which means that $\partial_{\bt} \left(\mathcal{L}^{\text{FE}} - \mathcal{L}^{\text{KL}}\right)$ depends only on $\bt$, not $\mathbf{p}$.  As a result, the update rule in Equation \ref{eqtraincp1} leads to the same result for both of these variational quantities:
\begin{equation}
	\Delta \bt = -\sum_k \frac{\pi_k(\bt)}{\pi_k(\bt)}\frac{\partial \pi_k(\bt)}{\partial \bt} + \sum_k \frac{\pi_k'}{\pi_k(\bt)}\frac{\pi_k(\bt)}{\partial \bt} = \sum_k \frac{\pi_k'}{\pi_k(\bt)}\frac{\partial \pi_k(\bt)}{\partial \bt} \label{eqtraincp2}
\end{equation}
where the first term vanishes due to normalization of $\pi_k(\bt)$.

\subsection{Computing the update derivatives}

The derivatives $\partial_{\bt} \pi_k(\bt)$ have received recent attention \cite{owen2020universal, fernandes2023topologically, aslyamov2024nonequilibrium, floyd2024learning}.  For $\theta_m \in \{E_j\}_{j=1}^{N_\text{n}}$, the derivative is simply \cite{owen2020universal}
\begin{equation}
	\frac{\partial \pi_k(\bt)}{\partial E_j} = \begin{cases} 
		-\pi_k(1-\pi_k) & \text{if}\ j=k \\
		\pi_k\pi_j & \text{if}\ j\neq k.
	\end{cases}
\end{equation}
Inserting this into Equation \ref{eqtraincp2} gives, after simplification, the local update rule
\begin{equation}
	\Delta E_j = \pi_j(\bt) - \pi_j'.
\end{equation}
For parameters $\theta_m \in \{B_{ij}\}_{ij\in\mathcal{E}}$ and $\theta_m \in \{F_{ij}\}_{ij\in\mathcal{E}}$, the derivatives do not simplify as cleanly.  Recent work has shown how the derivatives $\partial_{\bt}\pi_k(\bt)$ with respect to these parameters can be bounded in magnitude \cite{owen2020universal,fernandes2023topologically}, but they do not apparently admit simple expressions (but see Ref. \citenum{aslyamov2024nonequilibrium} for expressions in terms of matrix minors of $\mathbf{W}$).  As a result, we resort here to numerical approximation of these derivatives using finite differences, by evaluating 
\begin{equation}
	\frac{\partial \pi_k(\bt)}{\partial \theta_m} \approx \frac{\pi_k(\theta_m + \delta) - \pi_k(\theta_m - \delta)}{2\delta}
\end{equation}
with $\delta = 10^{-3}$ and the remaining parameters $\theta_n, \ n\neq m$, fixed.  We numerically compute $\bp$ as the normalized leading eigenvector of $\mathbf{W}$.

\subsection{Nudging}
To apply nudging, we first present an example of an input pattern $\mathbf{F}$ and find the steady state $\bp(\mathbf{F};\bt)$.  If we want this input to produce high probability at node $j$, and low probability at node $k$, for example, then we nudge $E'_j \leftarrow E_j - \epsilon$ (which will have the effect of raising $\pi_j$) and $E'_k \leftarrow E_k + \epsilon$ with $\epsilon \sim 1$.  We then recompute the nudged steady state $\bp' = \bp(\mathbf{F};\bt')$ under these new parameters.  Note that we are not clamping the steady-state values through this nudging procedure, only indirectly encouraging it to be higher or lower by adjusting the $\{E_j\}_{j=1}^{N_\text{n}}$ parameters.  To solve the classification problem, we encourage node $i$ to be high and other nodes $j \in \mathcal{O}$ in the output set to be low when an example from class $i$ is presented.   After presenting one example during the $n^\text{th}$ training iteration, we nudge and then apply the update 
\begin{equation}
	\bt^{n+1} = \bt^{n} + \eta \Delta \bt^{n}    
\end{equation} 
with a scalar learning rate $\eta \sim 1$.  We train for $10^3$ iterations or until convergence.  

\section{Visualization of trained networks}
Here we illustrate the training process and resulting network parameters for two graphs trained to solve a binary classification problem.  In Figure \ref{SI_TrainedGraphsHexagon} we show a fully-connected network with six nodes.  The training data and learned decision boundaries are shown in Figure \ref{SI_TrainedGraphsHexagon}A, and the convergence during training is shown in Figure \ref{SI_TrainedGraphsHexagon}B.  Figure \ref{SI_TrainedGraphsHexagon}C shows the assigned input edges and output nodes, and it further depicts the learned parameters $\bt$ that resulted from training.  One can see by inspection how the learned edge weights allow the network to shunt probability toward node $1$ when a force $\mathbf{F}^1$ from  in class 1 is presented, and how it directs probability toward node $2$ when a force $\mathbf{F}^2$ from  in class 2 is presented.  Figure \ref{SI_TrainedGraphsHexagon}D depicts the steady state of the network in each of these two cases, confirming that the classification task is successfully solved.  In Figure \ref{SI_TrainedGraphsRandom} we show the same information for a random network.  

\begin{figure*}[ht!]
	\begin{center}
		\includegraphics[width=\textwidth]{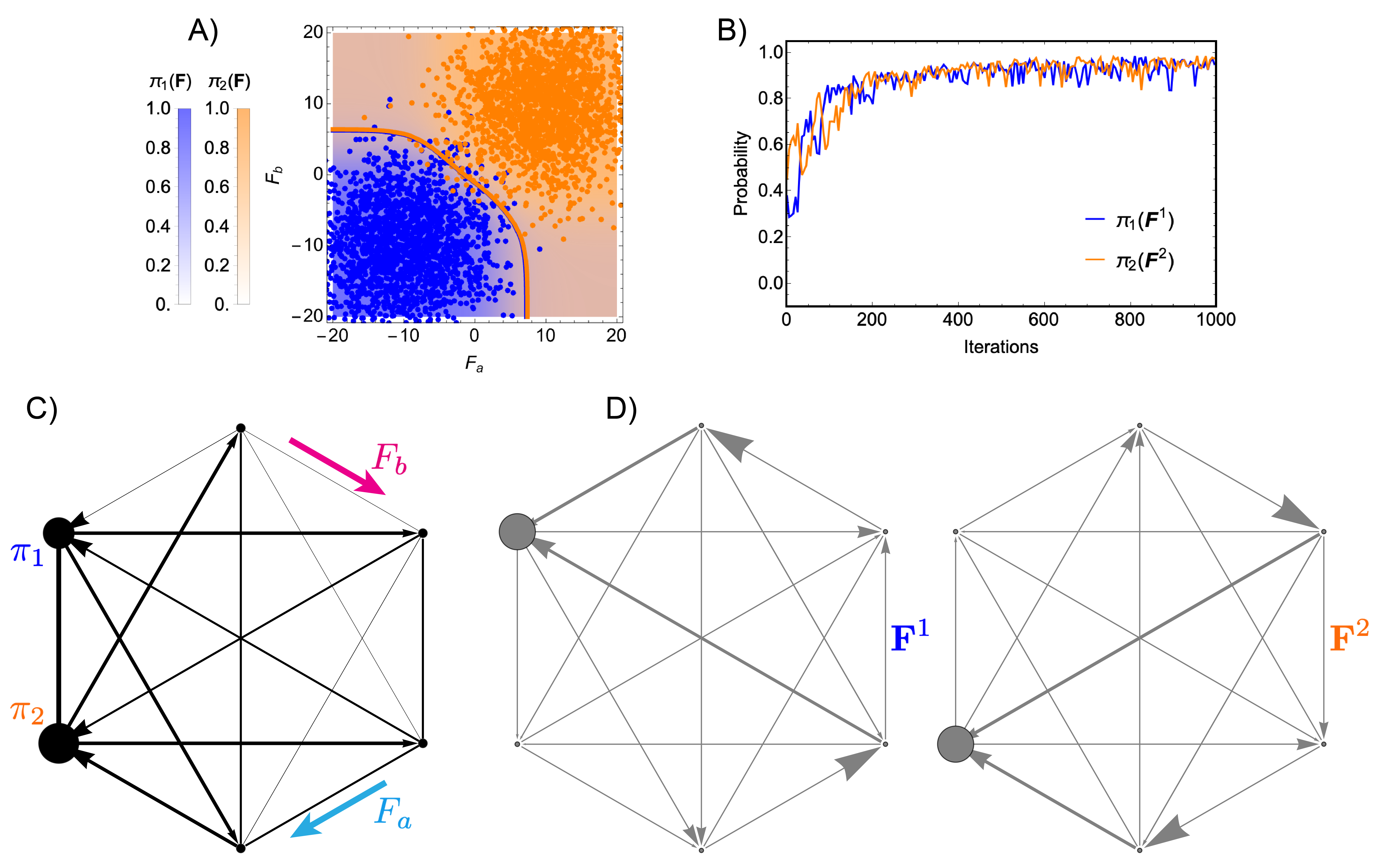}
		\caption{Visualization of training a fully-connected 6 node Markov network for binary classification.  A)  Plot of the learned classification functions $\pi_1(\mathbf{F})$ and $\pi_2(\mathbf{F})$ shown as colored density plots over the input force space.  On top of this, scatter plots show the data set, colored by assigned class, which was used to train the network.   Solid lines show the contour $\pi_1(\mathbf{F}) = 1/2$ and $\pi_2(\mathbf{F}) =1/2$.  B)  Plot illustrating convergence during training.  Convergence is measured as the steady-state probability $\pi_\rho(\mathbf{F}^\rho)$ of node $\rho$ when presented with examples $\mathbf{F}^\rho$ from its assigned class.  C)  Visualization of the learned parameters in the network.  The output nodes and input edges are labeled as in Figure \ref{MarkovDefinition}E of the main text.  The node sizes are a linear function of $e^{-E_j}$.  The edge widths are a linear function of $B_{ij}$, so that smaller values of $B_{ij}$ (faster edge rates) are thicker.  The arrowhead directions depend on the sign of $F_{ij}$ and their sizes are a linear function of $|F_{ij}|$.  D)  Visualization of the steady states for this network when presented with inputs from class 1 (left) and from class $2$ (right).  The node sizes are a linear function of $\pi_i$.  The edge widths are a linear function of the frenesy $\pi_j W_{ij} + \pi_i W_{ji} $. The arrowhead directions depend on the sign of the probability flux $\pi_j W_{ij} - \pi_i W_{ji} $ and their sizes are a linear function of the flux magnitude.}
		\label{SI_TrainedGraphsHexagon}
	\end{center}
\end{figure*}

\begin{figure*}[ht!]
	\begin{center}
		\includegraphics[width=\textwidth]{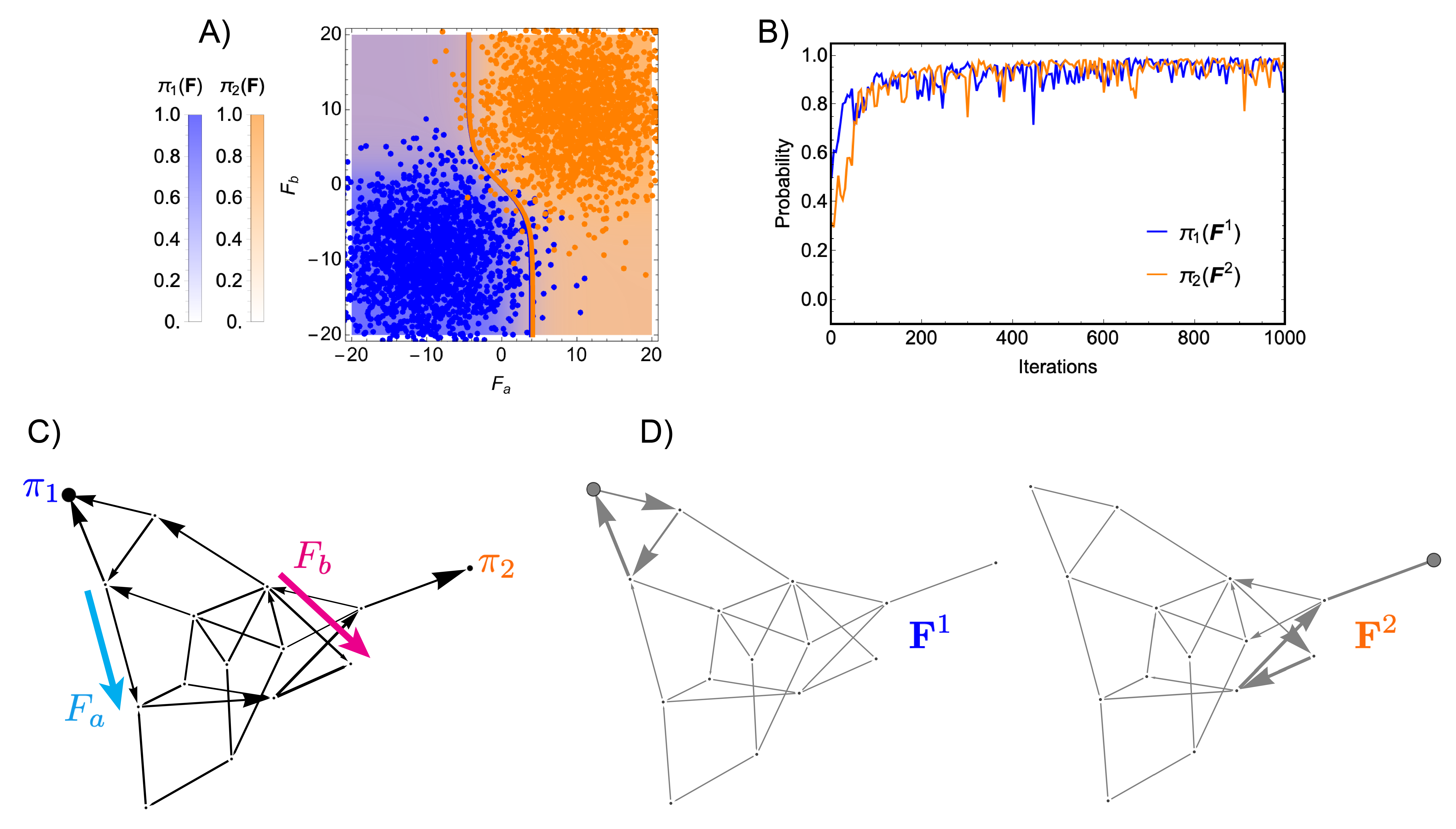}
		\caption{Visualization of training a random Markov network for binary classification.  This graph is the same as used in Figures \ref{MarkovDefinition}E, \ref{BinaryClass}, \ref{MultiClassFrustration}, and \ref{NonEqDriving} in the main text.  See the caption for Figure \ref{SI_TrainedGraphsHexagon} above for a description of the plots in this figure.}
		\label{SI_TrainedGraphsRandom}
	\end{center}
\end{figure*}

\section{Estimating the VC dimension for $D=1$}
In previous work we showed that if all other parameters are held fixed, then the partial derivative $\partial \pi_k / \partial F_{ij}$ has a fixed sign across the entire range of $F_{ij}$ \cite{floyd2024learning}.  This imposes a limitation on the expressivity of the possible decision boundaries.  To overcome this limitation, we can consider a new parameterization in which the input variables affect multiple edges simultaneously.  We denote by $F_a$ a component of the input vector $\mathbf{F}$ and write
\begin{equation}
	\frac{\partial \pi_k}{\partial F_a} = \sum_{ij \in \mathcal{E}_a }\frac{\partial \pi_k }{\partial \tilde{F}_{ij}}\frac{\partial \tilde{F}_{ij} }{\partial F_a}
\end{equation}
where $\mathcal{E}_a$ denotes the set of edges affected by the input variable $x_a$.  We assume that each of the $D$ components of the input vector affect the same number $M$ of edges, i.e., $|\mathcal{E}_a| = M, \ a\in\mathcal{A}$, so that the total number of affected edges is $MD$.  We model the input $F_a$ as directly adding to the learned value of $F_{ij}$ so that $\tilde{F}_{ij} = F_{ij} + F_a$ and $\partial \tilde{F}_{ij} / \partial F_a = 1$.  Unlike the partial derivative $\partial \pi_k / \partial F_{ij}$, the partial derivative $\partial \pi_k / \partial F_a$ is not constrained to have a fixed sign because it is the sum of several terms whose signs can differ, and, furthermore, each term $\partial \pi_k / \partial \tilde{F}_{ij}$ can change sign as $F_a$ changes because $\tilde{F}_{ij}$ is not the only parameter being varied.

How many times can the sign of $\partial \pi_k / \partial F_a$ change as a function of $M$?  We consider here the case $D=1$, meaning that there is one input variable $F$ which affects $M$ edges.  If $M=1$, then we can write $\pi_k$ as (cf. Equation \ref{eqboundcp-1} below)
\begin{equation}
	\pi_k = \frac{\zeta^k_{1} y + \zeta^k_{-1} y^{-1} + \zeta^k_0 }{\bar{\zeta}_1 y + \bar{\zeta}_{-1} y^{-1} + \bar{\zeta}_0}. \label{eqpikM1}
\end{equation}
This is a rational polynomial in $y = e^{F/2} >0$.  For general $M$, the numerator and denominator can contain tree factors whose dependence on $F$ ranges from $e^{-MF/2}, \ e^{-(M-1)F/2}, \ \ldots , e^{(M-1)F/2}, \ e^{MF/2}$ depending on if a tree contains all of the affected edges running with (incurring $e^{MF/2}$) or against (incurring a factor $e^{-MF/2}$) the positive orientation, or if there is a mixture.  Multiplying so that the lowest order term is proportional to $y^0 = 1$, the numerator and denominator of $\pi_k$ for general $M$ will be polynomials in $y$ of degree up to $2M$.  We write this schematically as 
\begin{equation}
	\pi_k = \frac{f(y)}{g(y)} \label{pikpolyfrac}
\end{equation}
where $f(y)$ and $g(y)$ are each of degree $2M$ (though possibly having some coefficients equal to zero).  To determine how many times $\partial \pi_k / \partial F $ can change sign we seek its roots.  Because $\partial \pi_k / \partial F = (y/2) \partial \pi_k / \partial y$, $\partial \pi_k / \partial F $ will have the same number of positive roots as $\partial \pi_k / \partial y$.  This latter derivative is
\begin{equation}
	\frac{\partial \pi_k}{\partial y} = \frac{f'(y)g(y) - g'(y)f(y)}{(g(y))^2}
\end{equation}
whose roots are given by solutions to 
\begin{equation}
	f'(y)g(y) - g'(y)f(y) = 0
\end{equation}
or 
\begin{equation}
	\frac{f(y)}{f'(y)} =  \frac{g(y)}{g'(y)}.
\end{equation}
Because $f(y)$ and $g(y)$ are polynomials, these fractions can be written as
\begin{equation}
	\frac{f(y)}{f'(y)} = y \frac{f'(y)}{f'(y)} + \frac{C_f}{f'(y)} = y + \frac{C_f}{f'(y)}
\end{equation}
where $C_f$ is the constant term in $f(y)$.  A similar expression holds for $g(y)/g'(y)$, and equating the two yields
\begin{equation}
	C_f g'(y) = C_g f'(y).
\end{equation}
The polynomials $f'(y)$ and $g'(y)$ are of degree $2M - 1$, and the polynomial $q(y) \equiv C_f g'(y) - C_g f'(y)$ is also of this degree.  Depending on the coefficients, $q(y)$ and hence $\partial \pi_k / \partial y$ can thus have up to $2M -1$ positive solutions.

For the case $M=1$, the monotonicity constraint imposes that there are zero roots of $\partial \pi_k / \partial F$, which falls under the allowed limit of $2M - 1 = 1$ for general rational polynomials of degree $M$.  As detailed in Ref.\ \citenum{floyd2024learning}, this fact for $M=1$ is due to a non-trivial set of inequalities satisfied by the terms $\zeta^k_1/\bar{\zeta}_1$, $\zeta^k_{-1}/\bar{\zeta}_{-1}$, and $\zeta^k_0/\bar{\zeta}_0$ in Equation \ref{eqpikM1}.  For $M>1$, however, there is no monotonicity constraint and the number of sign changes may saturate the algebraic bound $2M-1$.  Numerically, we find through random numerical search that the bound can indeed be saturated for $M=2$ and $M=3$.  For larger $M$, parameters saturating the bound $2M-1$ become rarer and harder to find through random search.  Assuming that there are sufficiently many degrees of freedom in the network which are not constrained, then the bound $2M-1$ can also be saturated for $M>3$. In summary, for $D=1$, the number of times the sign of $\partial \pi_k / \partial F$ can change as a function of $F$ is $0$ for $M=1$ and at most $2M-1$ for $M>1$.

This result informs about the Vapnik-Chervonenkis (VC) dimension of the classifier $\pi_k(F)$.  The VC dimension is the largest number $N_{VC}$ of points that the classifier can shatter for any of the $2^{N_{VC}}$ binary assignments of labels to the points.  For $D=1$, $N_{VC}$ can be estimated by determining the number of points which can be separated in the most challenging of the $2^{N_{VC}}$ label assignments.  The most challenging arrangement is the one in which each point's label differs from those of its nearest neighbors.  To separate such a set of labeled points, $\partial \pi_k / \partial F$ must be able to change signs at least $N_{VC}-2$ times.  For the classifier $\pi_k(F)$, then, the VC dimension is 2 for $M=1$ and $2M + 1$ for $M>1$.  This is consistent with the more general estimate for any $D$ based on the theorem by Dudley discussed in the main text.  

\section{Reduced degrees of freedom from equality constraints}\label{inputfrust}

\subsection{Equality constraint among directed tree weights}
In this section we present analytical and numerical arguments based on counting degrees of freedom with constraints which illustrate why certain multi-class classification tasks cannot be solved by networks with $M=1$.  The main result we rely on is an equality we derived in previous work \cite{floyd2024learning} that relates the products of directed tree weights conditioned on containing edge $i\leftarrow j$, $j \leftarrow i$, or neither of these, for trees rooted on nodes $i$, $j$, and an arbitrary node $p$.  We first describe this equality and then use it to show that two-class classification is possible for $D=1$ but, somewhat surprisingly, four-class classification is not possible for $D=2$.  We leave a general treatment for arbitrary $M$ and $D$ to future work.

We let $\zeta^p_{1}$ denote the sum over all directed tree weights for directed trees rooted at node $p$ and which include the directed edge $i \leftarrow j$, $\zeta^p_{-1}$ denote the same for directed trees which include the directed edge $j\leftarrow i$, and $\zeta^p_{0}$ denote the same for directed trees which do not include either $i\leftarrow j$ or $j \leftarrow i$.  Previous work of ours shows that \cite{floyd2024learning}
\begin{equation}
	\frac{\zeta^p_{1}}{W_{ij}}\zeta^i_{0} + \frac{\zeta^p_{-1}}{W_{ji}}\zeta^j_{0} = \frac{\zeta^i_{1}}{W_{ij}}\zeta^p_{0} = \frac{\zeta^j_{-1} }{W_{ji}}\zeta^p_{0}. \label{eqconstraint}
\end{equation}
Applying this for $p = i$ or $p = j$ yields the equations
\begin{equation}
	\frac{\zeta^i_{1}}{W_{ij}} = \frac{\zeta^j_{-1}}{W_{ji}} \label{eqconstraintij}
\end{equation} 
and 
\begin{equation}
	\zeta^i_{-1} = \zeta^j_{1}  = 0. \label{eqconstraint0}
\end{equation}  
In Ref.\ \citenum{floyd2024learning} we proved Equation \ref{eqconstraint} by establishing a one-to-one correspondence between the product of tree weights for each pair of trees in the products of sums on the left hand and right hand sides of the equality.  We established the correspondence using a recent theoretical tool called ``tree surgery,'' introduced in Ref.\ \citenum{owen2020universal}.  The special case Equation \ref{eqconstraintij} implies that the sum of all tree weights for directed trees rooted on node $i$ and which pass through the edge $i\leftarrow j$ is the same as that for directed trees rooted on node $j$ and which pass through edge $j \leftarrow i$, after dividing out $W_{ij}$ and $W_{ji}$.  Equation \ref{eqconstraint0} simply says that no directed tree can be rooted on node $i$ and pass through the edge $j \leftarrow i$; similarly no tree can be rooted on node $j$ and pass through the edge $i \leftarrow j$.

\subsection{Degrees of freedom for $M=1$, $D=1$}
\subsubsection{Counting degrees of freedom}
We first explicitly write out Equation \ref{eqmttzeta} in the main text for $M=1$ and $D=1$, 
\begin{equation}
	\pi_p = \frac{\zeta^p_1 e^{F_a/2}  + \zeta^p_{-1} e^{-F_a/2} +\zeta^p_0}{\bar{\zeta}_1 e^{F_a/2}  + \bar{\zeta}_{-1} e^{-F_a/2} +\bar{\zeta}_0},
\end{equation}
and ask how many degrees of freedom are accessible among the $3N_\text{n}$ parameters in the set $\mathcal{S}^\text{orig} \equiv \{\{\zeta^p_s\}_{s\in\{-1,0,1\}}\}_{p=1}^{N_\text{n}}$.  

First, by Equation \ref{eqconstraint0} we know that $\zeta^i_{-1} = \zeta^j_{1} = 0$, where $i$ and $j$ are the nodes adjacent to the input edge.  This reduces our degrees of freedom by $2$.  Next, we note that whether an equation counts as a constraint on the number of degrees of freedom in a set depends on whether the only functions appearing in the equation come from this set.  If they do, then the equation acts as a constraint and reduces the number of degrees of freedom in the set, but if there are functions not found in the set which appear in the equation, then these unaccounted for functions may be independently tuned and prevent the equation from constraining the number of degrees of freedom in the set.  

The edge rates $W_{ij}$ and $W_{ji}$ appear in Equations \ref{eqconstraint} and \ref{eqconstraintij} but are not contained in $\mathcal{S}^\text{orig}$.  To obtain constraints, we rearrange Equation \ref{eqconstraintij} as
\begin{equation}
	\frac{W_{ij}}{W_{ji}} = \frac{\zeta^i_{1}}{\zeta^j_{-1}} \label{eqsub}
\end{equation}
and, noticing that the dependence on $W_{ij}$ and $W_{ji}$ in Equation \ref{eqconstraint} occurs only through the ratio $W_{ij}/W_{ji}$, we substitute Equation \ref{eqsub} into each instance of Equation \ref{eqconstraint}:
\begin{equation}
	\zeta^p_{1}\zeta^i_{0} + \frac{\zeta^i_{1}}{\zeta^j_{-1}}\zeta^p_{-1}\zeta^j_{0} = \zeta^i_{1}\zeta^p_{0} \ \forall p \neq i,j.
\end{equation}
There are $N_\text{n} - 2$ instances of this equation for each $p \neq i,j$, and each instance represents a constraint on the degrees of freedom in $\mathcal{S}^\text{orig}$.  Accounting for the $2$ functions which are zero, there are thus $3 N_\text{n} - 2 - (N_\text{n} - 2) = 2 N_\text{n}$ degrees of freedom in $\mathcal{S}^\text{orig}$.

\subsubsection{Classification of two classes}
We now ask how this type of counting argument can be used to assess whether it is possible to classify two classes for the case $M=1$, $D=1$.  We assume here that class $1$ lies in the region $F_a > 0$ and class $2$ in the region $F_a < 0$.  We can place a condition for solving this classification task on the limiting values of $\pi_p$ in each of these spaces.  Specifically, we require that
\begin{eqnarray}
	\lim_{F_a \rightarrow \infty} \pi_1 &>&  \lim_{F_a \rightarrow \infty} \pi_p \ \forall \ p \neq 1 \\
	\lim_{F_a \rightarrow -\infty} \pi_2 &>&  \lim_{F_a \rightarrow \infty} \pi_p \ \forall \ p \neq 2. 
\end{eqnarray}
It is straightforward to show that these inequalities in turn place the following conditions on the functions
\begin{eqnarray}
	\zeta^1_{1} &>& \zeta^p_{1} \ \forall \ p \neq 1 \label{eqzineq1} \\
	\zeta^2_{-1} &>& \zeta^p_{-1} \ \forall \ p \neq 2. \label{eqzineq2}
\end{eqnarray}
These are $2N_\text{n} - 2$ inequalities which must be satisfied by the $2 N_\text{n}$ functions in $\mathcal{S}^\text{ineq} \equiv \{\{\zeta^p_s\}_{s\in\{-1,1\}}\}_{p=1}^{N_\text{n}} \subset \mathcal{S}^\text{orig}$.  At first glance, because $2N_\text{n} > 2 N_\text{n}-2$, we appear to have enough degrees of freedom to satisfy all of the inequalities.  It will turn out that we do have enough, but the situation is slightly more complicated because we have not yet accounted for the constraints which exist among the functions in $\mathcal{S}^\text{ineq}$.  

First, the functions $\zeta^i_{-1}$ ane $\zeta^j_{1}$ are zero, which trivially satisfies the inequalities in which they appear, assuming that $i\neq 2$ and $j \neq 1$.  We can thus subtract $2$ from the number of non-trivial inequalities which remain, bringing the number to $2N_\text{n} - 4$, and which need to be satisfied by the remaining $2N_\text{n} - 2$ non-zero functions.  

Additionally, the inequalities in Equations \ref{eqzineq1} and \ref{eqzineq2} cannot depend on $W_{ij}$ or $W_{ji}$.  This is because Equation \ref{eqzineq1} cannot include $W_{ji}$ as a factor (since all trees pass through $i\leftarrow j$), and both sides contain $W_{ij}$ as a common factor so that it divides out.  Similar reasoning holds for Equation \ref{eqzineq2}.  We may thus treat $W_{ij}$ and $W_{ji}$ as arbitrary constants when counting remaining degrees of freedom, and we no longer need to use Equation \ref{eqconstraintij} to eliminate $W_{ij}$ or $W_{ji}$ where they appear.  

Considering the case $N_\text{n} = 2$, there are 2 non-zero functions in $\mathcal{S}^\text{ineq} = \{\zeta_{1}^i, \zeta_{-1}^j, 0, 0\}$, and there is only 1 degree of freedom among these because of Equation \ref{eqconstraintij}.  This 1 degree of freedom is sufficient to satisfy the only requirement for two-class classification, which is that $\zeta^i_{1}$ and $\zeta^j_{-1}$ are both non-zero.  Both non-zero functions in the set $\mathcal{S}^{\o} \equiv \mathcal{S}^\text{orig} \setminus \mathcal{S}^\text{ineq} = \{\zeta_{0}^i, \zeta_{0}^j\}$ are degrees of freedom because no equalities constrain them.  

For $N_\text{n} > 2$, every additional node $p$ introduces 2 new functions $\{\zeta_{\pm1}^p\}$ in $\mathcal{S}^\text{ineq}$ and 1 new function $\{\zeta_{0}^p\}$ in $\mathcal{S}^{\o}$.  Equation \ref{eqconstraint} can be used to solve for $\zeta_{0}^p$ and eliminate it as a degree of freedom.  Although this equation also involves the 2 elements $\{\zeta_{0}^i, \zeta_{0}^j\} \subset \mathcal{S}^{\o}$, the values of these functions do not affect the inequality conditions and we may use the 2 degrees of freedom which they represent to set their values arbitrarily.  Then Equation \ref{eqconstraint} can be used to fix $\zeta_{0}^p$ so that the 2 functions $\{\zeta_{\pm1}^p\}$ remain as degrees of freedom.  We therefore have 1 degree of freedom in $\mathcal{S}^\text{ineq}$ for $N_\text{n} = 2$ and 2 degrees of freedom for every node beyond this, so that the total number of degrees of freedom in this set (having treated $W_{ij}$ and $W_{ji}$ as fixed) is $1 + 2(N_\text{n} - 2) = 2N_\text{n} - 3$.  Because this is greater than the number of non-trivial inequalities to satisfy, two-class classification is possible for any $N_\text{n}$.

\subsubsection{Numerical verification}
To numerically verify these counting arguments, we form the Jacobian matrices $J^\mathcal{S^\text{orig}}$, whose elements are the derivatives of functions in $\mathcal{S^\text{orig}}$ with respect to all edge rates, and $J^{\mathcal{S}^\text{ineq}}$, whose elements are the derivatives of functions in $\mathcal{S}^\text{ineq}$ with respect to all edge rates except $W_{ij}$ or $W_{ji}$.  The rank of these matrices indicates the corresponding number of degrees of freedom in the arguments presented above.  Numerical evaluation for fully connected graphs with increasing numbers of nodes confirms that $ \text{rank}(J^{\mathcal{S}^\text{orig}}) = 2N_\text{n}$ and $\text{rank}(J^{\mathcal{S}^\text{ineq}}) = 2N_\text{n} -3$.

\subsection{Degrees of freedom for $M=1$, $D=2$}
\subsubsection{Counting degrees of freedom}
We next add an additional input edge $kl$ and explicitly write out Equation \ref{eqmttzeta} in the main text for $M=1$ and $D=2$: 
\begin{equation}
	\pi_p = \frac{\zeta^p_{1,1}e^{F_a/2}e^{F_b/2} + \zeta^p_{-1,1}e^{-F_a/2}e^{F_b/2} + \zeta^p_{-1,-1}e^{-F_a/2}e^{-F_b/2} + \zeta^p_{1,-1}e^{F_a/2}e^{-F_b/2} + \ldots }{\bar{\zeta}_{1,1}e^{F_a/2}e^{F_b/2} + \bar{\zeta}_{-1,1}e^{-F_a/2}e^{F_b/2} + \bar{\zeta}_{-1,-1}e^{-F_a/2}e^{-F_b/2} + \bar{\zeta}_{1,-1}e^{F_a/2}e^{-F_b/2} + \ldots },
\end{equation}
where the ellipses represent terms that are less than second order in $e^{F_a/2}$ and $e^{F_b/2}$.  The coefficient $\zeta^p_{s,t}$ represents the sum of all directed tree weights for trees rooted at node $p$ and which contain the input $F_a$ (along edge $i\leftarrow j$) in its $s\in \{-1,0,1\}$ orientation and the input $F_b$ (along edge $k\leftarrow l$) in its $t\in \{-1,0,1\}$ orientation.  We ask how many degrees of freedom are accessible among the $9N_\text{n}$ in functions in the set $\mathcal{S}^\text{orig} \equiv \{\{\zeta^p_{s,t}\}_{s,t\in\{-1,0,1\}}\}_{p=1}^{N_\text{n}}$ by varying all the edge rates.

To understand how many constraints we can write down, we consider Equation \ref{eqconstraint} for node $p$ and input edge $ij$.  A direct application yields
\begin{equation}
	\frac{\sum_{n\in\{-1,0,1\}}\zeta^p_{1,n}}{W_{ij}}\left(\sum_{n'\in\{-1,0,1\}}\zeta^i_{0,n'}\right) + \frac{\sum_{n\in\{-1,0,1\}}\zeta^p_{-1,n}}{W_{ij}}\left(\sum_{n'\in\{-1,0,1\}}\zeta^j_{0,n'}\right) = \frac{\sum_{n\in\{-1,0,1\}}\zeta^i_{1,n}}{W_{ij}}\left(\sum_{n'\in\{-1,0,1\}}\zeta^p_{0,n'}\right). \label{eqdef1}
\end{equation}
Here we have summed over the possible occurrences at edge $kl$.  Each product of the form $\sum_{n}f_n \sum_{n'} g_{n'}$ involves 9 terms, and Equation \ref{eqdef1} involves summing over every term.  In fact, we can obtain more equations by conditioning, rather than summing, over the possible occurrences at edge $kl$.  For example, we can condition on $n,n'=1$ to write
\begin{equation}
	\frac{\zeta^p_{1,1}}{W_{ij}}\zeta^i_{0,1} + \frac{\zeta^p_{-1,1}}{W_{ji}}\zeta^j_{0,1} = \frac{\zeta^i_{1,1}}{W_{ij}}\zeta^p_{0,1}.
\end{equation}
We can similarly write ($n,n'=0$)
\begin{equation}
	\frac{\zeta^p_{1,0}}{W_{ij}}\zeta^i_{0,0} + \frac{\zeta^p_{-1,0}}{W_{ji}}\zeta^j_{0,0} = \frac{\zeta^i_{1,0}}{W_{ij}}\zeta^p_{0,0}
\end{equation}
and ($n,n'=-1$)
\begin{equation}
	\frac{\zeta^p_{1,-1}}{W_{ij}}\zeta^i_{0,-1} + \frac{\zeta^p_{-1,-1}}{W_{ji}}\zeta^j_{0,-1} = \frac{\zeta^i_{1,-1}}{W_{ij}}\zeta^p_{0,-1}.
\end{equation}
It can be shown that these ``diagonal'' terms of the double sum $\sum_{n}f_n \sum_{n'} g_{n'}$ are supplemented by symmetrized off-diagonal combinations.  We therefore also have ($n=1,n'=-1 \cup n=-1,n'=1$)
\begin{equation}
	\frac{\zeta^p_{1,1}}{W_{ij}}\zeta^i_{0,-1} + \frac{\zeta^p_{-1,1}}{W_{ji}}\zeta^j_{0,-1} + \frac{\zeta^p_{1,-1}}{W_{ij}}\zeta^i_{0,1} + \frac{\zeta^p_{-1,-1}}{W_{ji}}\zeta^j_{0,1} = \frac{\zeta^i_{1,1}}{W_{ij}}\zeta^p_{0,-1} + \frac{\zeta^i_{1,-1}}{W_{ij}}\zeta^p_{0,1}
\end{equation}
and ($n=1,n'=0 \cup n=0,n'=1$)
\begin{equation}
	\frac{\zeta^p_{1,1}}{W_{ij}}\zeta^i_{0,0} + \frac{\zeta^p_{-1,1}}{W_{ji}}\zeta^j_{0,0} + \frac{\zeta^p_{1,0}}{W_{ij}}\zeta^i_{0,1} + \frac{\zeta^p_{-1,0}}{W_{ji}}\zeta^j_{0,1} = \frac{\zeta^i_{1,1}}{W_{ij}}\zeta^p_{0,0} + \frac{\zeta^i_{1,0}}{W_{ij}}\zeta^p_{0,1}
\end{equation}
and ($n=-1,n'=0 \cup n=0,n'=-1$)
\begin{equation}
	\frac{\zeta^p_{1,-1}}{W_{ij}}\zeta^i_{0,0} + \frac{\zeta^p_{-1,-1}}{W_{ji}}\zeta^j_{0,0} + \frac{\zeta^p_{1,0}}{W_{ij}}\zeta^i_{0,-1} + \frac{\zeta^p_{-1,0}}{W_{ji}}\zeta^j_{0,-1} = \frac{\zeta^i_{1,-1}}{W_{ij}}\zeta^p_{0,0} + \frac{\zeta^i_{1,0}}{W_{ij}}\zeta^p_{0,-1}.
\end{equation}

In addition to these $6$ equations for node $p$ based on edge $ij$, we have $6$ equations for node $p$ based on edge $kl$, which are
\begin{eqnarray}
	\frac{\zeta^p_{1,1}}{W_{kl}}\zeta^k_{1,0} + \frac{\zeta^p_{1,-1}}{W_{lk}}\zeta^l_{1,0} &=& \frac{\zeta^k_{1,1}}{W_{kl}}\zeta^p_{1,0}\\
	\frac{\zeta^p_{0,1}}{W_{kl}}\zeta^k_{0,0} + \frac{\zeta^p_{0,-1}}{W_{lk}}\zeta^l_{0,0} &=& \frac{\zeta^k_{0,1}}{W_{kl}}\zeta^p_{0,0}\\
	\frac{\zeta^p_{-1,1}}{W_{kl}}\zeta^k_{-1,0} + \frac{\zeta^p_{-1,-1}}{W_{lk}}\zeta^l_{-1,0} &=& \frac{\zeta^k_{-1,1}}{W_{kl}}\zeta^p_{-1,0}\\
	\frac{\zeta^p_{1,1}}{W_{kl}}\zeta^k_{-1,0} + \frac{\zeta^p_{1,-1}}{W_{lk}}\zeta^l_{-1,0} + \frac{\zeta^p_{-1,1}}{W_{kl}}\zeta^k_{1,0} + \frac{\zeta^p_{-1,-1}}{W_{lk}}\zeta^l_{1,0} &=& \frac{\zeta^k_{1,1}}{W_{kl}}\zeta^p_{-1,0} + \frac{\zeta^k_{-1,1}}{W_{kl}}\zeta^p_{1,0}\\
	\frac{\zeta^p_{1,1}}{W_{kl}}\zeta^k_{0,0} + 
	\frac{\zeta^p_{1,-1}}{W_{lk}}\zeta^l_{0,0} + 
	\frac{\zeta^p_{0,1}}{W_{kl}}\zeta^k_{1,0} + 
	\frac{\zeta^p_{0,-1}}{W_{lk}}\zeta^l_{1,0} &=& 
	\frac{\zeta^k_{1,1}}{W_{kl}}\zeta^p_{0,0} 
	+ \frac{\zeta^k_{0,1}}{W_{kl}}\zeta^p_{1,0}\\
	\frac{\zeta^p_{-1,1}}{W_{kl}}\zeta^k_{0,0} + 
	\frac{\zeta^p_{-1,-1}}{W_{lk}}\zeta^l_{0,0} + 
	\frac{\zeta^p_{0,1}}{W_{kl}}\zeta^k_{-1,0} + 
	\frac{\zeta^p_{0,-1}}{W_{lk}}\zeta^l_{-1,0} &=& 
	\frac{\zeta^k_{-1,1}}{W_{kl}}\zeta^p_{0,0} 
	+ \frac{\zeta^k_{0,1}}{W_{kl}}\zeta^p_{-1,0}.
\end{eqnarray}

If we consider $p=i$ or $p =j$ at edge $ij$ we are led to the equations (cf. Equation \ref{eqconstraintij})
\begin{eqnarray}
	\frac{\zeta^i_{1,1}}{W_{ij}} &=& \frac{\zeta^j_{-1,1}}{W_{ij}} \label{eqijeq1} \\
	\frac{\zeta^i_{1,0}}{W_{ij}} &=& \frac{\zeta^j_{-1,0}}{W_{ij}} \\
	\frac{\zeta^i_{1,-1}}{W_{ij}} &=& \frac{\zeta^j_{-1,-1}}{W_{ij}} \label{eqijeq2}
\end{eqnarray}
and (cf. Equation \ref{eqconstraint0})
\begin{eqnarray}
	\zeta^i_{-1,1} &=& 0 \label{eqi01} \\
	\zeta^i_{-1,0} &=& 0\\
	\zeta^i_{-1,-1} &=& 0\\
	\zeta^j_{1,1} &=& 0\\
	\zeta^j_{1,0} &=& 0\\
	\zeta^j_{1,-1} &=& 0. \label{eqi0l}
\end{eqnarray}
Similarly, considering $p=k$ or $p =l$ at edge $kl$ leads to
\begin{eqnarray}
	\frac{\zeta^k_{1,1}}{W_{kl}} &=& \frac{\zeta^l_{1,-1}}{W_{lk}} \label{eqkleq1} \\
	\frac{\zeta^k_{0,1}}{W_{kl}} &=& \frac{\zeta^l_{0,-1}}{W_{lk}} \\
	\frac{\zeta^k_{-1,1}}{W_{kl}} &=& \frac{\zeta^l_{-1,-1}}{W_{lk}} \label{eqkleq2}
\end{eqnarray}
and 
\begin{eqnarray}
	\zeta^k_{1,-1} &=& 0 \label{eqk01}\\
	\zeta^k_{0,-1} &=& 0\\
	\zeta^k_{-1,-1} &=& 0\\
	\zeta^l_{1,1} &=& 0\\
	\zeta^l_{0,1} &=& 0\\
	\zeta^l_{-1,1} &=& 0. \label{eqk0l}
\end{eqnarray}

We have listed the various equations derived from Equation \ref{eqconstraint} which may reduce the number of degrees of freedom in $\mathcal{S}^\text{orig}$.  Equations \ref{eqi01}-\ref{eqi0l} and \ref{eqk01}-\ref{eqk0l} constrain $12$ functions to be zero.  We use Equations \ref{eqijeq1} and \ref{eqkleq1} to substitute $W_{ij} / W_{ji}$ and $W_{kl}/W_{lk}$ where they appear in the remaining equations so that they are closed in the functions in $\mathcal{S}^\text{orig}$.  

An additional complication remains, however, which is that these equations are not all independent of each other.  We first count how many equations we can write down and then consider how to find the number of independent constraints among these:  
\begin{itemize}
	\item Considering $p = i$ or $p = j$ at edge $ij$ yields 2 equations (not counting Equation \ref{eqijeq1} which allowed eliminating $W_{ij}/W_{ji}$, and not counting Equations \ref{eqi01}-\ref{eqi0l} which allowed zeroing out $6$ functions)
	\item Considering $p = k$ or $p = l$ at edge $kl$ yields 2 equations (not counting Equation \ref{eqkleq1} which allowed eliminating $W_{kl}/W_{lk}$, and not counting Equations \ref{eqk01}-\ref{eqk0l} which allowed zeroing out $6$ functions)
	\item Considering $p = i$ at edge $kl$ yields 6 equations
	\item Considering $p = j$ at edge $kl$ yields 6 equations
	\item Considering $p = k$ at edge $ij$ yields 6 equations
	\item Considering $p = l$ at edge $ij$ yields 6 equations
	\item Considering arbitrary $p \neq i,j,k,l$ at edge $ij$ yields 6 equations for each $p$
	\item Considering arbitrary $p \neq i,j,k,l$ at edge $kl$ yields 6 equations for each $p$
\end{itemize}
We thus have $28 + 12(N_\text{n} - 4)$ equalities for $N_\text{n} \geq 4$.  We collect these equations, of the form $C(\mathcal{S}^\text{orig}) = 0$, in a set $\mathcal{C}$.  To determine how many of these equations are independent, we form the Jacobian matrix $J^\mathcal{C}$ whose elements are the derivatives of the equations with respect to the elements of $\mathcal{S}^\text{orig}$.  

To understand the scaling with $N_\text{n}$, we first ignore node $p\neq i,j,k,l$ and consider the submatrix of $J^\mathcal{C}$ formed by excluding the equations that involve node $p$ and excluding the functions $\zeta^p_{s,t}$ from $\mathcal{S}^\text{orig}$; this is a closed set of equations on functions defined at nodes $i$, $j$, $k$, and $l$.  We find that the rank of this matrix is $12$, far fewer than the $28$ equations involving these functions listed above.  These $28$ equations thus represent $12$ independent constraints on the $36 - 12 = 24$ non-zero functions in $\mathcal{S}^{\text{orig}} \setminus \{\zeta^p_{s,t}\}_{s,t\in\{-1,0,1\}}$, leaving $12$ degrees of freedom.  

We now include the equations and functions for node $p$.  Note that node $p$ will only be coupled via the equations to nodes $i,j,k$, and $l$, not to any other node $p'$.  We may thus view the degrees of freedom per extra node as adding linearly. The rank of the matrix $J^\mathcal{C}$ including the fifth node $p$ is 18, representing $6$ new independent constraints.  Every node beyond the fourth thus introduces $9$ new functions and $6$ new constraints, leaving 3 degrees of freedom per node.  This means that the total number of degrees of freedom $\mathcal{S}^\text{orig}$ is $12 + 3(N_\text{n} - 4) = 3 N_\text{n}$.  Numerical evaluation of the rank of the Jacobian matrix $J^{\mathcal{S}^\text{orig}}$ for fully connected graphs with increasing $N_\text{n}$ confirms this predicted scaling: $\text{rank}(J^{\mathcal{S}^\text{orig}}) = 3 N_\text{n}$.  This shows that equations derived from Equation \ref{eqconstraint}, after accounting for dependencies among these equations, correctly predicts how many degrees of freedom are available in the set of polynomial coefficients $\mathcal{S}^\text{orig}$.

\subsubsection{Prevented classification of four classes}
We now assess whether it is possible to classify two classes for the case $M=1$, $D=2$.  We assume that we have a problem with class $1$ in quadrant I ($F_a, \ F_b >0$), class $2$ in quadrant II ($F_a<0, \ F_b > 0$), class $3$ in quadrant III ($F_a, \ F_b <0$), and class $4$ in quadrant IV ($F_a>0, \ F_b < 0$).  We can place a condition for solving this classification task on the limiting values of $\pi_p$ in each of these quadrants.  Specifically, we require that
\begin{eqnarray}
	\lim_{F_a \rightarrow \infty, F_b \rightarrow \infty} \pi_1 &>&  \lim_{F_a \rightarrow \infty, F_b \rightarrow \infty} \pi_p \ \forall \ p \neq 1 \\
	\lim_{F_a \rightarrow -\infty, F_b \rightarrow \infty} \pi_2 &>&  \lim_{F_a \rightarrow -\infty, F_b \rightarrow \infty} \pi_p \ \forall \ p \neq 2 \\
	\lim_{F_a \rightarrow -\infty, F_b \rightarrow -\infty} \pi_3 &>&  \lim_{F_a \rightarrow -\infty, F_b \rightarrow -\infty} \pi_p \ \forall \ p \neq 3 \\
	\lim_{F_a \rightarrow \infty, F_b \rightarrow -\infty} \pi_4 &>&  \lim_{F_a \rightarrow \infty, F_b \rightarrow -\infty} \pi_p \ \forall \ p \neq 4.
\end{eqnarray}
These inequalities in turn place the following conditions on the functions
\begin{eqnarray}
	\zeta^1_{1,1} &>& \zeta^p_{1,1} \ \forall \ p \neq 1 \label{eq2di1} \\
	\zeta^2_{-1,1} &>& \zeta^p_{-1,1} \ \forall \ p \neq 2 \label{eq2di2} \\
	\zeta^3_{-1,-1} &>& \zeta^p_{-1,-1} \ \forall \ p \neq 3 \label{eq2di3} \\
	\zeta^4_{1,-1} &>& \zeta^p_{1,-1} \ \forall \ p \neq 4.\label{eq2di4}
\end{eqnarray}
These are $4N_\text{n} - 4$ inequalities which must be satisfied by the $4 N_\text{n}$ functions in $\mathcal{S}^\text{ineq} \equiv \{\{\zeta^p_{s,t}\}_{s,t\in\{-1,1\}}\}_{p=1}^{N_\text{n}} \subset \mathcal{S}^\text{orig}$.  From Equations \ref{eqi01}-\ref{eqi0l} and \ref{eqk01}-\ref{eqk0l}, $8$ of these functions are zero and trivially satisfy the inequalities in which they appear, assuming that $i\neq2$, $j\neq1$, $k\neq4$, and $l\neq 3$.  We note that inequalities in Equations \ref{eq2di1}-\ref{eq2di4} cannot depend on $W_{ij}$, $W_{ji}$, $W_{kl}$ or $W_{lk}$ for similar reasons as described in the previous section (either both sides of an inequality contain no dependence on an edge rate or are proportional to it).  We are thus led to ask how many non-zero degrees of freedom are available in $\mathcal{S}^\text{ineq}$ through variation in all edges rates except $W_{ij}$, $W_{ji}$, $W_{kl}$ and $W_{lk}$.

Considering the case $N_\text{n} = 4$, there are $8$ non-zero functions in $\mathcal{S}^\text{ineq}$.  There are now 14 independent equations among all the function in $\mathcal{S}^\text{orig}$ (because we no longer use Equations \ref{eqijeq1} and \ref{eqkleq1} to eliminate $W_{ij}/W_{ji}$ and $W_{kl}/W_{lk}$), and 4 of these equations (Equations \ref{eqijeq1}, \ref{eqijeq2}, \ref{eqkleq1}, and \ref{eqkleq2}) constrain elements in $\mathcal{S}^\text{ineq}$, leaving 4 degrees of freedom in this set.  There are thus exactly as many degrees in freedom in $\mathcal{S}^\text{ineq}$ as there are inequality conditions to satisfy for $N_\text{n} = 4$, but we show in the next section that four-class classification is still impossible because these inequalities are mutually contradictory.  There are 16 non-zero functions in $\mathcal{S}^{\o} \equiv \mathcal{S}^{\text{orig}} \setminus \mathcal{S}^\text{ineq}$, and the remaining 10 constraints reduce the number of degrees of freedom in this set to 6.  

For $N_\text{n} > 4$, every additional node $p$ introduces 4 new functions $\{\zeta^p_{\pm1, \pm1} \}$ in $\mathcal{S}^\text{ineq}$ and 5 new functions in $\mathcal{S}^{\o}$.  As shown in the previous section we also introduce 6 new constraints among all 9 functions for this node and the functions for nodes $i,j,k,$ and $l$.  Although these constraints also involve the 6 degrees of freedom in $\mathcal{S}^{\o}$ for nodes $i,j,k,$ and $l$, the values of the functions in in $\mathcal{S}^{\o}$ do not affect the inequality conditions, and we may use up their degrees of freedom to set their values arbitrarily.  We may then use the 6 independent constraints for node $p$ to solve for all 5 new functions in $\mathcal{S}^{\o}$ and for 1 new function in $\mathcal{S}^\text{ineq}$.  This leaves only 3 new degrees of freedom in $\mathcal{S}^\text{ineq}$ for every node $p$.  The total number of degrees of freedom in this set (having treated $W_{ij}$, $W_{ji}$, $W_{kl}$, and $W_{lk}$ as fixed) is thus $4 + 3(N_\text{n} - 4) = 3 N_\text{n} - 8$.  We numerically confirm this by evaluating $\text{rank}(J^{\mathcal{S}^\text{ineq}})  = 3 N_\text{n} -8$ for fully connected graphs with increasing $N_\text{n}$ (again not including Jacobian columns for derivatives with respect to $W_{ij}$, $W_{ji}$, $W_{kl}$, or $W_{lk}$).  Comparing $3 N_\text{n} -8$ to the number of non-trivial inequalities, $4 N_\text{n} - 12$, we conclude that there will be insufficient degrees of freedom to achieve four-class classification of freedom for $N_\text{n} > 4$.  The case $N_\text{n} = 4$ is discussed next.  We note that to achieve three-class classification there are are only $3 N_\text{n} - 8$ non-trivial inequalities to satisfy, so this is indeed possible for $M=1$, $D=2$.

\subsubsection{Frustrated inequalities}
An interesting situation arises when considering classification with $M=1,D=2$, and $N_\text{n} = 4$.  Assuming that the input edges are not co-incident on any node, it is necessary that the four output nodes be adjacent to the input edges since there are no other nodes in the network.  The situation we will describe in fact happens whenever the output nodes and input edges are adjacent, so can consider a more general setting (with arbitrary $N_\text{n}$) in which class $1$ is at node $i$, class $2$ at node $j$, class $3$ at node $k$, and class $4$ at node $l$.  Accounting for the fact that $\zeta^1_{-1,1} = \zeta^1_{-1,-1} = \zeta^2_{1,1} = \zeta^2_{1,-1} = \zeta^3_{1,-1} = \zeta^3_{-1,-1} = \zeta^4_{1,1} = \zeta^4_{-1,1} = 0$, the four inequalities which must be satisfied to solve the four-class classification problem are 
\begin{eqnarray}
	\zeta^1_{1,1} &>& \zeta^2_{1,1} \\
	\zeta^2_{-1,1} &>& \zeta^3_{-1,1} \\
	\zeta^3_{-1,-1} &>& \zeta^4_{-1,-1} \\
	\zeta^4_{1,-1} &>& \zeta^1_{1,-1}.
\end{eqnarray}
We may divide each of these inequalities by the product of edge rates which factor out on both sides, so that
\begin{eqnarray}
	\frac{\zeta^1_{1,1}}{W_{ij}W_{kl}} &>& \frac{\zeta^2_{1,1}}{W_{ij}W_{kl}} \\
	\frac{\zeta^2_{-1,1}}{W_{ji}W_{kl}} &>& \frac{\zeta^3_{-1,1}}{W_{ji}W_{kl}} \\
	\frac{\zeta^3_{-1,-1}}{W_{ji}W_{lk}} &>& \frac{\zeta^4_{-1,-1}}{W_{ji}W_{lk}} \\
	\frac{\zeta^4_{1,-1}}{W_{ij}W_{lk}} &>& \frac{\zeta^1_{1,-1}}{W_{ij}W_{lk}}.
\end{eqnarray}
Now we use the fact that since the output nodes are adjacent to the input edges they must obey Equations \ref{eqijeq1}, \ref{eqijeq2}, \ref{eqkleq1}, and \ref{eqkleq2}.  Substituting from these equations, we arrive at 
\begin{eqnarray}
	\frac{\zeta^3_{-1,1}}{W_{ji}W_{kl}} &>& \frac{\zeta^2_{1,1}}{W_{ij}W_{kl}} \\
	\frac{\zeta^2_{-1,1}}{W_{ji}W_{kl}} &>& \frac{\zeta^3_{-1,1}}{W_{ji}W_{kl}} \\
	\frac{\zeta^3_{-1,-1}}{W_{ji}W_{lk}} &>& \frac{\zeta^2_{-1,1}}{W_{ji}W_{kl}} \\
	\frac{\zeta^2_{1,1}}{W_{ij}W_{kl}} &>& \frac{\zeta^3_{-1,-1}}{W_{ji}W_{lk}}.
\end{eqnarray}
One can see that this set of inequalities is contradictory.  As a result, four-class classification is impossible for $N_\text{n} = 4$, or more generally when the output nodes are adjacent to the input edges.

\section{Sharpness of the decision boundary}
\subsection{Derivation of Equation \ref{eqsharpbound} in the main text}
\subsubsection{Maximizing $\partial \pi_S(F;\bt) / \partial F$}
In the case $D=1$, we can simplify Equation \ref{eqmttzeta} of the main text applied at the substrate node $S$ as
\begin{equation}
	\pi_S(F;\bt) = \frac{\sum_{m = M_\text{min}}^{M_\text{max}}\zeta^S_m(\bt)e^{mF/2} }{\sum_{m = M_\text{min}}^{M_\text{max}}\bar{\zeta}_m(\bt)e^{mF/2} }. \label{eqboundcp-1}
\end{equation}
Here, $M_\text{min}$ refers to the smallest power of $e^{F/2}$ which occurs in the numerator or denominator and $M_\text{max}$ refers to the greatest such power.  We multiply the numerator and denominator of Equation \ref{eqboundcp-1} by $e^{-M_\text{min} F /2}$, define $M_R \equiv M_\text{max} - M_\text{min}$, and re-define the index as $m \leftarrow m - M_\text{min}$, so that
\begin{equation}
	\pi_S(F;\bt) =\frac{\sum_{m=0}^{M_R}\zeta_m^S(\bt)e^{m F/2}}{\sum_{m=0}^{M_R}\bar{\zeta}_m(\bt)e^{m F/2}}.
\end{equation}
We are interested in the magnitude of the gradient $\partial \pi_S / \partial F$ evaluated at the decision boundary $F=0$, which is (suppressing the $\bt$ dependence of the coefficients)
\begin{equation}
	\frac{\partial \pi_S(F;\bt)}{\partial F} = \frac{\sum_{m=0}^{M_R}\sum_{m'=0}^{M_R}(m-m')\zeta^S_m\bar{\zeta}_{m'} e^{mF/2} }{2\left(\sum_{m=0}^{M_R}\bar{\zeta}_{m} e^{mF/2}\right)^2 }.
\end{equation}
We note that an offset of the decision boundary location to $F_0$ could be absorbed into the coefficients by redefining $F\leftarrow F-F_0$.  At $F=0$ we have
\begin{equation}
	\frac{\partial \pi_S(0;\bt)}{\partial F} = \frac{\sum_{m=0}^{M_R}\sum_{m'=0}^{M_R}(m-m')\zeta^S_m\bar{\zeta}_{m'} }{2\left(\sum_{m=0}^{M_R}\bar{\zeta}_{m}\right)^2}.
\end{equation}
We can rewrite the numerator of this expression as
\begin{eqnarray}
	\sum_{m=0}^{M_R}\sum_{m'=0}^{M_R}(m-m')\zeta^S_m\bar{\zeta}_{m'} &=& \left(\sum_{m'=0}^{M_R}\bar{\zeta}_{m'}\right)\sum_{m=0}^{M_R}m\zeta^S_m - \left(\sum_{m=0}^{M_R}\zeta^S_m\right)\sum_{m'=0}^{M_R}m'\bar{\zeta}_{m'} \nonumber \\
	&\equiv& \bar{\zeta}_\text{tot}\sum_{m=0}^{M_R}m\zeta^S_m - \zeta^S_\text{tot}\sum_{m'=0}^{M_R}m'\bar{\zeta}_{m'}
\end{eqnarray}
and the denominator as $2\bar{\zeta}_\text{tot}^2$.  We thus have
\begin{equation}
	\frac{\partial \pi_S(0;\bt)}{\partial F} = \frac{\zeta^S_\text{tot}}{2\bar{\zeta}_\text{tot}}\left(\frac{\sum_{m=0}^{M_R}m\zeta^S_m}{\zeta^S_\text{tot}} - \frac{\sum_{m=0}^{M_R}m\bar{\zeta}_m}{\bar{\zeta}_\text{tot}}\right).\label{eqboundcp0}
\end{equation}
We next aim to maximize this function to understand its dependence on $M_R$, the range of powers of $e^{F/2}$. We first maximize with respect to the functions $\{\zeta^S_m\}_{m=0}^{M_R}$.  It is straightforward to show that
\begin{equation}
	\frac{\sum_{m=0}^{M_R}m\zeta^S_m}{\zeta^S_\text{tot}} \leq M_R
\end{equation}
with equality when $\zeta^S_{M_R} = \zeta^S_\text{tot}$ and the other $\zeta^S_m$ are zero.  Thus,
\begin{equation}
	\frac{\partial \pi_S(0;\bt)}{\partial F} \leq \frac{\zeta^S_\text{tot}}{2\bar{\zeta}_\text{tot}}\left(M_R - \frac{\sum_{m=0}^{M_R}m\bar{\zeta}_m}{\bar{\zeta}_\text{tot}}\right). \label{eqboundcp1}
\end{equation}
We next maximize this with respect to the quantities $\{\bar{\zeta}_m\}_{m=0}^{M_R}$.  To encode the fact that $\bar{\zeta}_m \geq \zeta^S_m \ \forall \ m$, we write $\bar{\zeta}_m = \zeta^S_m + \tilde{\zeta}_m^S$ and $\bar{\zeta}_\text{tot} = \zeta^S_\text{tot} + \tilde{\zeta}^S_\text{tot}$, with $\tilde{\zeta}_m^S \geq 0$ and $\tilde{\zeta}^S_\text{tot} \geq 0$.  We have
\begin{equation}
	\frac{\zeta^S_\text{tot}}{2\bar{\zeta}_\text{tot}}\left(M_R - \frac{\sum_{m=0}^{M_R}m\bar{\zeta}_m}{\bar{\zeta}_\text{tot}}\right) = \frac{\zeta^S_\text{tot}}{2(\zeta^S_\text{tot} + \tilde{\zeta}_\text{tot})}\left(M_R - \frac{M_R\zeta^S_\text{tot} +  \sigma}{\zeta^S_\text{tot} + \tilde{\zeta}_\text{tot}}\right)
\end{equation}
where $\sigma \equiv \sum_{m=0}^{M_R}m\bar{\zeta}_m - M_R\zeta^S_\text{tot}$.  Standard calculus techniques allow showing that this expression is optimized with respect to $\tilde{\zeta}_\text{tot}$ when
\begin{equation}
	\tilde{\zeta}_\text{tot} = \zeta^S_\text{tot} + \frac{2 \sigma}{M_R},
\end{equation}
i.e., when
\begin{equation}
	\bar{\zeta}_\text{tot} = 2\left(\zeta^S_\text{tot} + \frac{\sigma}{M_R}\right). \label{eqboundcp2}
\end{equation}
Inserting the solution into Equation \ref{eqboundcp1} gives
\begin{equation}
	\frac{\partial \pi_S(0;\bt)}{\partial F} \leq \frac{\zeta^S_\text{tot} M^2_R}{8(\zeta^S_\text{tot} M_R + \sigma)}. \label{eqboundcp3}
\end{equation}
We finally aim to maximize this expression with respect to the $\{\bar{\zeta}_m\}_{m=0}^{M_R}$ that enter the variable $\sigma$, subject to Equation \ref{eqboundcp2}.  Equation \ref{eqboundcp3} will be maximized when $\sigma = 0$, when Equation \ref{eqboundcp2} imposes $\bar{\zeta}_\text{tot} = 2 \zeta^S_\text{tot}$.  Recalling that maximization with respect to the quantities $\{\zeta^S_m\}_{m=0}^{M_R}$ yielded $\zeta^S_\text{tot} = \zeta^S_{M_R}$, we see that these conditions will be achieved if $\bar{\zeta}_0 = \bar{\zeta}_{M_R} = \zeta^S_{M_R}$ and all other $\bar{\zeta}_m = 0$.  At this point, we have the desired result
\begin{equation}
	\text{max}_{\{\zeta^S_m\},\{\bar{\zeta}_m\}}\frac{\partial \pi_S(0;\bt)}{\partial F} =\frac{M_R}{8}.
\end{equation}
Thus, the bound of the derivative's magnitude is proportional to the range of exponential powers.  We verified this expression numerically using conjugate gradient optimization under the assumption that all functions $\zeta_m^S$ and $\tilde{\zeta}^S_m$ are free and independent.

\subsubsection{Minimizing $\partial \pi_S(F;\bt) / \partial F$}
Let us now minimize Equation \ref{eqboundcp0}.  We have that
\begin{equation}
	\frac{\sum_{m=0}^{M_R}m\zeta^S_m}{\zeta^S_\text{tot}} \geq 0
\end{equation}
with equality when $\zeta^S_{0} = \zeta^S_\text{tot}$ and the other $\zeta^S_m$ are zero.  This implies
\begin{equation}
	\frac{\partial \pi_S(0;\bt)}{\partial F} \geq -\frac{\zeta^S_\text{tot}}{2\bar{\zeta}_\text{tot}}\frac{\sum_{m=0}^{M_R}m\bar{\zeta}_m}{\bar{\zeta}_\text{tot}} = -\frac{\zeta^S_\text{tot}}{2(\zeta^S_\text{tot} + \tilde{\zeta}_\text{tot})^2} \sum_{m=0}^{M_R}m\tilde{\zeta}_m^S, \label{eqboundcp4}
\end{equation}
where the last equality results from the condition that only $\zeta^S_0$ is non-zero when writing $\bar{\zeta}_m = \zeta^S_m + \tilde{\zeta}^S_m$.  Equation \ref{eqboundcp4} is minimized when $\tilde{\zeta}^S_{M_R} = \tilde{\zeta}_\text{tot}$ and when $\tilde{\zeta}_\text{tot} = \zeta^S_\text{tot}$, i.e. when $\bar{\zeta}_0 = \bar{\zeta}_{M_R} = \zeta^S_{0} $.  Inserting these values into Equation \ref{eqboundcp0} yields
\begin{equation}
	\text{min}_{\{\zeta^S_m\},\{\bar{\zeta}_m\}}\frac{\partial \pi_S(0;\bt)}{\partial F} =-\frac{M_R}{8}. 
\end{equation}
We again verified this result numerically using conjugate gradient optimization.

To summarize, the magnitude of the derivative is bounded as 
\begin{equation}
	\left| \frac{\partial \pi_S(0;\bt)}{\partial F} \right| \leq \frac{M_R}{8}. \label{eqboundcp5}
\end{equation}
Returning to the original definition of the index $m$ which runs from $M_\text{min}$ to $M_\text{max}$, the lower bound is saturated when $\bar{\zeta}_{M_\text{min}} = \bar{\zeta}_{M_\text{max}} = \zeta^S_{M_\text{min}}$, and the upper bound is saturated when $\bar{\zeta}_{M_\text{min}} = \bar{\zeta}_{M_\text{max}} = \zeta^S_{M_\text{max}}$.

\subsubsection{Tightening the bound}
The exponential range $M_R$ in Equation \ref{eqboundcp5} refers to the difference between the maximum and minimum powers of $e^{F/2}$ in the terms of Equation \ref{eqboundcp-1} whose coefficients $\bar{\zeta}_m $ are non-zero.  These functions $\bar{\zeta}_m = \sum_i \zeta^i_m$ sum over all nodes in the network, but in order for the output function to saturate near $\pi_i \approx 1$ in its assigned input region for class $i$ the terms $\zeta^j_me^{mF/2}$ for $j\neq i$ should be small in this region.  The trained network will thus tend to have large values of $\zeta^i_m$ only for nodes $i$ in the set of designated output nodes $\mathcal{O}$.  If the non-output nodes had large values then they would compete with the output nodes and prevent their probabilities from saturating in their designated input regions.  As a result, we can approximate $\pi_S(F)$ near the decision boundary as (cf. Equation \ref{eqboundcp-1})
\begin{equation}
	\pi_S(F;\bt) \approx \frac{\sum_{m = M^\mathcal{O}_\text{min}}^{M^\mathcal{O}_\text{max}}\zeta^S_m(\bt)e^{mF/2} }{\sum_{m = M^\mathcal{O}_\text{min}}^{M^\mathcal{O}_\text{max}}\bar{\zeta}^\mathcal{O}_m(\bt)e^{mF/2} }
\end{equation}
where $\bar{\zeta}^\mathcal{O}_m(\bt) \equiv \sum_{i\in \mathcal{O}} \zeta^i_m(\bt)$ and $M^\mathcal{O}_\text{min}$ (resp. $M^\mathcal{O}_\text{max}$) is the minimum (resp. maximum) power of $e^{F/2}$ whose coefficient $\bar{\zeta}^\mathcal{O}_m(\bt)$ is non-zero.  From this, we can effectively tighten the bound in Equation \ref{eqboundcp5} as 
\begin{equation}
	\left| \frac{\partial \pi_S(0;\bt)}{\partial F} \right| \lesssim \frac{M_R^\mathcal{O}}{8} \leq \frac{M_R}{8} \label{eqcoundcp6}
\end{equation}
where $M_R^\mathcal{O} \equiv M^\mathcal{O}_\text{max} - M^\mathcal{O}_\text{min} \leq M_R$.   There will be at least as large a range $M_R$ of powers among all nodes as the range $M_R^\mathcal{O}$ among the output nodes.  Because of this, at extremal values of $F \rightarrow \pm \infty$ the highest and lowest powers of $e^{F/2}$ will dominate, which may not correspond to the large probabilities at the output nodes.  Near the decision boundary, however, and for the finite values of $F$ in the data on which the network was trained, the approximation $\bar{\zeta}_m \approx \bar{\zeta}^\mathcal{O}_m$ should hold due to the training goal of saturating $\pi_i \approx 1$ for $i \in \mathcal{O}$ in these regions of input space.  

\subsection{Analysis of trees in the extended push-pull networks}
How does this analysis help to understand the difference between the parallel and series extensions of the push-pull network in Figure \ref{Sharpness} of the main text?  We first consider the parallel extension.  The range in exponential powers $M^\mathcal{O}_R$ depends on the set of directed spanning trees.  A maximum power of the exponent  will occur for a tree rooted on node $S$ or $S^*$ in which the path connecting $S$ and $S^*$ occurs among the un-driven edges (black edges in Figure \ref{Sharpness}).  The directed edges for these rooted spanning trees will be oriented in the direction of $F>0$ for $n+1$ of the driven edges, contributing a term of $e^{(n+1)F/2}$.  A minimum power of the exponent will occur for trees rooted on node $S$ or $S^*$ in which the path connecting $S$ and $S^*$ occurs among the driven edges.  The directed edges for these rooted spanning trees will have two edges oriented in opposite directions relative to $F>0$, cancelling, and $n$ edges oriented in the direction of $F>0$.  These trees thus contribute a term of $e^{nF/2}$.  The range $M^\mathcal{O}_R = 1$ for the parallel extension does not grow with $n$, explaining why the decision boundary cannot get sharper as $n$ increases.

By contrast, $M^\mathcal{O}_R$ can grow with $n$ for the serial extension of the push-pull network.  A maximum power of the exponent will occur for trees rooted on node $S$ or $S^*$ in which the removed edge adjacent to the central top node.  This will contribute a term $e^{(n+1)F/2}$.  A minimum power will occur for trees in which the removed edge is taken from among the un-driven edges.  These trees will contribute a term $e^{0}$, implying that $M^\mathcal{O}_R = n+1$, which grows with $n$ as desired.  Note that $M^\mathcal{O}_R = 1$ for $n=0$ for both serial and parallel. 

In Figure \ref{SI_PushPullzetas} we plot the various functions $\zeta^S_m$, $\zeta^\mathcal{O}_m$, and $\bar{\zeta}_m$ which have been learned during training of the extended push-pull networks.  These numerical results validate the approximation $\zeta^\mathcal{O}_m \approx \bar{\zeta}_m$ and the strategy of trying to set $\bar{\zeta}_{M_\text{min}^\mathcal{O}} = \bar{\zeta}_{M_\text{max}^\mathcal{O}} = \zeta^S_{M_\text{min}^\mathcal{O}}$ to minimize $\partial_S \pi_S (0,F)$.

\begin{figure*}[ht!]
	\begin{center}
		\includegraphics[width=\textwidth]{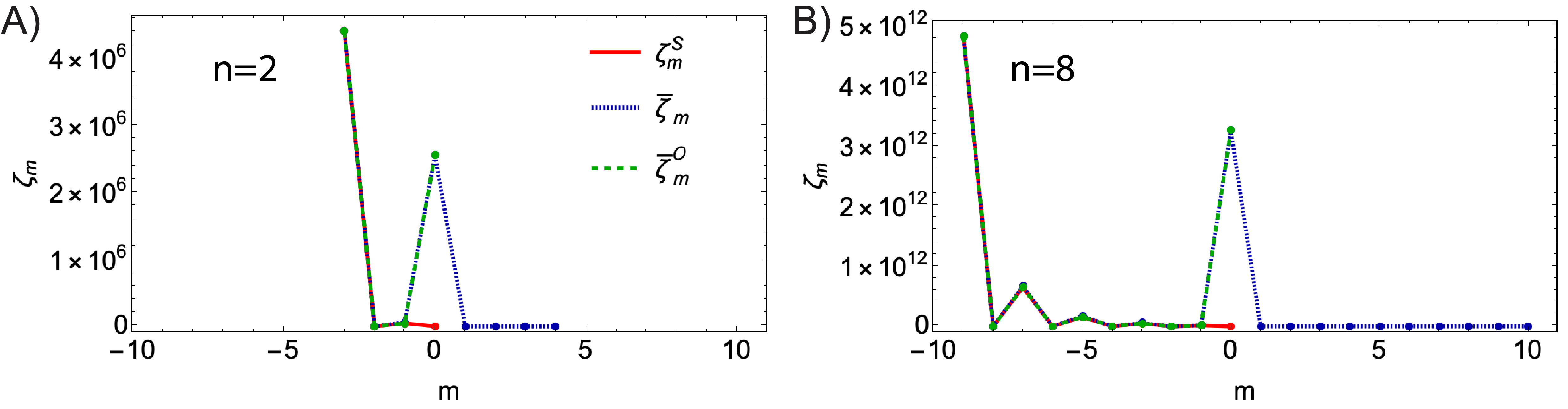}
		\caption{Plots of the learned values $\{\zeta^S_m\}_{m=M_\text{min}}^{M_\text{max}}$, $\{\bar{\zeta}_m\}_{m=M_\text{min}}^{M_\text{max}}$, and $\{\bar{\zeta}^{\mathcal{O}}_m\}_{m=M_\text{min}}^{M_\text{max}}$ for the serially extended push-pull networks in Figure \ref{Sharpness} of the main text.  }
		\label{SI_PushPullzetas}
	\end{center}
\end{figure*}

\begin{figure*}[ht!]
	\begin{center}
		\includegraphics[width=\textwidth]{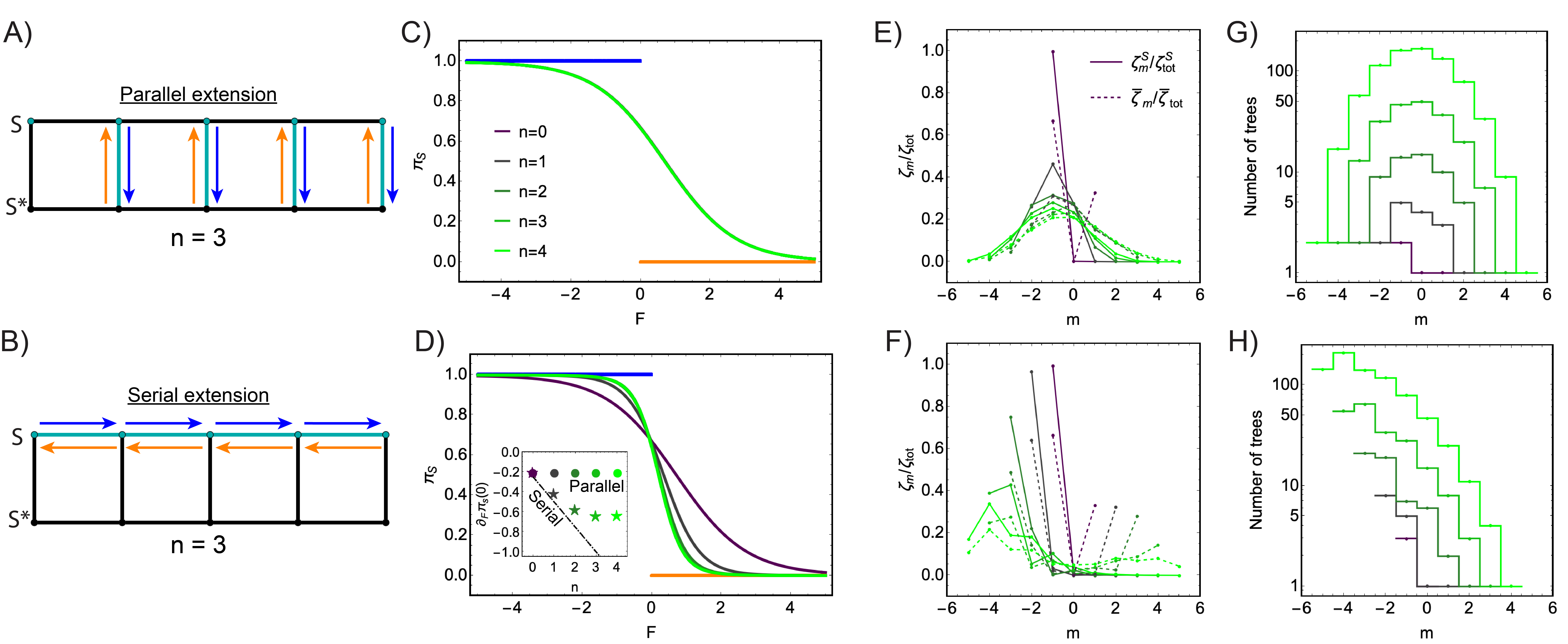}
		\caption{Sharpening decision boundaries for common biochemical motifs. A,B) Illustrations of a ladder network with edges driven in parallel (A) or in series (B).  C,D)  Plots of the trained $\pi_S(F;\bt)$ curves for increasing numbers of nodes in the parallel (C) or serial (D) extensions of the ladder network.  The inset is the same as in Figure \ref{Sharpness}C of the main text; here the dashed and dotted lines overlap.  E,F)  Plots of the trained functions for the parallel (E) or serial (F) extensions $\{\zeta_m^S\}_{m=M_\text{min}}^{M_\text{max}}$ and $\{\bar{\zeta}_m\}_{m=M_\text{min}}^{M_\text{max}}$ as fractions of their summations over $m$.  G,H)   Histograms of trees in the parallel (G) or serial (H) extensions rooted at node $S$ which have a net number $m$ of contributions from the driven edges.  
		}
		\label{SI_Sharpness}
	\end{center}
\end{figure*}

\subsection{Analysis of trees in the extended ladder networks}

We consider another classic biochemical motif, a ladder network, which is often used to describe sequential conformational states proteins such as the flagellar motor \cite{tu2008nonequilibrium, owen2023size}.  We consider extending these motifs as before in a parallel and a serial fashion (Figures \ref{SI_Sharpness}A,B).  
\begin{wrapfigure}{l}{0.5\textwidth}
	\begin{center}
		\includegraphics[width=0.45\textwidth]{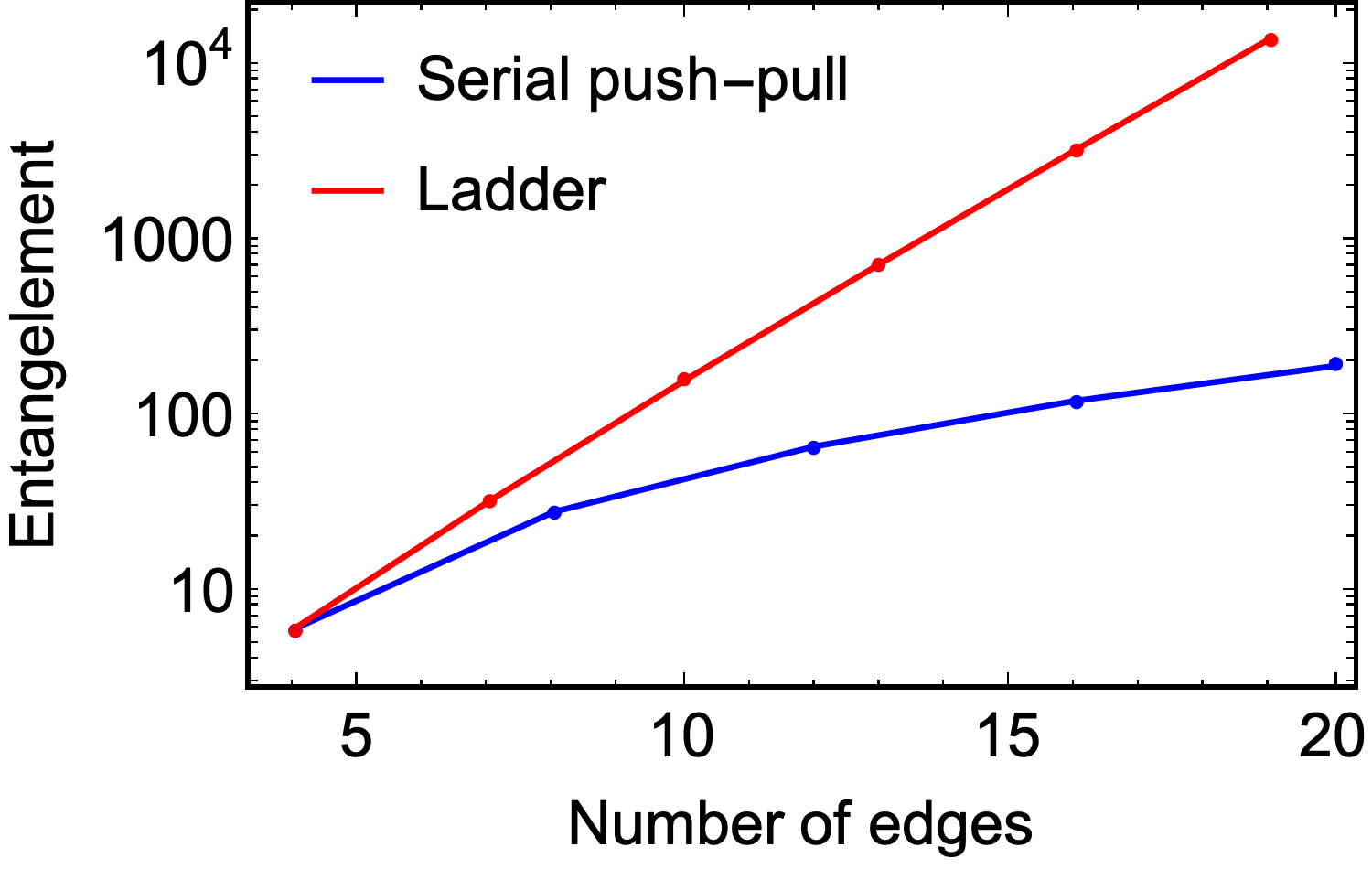}
	\end{center}
	\caption{Entanglement, defined as the average number of directed trees which a given directed edge is a part of, as a function of network size for the serial push-pull and ladder networks.}
	\label{Entanglements}
\end{wrapfigure}
Training these networks for increasing $n$, we find as before that the parallel strategy does not allow sharpening the boundary whatsoever (Figures \ref{SI_Sharpness}C).  In contrast to the previous example, however, the gain in sharpness from adding edges in a serial manner begins to diminish after $n\sim 3$ (Figures \ref{SI_Sharpness}D).  As for the push-pull networks, one can analyze the spanning trees in the ladder networks to show that for both the parallel and serial extensions, $M^\mathcal{O}_R = 2(n+1)$.  To understand why this bound is not tight, recall how saturating Equation \ref{eqboundcp5} requires the concentrating all the ``weight'' of the $\zeta_m^S$ functions to $\zeta^S_{M^\mathcal{O}_\text{min}}$ and the weight of the $\bar{\zeta}_m$ functions evenly to $\bar{\zeta}_{M^\mathcal{O}_\text{min}}$ and $\bar{\zeta}_{M^\mathcal{O}_\text{max}}$.  As discussed in Section \ref{inputfrust} above, equality constraints among the coefficients reduce the number of degrees of freedom. We are prevented from freely setting these functions as desired since they are themselves complicated and interrelated polynomials of the adjustable $W_{ij}$ parameters.  Each $\zeta^S_m$ involves a sum over all directed trees rooted at node $S$ which include a net number $m$ of driven edges (counted with sign according to the directed edge orientation relative to positive driving).  In Figures \ref{SI_Sharpness}E,F we show the values of $\zeta^S_m$ and $\bar{\zeta}_m$ for the parallelly and serially extended ladder networks, and in Figures \ref{SI_Sharpness}G,H we show histograms of the number of trees which contribute to the sums for each $\zeta^S_m$ term for these networks.  We see that the the multiplicity of trees contributing to $\zeta^S_m$ for intermediate values of $m$ for serially extended networks continues to grow with $n$, increasing the corresponding intermediate values of $\zeta^S_m$ and preventing the network from pushing weight toward $\zeta^S_{M^\mathcal{O}_\text{min}}$.  In other words, even though the range of exponential powers and bound on sharpness continues to grow, there are too many trees competing with the extremal ones which weakens the allowable gain in sharpness.  For these ladder networks there are many entanglements of the degrees of freedom, and the sharpness optimization point likely lies far from the reachable subspace spanned by the trainable degrees of freedom $W_{ij}$.  

A simple metric supports this picture.  For a given graph we count how many directed trees a given directed edge is a part of, and we average this number over all directed edges in the graph.  This gives a rough estimate of entanglement (corresponding to the leftmost two layers of the schematic in Figure \ref{MultiClassFrustration}B of the main text).  In Figure \ref{Entanglements} we show this metric defined for increasing sizes of the serial push-pull network and the ladder networks.  For the ladder networks, there are orders of magnitude more trees which a single degree of freedom $W_{ij}$ is typically a part of, making the optimization task more difficult to solve and preventing saturation of the bound in Equation \ref{eqcoundcp6}.  

\section{Low dimensionality of trained networks}

\begin{figure*}[ht!]
	\begin{center}
		\includegraphics[width= 0.9 \textwidth]{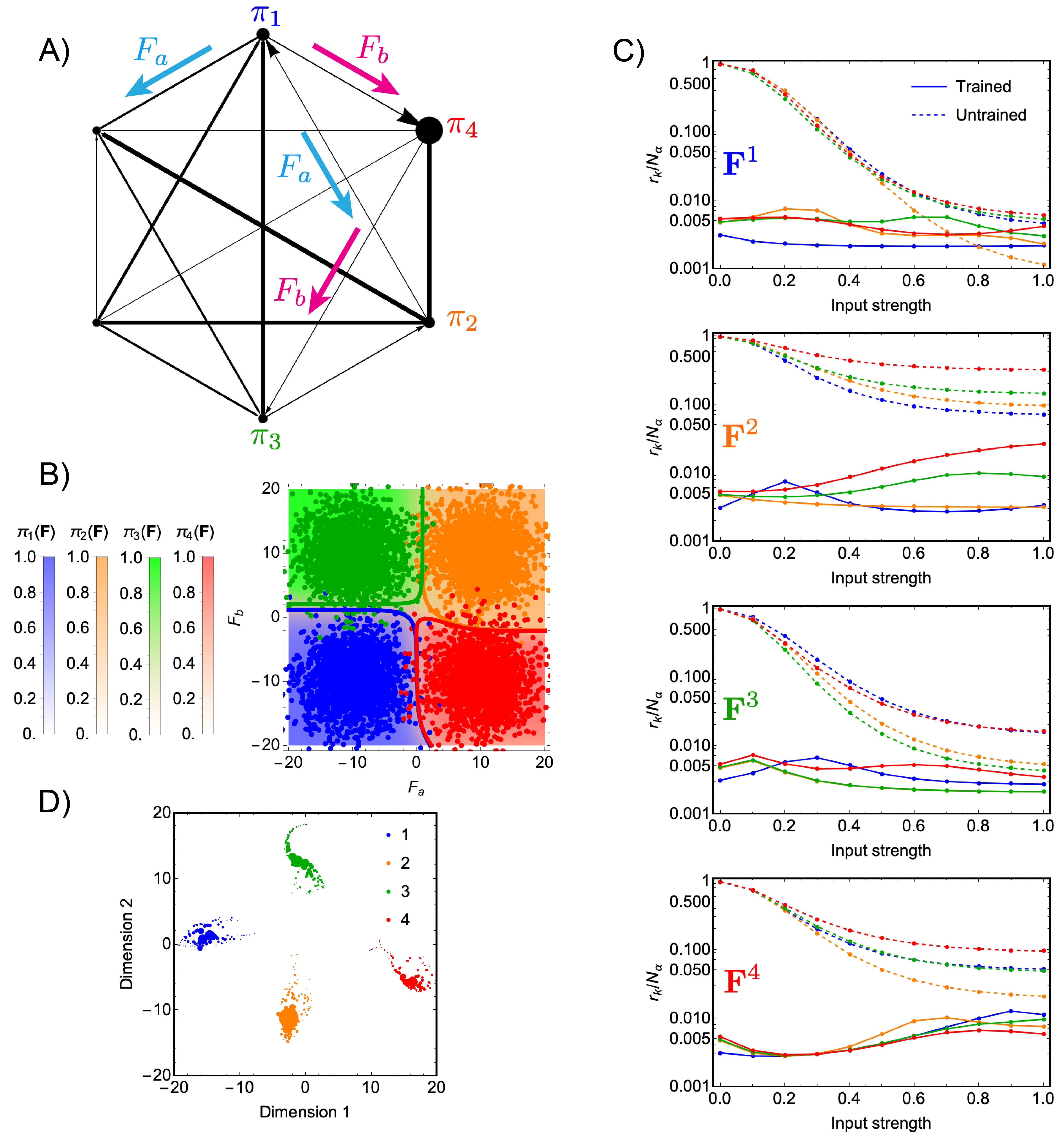}
		\caption{Low dimensionality of a trained network.  A) Illustration of the problem definition and the learned parameters (cf. Figure \ref{SI_TrainedGraphsHexagon} above).  B)   Plot of the learned decision boundaries which solve the four-class classification problem.  C)  Plots of the fractional inverse participation ratio $r_k / N_\alpha \in [1/N_\alpha,1]$ for $k=1,2,3,4$ in trained and untrained graphs, as input forces $f\mathbf{F}^\rho$ from patterns $\rho=1,2,3,4$ are presented for increasing strengths $f \in [0,1]$.  D)  A visualization showing clustering of the high dimensional $\mathbf{A}(\mathbf{F})$ which record the frenesy for each tree in the tree space, projected into two reduced dimensions using t-SNE.  100 samples of $\mathbf{F}^\rho$ from each class $\rho = 1,2,3,4$ (the data clouds in panel B) are drawn, and the projection $\mathbf{A}(\mathbf{F}^\rho)$ for each input is a point in the graph colored by $\rho$.  The points are additionally sized so that the projected vectors $\mathbf{A}(\mathbf{F})$ is larger if $\mathbf{F}^\rho$ is close to the center of the class $\rho$. }
		\label{DimensionalityPlots}
	\end{center}
\end{figure*}

Here we ask how networks trained to perform classification differ from networks with uniform weights.  It is known that neural networks trained to perform classification have a subset of weights which are important for accurate performance and a different, less important subset which can actually be pruned from the network without significantly decreasing the accuracy \cite{han2015learning}.  Relatedly, physical networks trained to perform classification have heterogeneous weights and more localized normal modes \cite{stern2024physical}.  To explore this in the context of Markov networks, we consider a standard measure of mode delocalization called the inverse participation ratio (IPR).  For an eigenbasis $\{\mathbf{v}_k\}_{k=1}^{N_\text{n}}$, where each vector $\mathbf{v}_k$ has components $v_{k,\alpha}$ the IPR is defined as 
\begin{equation}
	r^H_k = \frac{\left(\sum_\alpha v_{k,\alpha}^2 \right)^2}{\sum_\alpha v_{k,\alpha}^4} = v_k^{4}\left(\sum_\alpha v_{k,\alpha}^4\right)^{-1} \label{eqiprs}
\end{equation}
where $v_k$ is the vector norm of $\mathbf{v}_k$.  It is straightforward to show that if the eigenmode is spread evenly over its $N_\alpha$ components (i.e., $v_{k,\alpha} \propto 1/N_\alpha$), then $r_k = N_\alpha$; if it is fully localized on one component, then $r_k = 1$.  Analogously, we define for node $k$ of the Markov network
\begin{equation}
	r_k = \frac{\left(\sum_\alpha w(T_k^\alpha) \right)^2}{\sum_\alpha w(T_k^\alpha)^2} = w_k^2 \left( \sum_\alpha w(T_k^\alpha)^2\right)^{-1}. \label{eqinvpart}
\end{equation}
To justify this definition, we note that $\pi_k \propto \sum_\alpha w(T_k^\alpha) \equiv w_k$.  If only one directed tree, say $\beta$, contributes to this sum, then $w(T_k^\alpha) = w_k \delta_{\alpha \beta}$, and $r_k = 1$.  If, instead, every tree contributes evenly, then $w(T_k^\alpha) = 1/ w_k$, and $r_k = N_\alpha$.  This definition thus has the desired properties to indicate how delocalized the contribution of spanning tree weights are to the steady-state probability $\pi_k$.  We use the exponent $2$ instead of $4$ because the vector components $w(T_k^\alpha)$ are all non-negative and their ``natural'' norm is $w_k = \sum_\alpha w(T_k^\alpha)$, not $\sum_\alpha w(T_k^\alpha)^2$.  Mathematically, though, using the exponent $4$ instead would also have the desired property that $r_k$ ranges from $1$ to $N_\alpha$. 

We next study the IPR in untrained ($\bt = \mathbf{0}$) and trained ($\bt = \bt^*$) graphs.  The graph shown in Figure \ref{DimensionalityPlots}A was trained to solve a four-class classification task, with the results shown in Figure \ref{DimensionalityPlots}B.  We find that the trained network with $\mathbf{F} = \mathbf{0}$ has a significantly lower value of $r_k$ than the untrained network for each of the output nodes, $r_1$, $r_2$, $r_3$, $r_4$ (Figure \ref{DimensionalityPlots}C).  Additionally, these values depend on which input pattern is presented to the network and the strength at which it is presented.  As the input forcing $\mathbf{F}$ is presented more strongly the values of $r_k$ generally decrease in the untrained graph, because certain trees aligned with the input forcing begin to dominate over others.  However, in the trained graph, the delocalization $r_i$ does not necessarily decrease when input $\mathbf{F}^j$ is presented because this input is exciting trees other than those which have been trained to contribute to node $i$.  As during development of neural pathways in the brain, training of these networks for a classification task graphs has, via updates to $\bt$, winnowed certain trees from contributing to $\pi_i$ and reinforced others that help to solve the task \cite{sanes2011development, gerstner2014neuronal}.  We note that these results are broadly consistent with those described for neural networks and resistor and spring networks in Refs.\ \citenum{han2015learning,stern2024physical}.  

To explore which susbets of trees contribute to the networks response, we define a measure of undirected tree activation in terms of the total frenesy $A(T_\alpha) = \sum_{ij \in T_\alpha}\left(\pi_j W_{ij} + \pi_i W_{ji}\right)$ along the edges of the tree $T_\alpha$ at steady state.  We computed the high dimensional vector over tree space $\mathbf{A}(\mathbf{F})$ for hundreds input forces drawn from the four input classes in Figure \ref{DimensionalityPlots}B.  We then use t-SNE \cite{van2008visualizing} to reduce the dimensionality of these vectors from $N_\alpha$ to 2 (Figure \ref{DimensionalityPlots}D).  This allows easy visualization of the grouping of vectors $\mathbf{A}(\mathbf{F}^\rho)$ according to class $\rho$, illustrating that the tree activation profiles are clustered in this high-dimensional space according to which class the input force is drawn from.  Thus, the trained network tends to allocate disjoint sets of trees which are separately responsible for coupling the input $\mathbf{F}^{\rho}$ to $\pi_\rho$.

\section{Softmax-like form of the classification function}
\subsection{Connection to transformers and modern Hopfield networks}
Here we explore connections between classification using Markov networks as treated in this paper and recent developments in machine learning that link ``modern Hopfield models'' of associative memory to transformer architectures, which are commonly used in technologies like ChatGPT.  We first summarize this link, and then discuss how our work may fit into this emerging picture.  

In the classic Hopfield model of associative memory \cite{hopfield1982neural}, one considers a trained Hamiltonian of spin degrees of freedom $\mathbf{x}$ which involves a sum over patterns $\boldsymbol{\xi}^\rho$ vectors:
\begin{equation}
	\mathcal{F}(\mathbf{x};\{\boldsymbol{\xi}^\rho\}) = -\sum_{\rho} (\boldsymbol{\xi}^\rho \cdot \mathbf{x})^2. \label{eqhopclassic}
\end{equation}
This expression is equivalent to the spin-glass energy
\begin{equation}
	\mathcal{F}(\mathbf{x};\{\boldsymbol{\xi}^\rho\}) = -\sum_{\alpha,\beta} x_\alpha J_{\alpha\beta} x_\beta
\end{equation}
where 
\begin{equation}
	J_{\alpha \beta} = \sum_\rho \xi^\rho_{\alpha} \xi^\rho_{\beta}
\end{equation}
is given by the Hebbian rule.  Equation \ref{eqhopclassic} has recently been generalized into a so-called modern Hopfield model as \cite{ramsauer2020hopfield, lucibello2024exponential}
\begin{equation}
	\mathcal{F}(\mathbf{x};\{\boldsymbol{\xi}^{\rho}\}) = -\ln \sum_{\rho} F(\boldsymbol{\xi}^{\rho} \cdot \mathbf{x}) \label{eqmhen}
\end{equation}
for some non-linear function $F$.  Taking $F(x) = e^{x}$, which can be viewed as including all orders of a polynomial expansion of $\boldsymbol{\xi}^{\rho} \cdot \mathbf{x}$, has been shown to provide a significantly larger capacity for pattern memory than in classical counterpart \cite{lucibello2024exponential}.  Additionally, this form leads to an energy gradient-based dynamical update rule for $\mathbf{x}$ which is equivalent to a softmax formula that compares $\mathbf{x}$ to the input patterns $\boldsymbol{\xi}^\rho$ \cite{lucibello2024exponential}.  This softmax update rule introduces an interesting connection between models of associative memory and transformers, which commonly use softmax functions to compare a query vector $\mathbf{x}$ to a set of key vectors $\boldsymbol{\xi}^\rho$.  

To connect to our work on classification tasks performed by Markov networks, we note that transformers can in some cases be made more computationally efficient using an approximation called linear attention:
\begin{equation}
	\text{Softmax}(\mathbf{x},\boldsymbol{\xi}^\rho) = \frac{e^{\boldsymbol{\xi}^\rho \cdot \mathbf{x}}}{\sum_{\rho'} e^{\boldsymbol{\xi}^{\rho'} \cdot \mathbf{x}}} \approx \frac{\boldsymbol{\phi}_\xi(\boldsymbol{\xi}^\rho) \cdot \boldsymbol{\phi}_x(\mathbf{x})}{\sum_{\rho'} \boldsymbol{\phi}_\xi(\boldsymbol{\xi}^{\rho'}) \cdot \boldsymbol{\phi}_x(\mathbf{x})}, \label{eqcrcp1}
\end{equation}
where $\boldsymbol{\phi}_\xi$ and $\boldsymbol{\phi}_x$ are non-linear vector functions which featurize the input data and patterns \cite{katharopoulos2020transformers}.  Interestingly, linear attention has recently been shown to appear naturally in a biophysical model for the computational role played by astrocytes in the brain \cite{kozachkov2023building}.  Comparison of Equation \ref{eqcrcp1} and Equation \ref{eqmtt} of the main text reveals the way in which classification using Markov networks implements a type of softmax-based attention on the input data $\mathbf{F}$ using the non-linear feature vectors $\boldsymbol{\psi}(i;\bt)$ and $\boldsymbol{\chi}(i,\mathbf{F})$.  We note, however, that our feature vectors functions are different functions for each node, i.e., $\boldsymbol{\psi}(i;\bt) \neq \boldsymbol{\psi}(j;\bt)$ and similarly $\boldsymbol{\chi}(j;\mathbf{F}) \neq \boldsymbol{\chi}(j;\mathbf{F})$.  We next show how Equation \ref{eqmtt} in the main text is derived from the matrix tree theorem.

\subsection{Derivation of Equation \ref{eqmtt} in the main text}
Starting from Equations \ref{eqmtt0} and \ref{eqmtt} of the main text, we first identify 
\begin{equation}
	\psi(T_i^\alpha;\bt)\chi(T_i^\alpha, \mathbf{F}) = w(T_i^\alpha, \mathbf{F};\bt)\label{eqsmcp0}
\end{equation} 
and can hence interpret the sum over $\alpha$ as a dot product among the ``vector elements'' $\psi(T_i^\alpha;\bt)$ and $\chi(T_i^\alpha, \mathbf{F})$.  This decomposition is possible because the directed tree weight $w(T_i,\mathbf{F};\bt)$ depends exponentially on sums involving the parameters $\bt$ and input forces $\mathbf{F}$.   Specifically, we have
\begin{equation}
	w(T_i^\alpha, \mathbf{F};\bt) = \text{exp}\left( \sum_k s^E(k,T_i^\alpha) E_k -  \sum_{jk} s^B(jk,T_i^\alpha) B_{jk} + \sum_{jk}s^F(jk,T_i^\alpha)F_{jk} \right)\text{exp}\left(\sum_a s^{F}(a,T_i^\alpha)F_{a} \right) \label{eqsmcp1}
\end{equation}
where $s^E(k,T^\alpha_i) \in \{0,1\}$ is a structural indicator of whether $E_k$ appears in the product edge rates for the directed tree $T^\alpha_i$, $s^B(jk,T_i^\alpha) \in \{0,1\}$ similarly encodes the structural dependence $B_{jk}$, $s^F(jk,T^\alpha_i) \in \{-1/2,0,1/2\}$ does so for the learned parameter $F_{ij}$, and $s^F(a,T_i^\alpha)\in \{-1/2,0,1/2\}$ does so for the input force $F_a$.  We can write this all more compactly as in Equation \ref{eqsmcp0} with  
\begin{equation}
	\psi(T_i^\alpha;\bt) = \exp\left(\mathbf{u}_i^\alpha \cdot \bt\right)
\end{equation}
and
\begin{equation}
	\chi(T_i^\alpha;\mathbf{F}) = \exp\left(\mathbf{v}_i^\alpha \cdot \mathbf{F}\right),
\end{equation}
Here, $\mathbf{u}_i^\alpha$ is a vector with the same dimension as the total number of parameters which contains the structural factors $s^E,\ s^B, \ s^F$ for directed tree $T_i^\alpha$.   Similarly, $\mathbf{v}_i^\alpha$ contains the structural factors $s^F$ for $T_i^\alpha$.  A dot product over trees $*$ can be defined in terms of the vector elements $\psi(T_i^\alpha;\bt)$ and $\chi(T_i^\alpha,\mathbf{F})$, and we can write
\begin{equation}
	\boldsymbol{\psi}(i;\bt) * \boldsymbol{\chi}(i,\mathbf{F}) = \sum_{\alpha} \psi(T_i^\alpha;\bt) \chi(T_i^\alpha,\mathbf{F}).
\end{equation}
Since the elements of $\boldsymbol{\psi}(i;\bt)$ and $\boldsymbol{\chi}(i,\mathbf{F})$ are non-negative, the dot product is as well.  With these definitions, we can rewrite the steady-state occupation as in Equation \ref{eqmtt} of the main text.  

\subsection{Creased landscape of $\ln Z$}
Dynamics in modern Hopfield networks are governed by relaxation down the gradient of the energy function (Equation \ref{eqmhen}), which is a quantity of central interest.  By analogy, we are led to consider the quantity
\begin{equation}
	\mathcal{F}(\mathbf{F};\bt) = -\ln Z (\mathbf{F};\bt)= -\ln \sum_k \boldsymbol{\psi}(k;\bt) * \boldsymbol{\chi}(k,\mathbf{F}).
\end{equation}
Here, we characterize $\mathcal{F}$ and explore how it differs between trained and untrained networks.  

We train the graph shown in Figure \ref{SI_CreasedLandscape}A to solve a four-class classification problem, and it learns the solution shown in Figure \ref{SI_CreasedLandscape}B; note that these data clouds are rotated in input space relative to other four-class classification tasks presented in this paper.  In Figure \ref{SI_CreasedLandscape}C we plot the entropy 
\begin{equation}
	S[\bp] = -\sum_k \pi_k \ln \pi_k
\end{equation}
of the steady-state distribution over input space.  Regions of high entropy tend to localize along the learned decision boundaries, and away from these regions the entropy is generally low because the output nodes dominate the steady state.  

In Figures \ref{SI_CreasedLandscape}D and E we plot $\mathcal{F}(\mathbf{F};\bt)$ for the untrained ($\bt = \mathbf{0}$) and trained ($\bt = \bt^*$) networks.  We see that in both cases it is characterized by flat sloping regions that are ``creased'' along strips of high curvature $\nabla_{\mathbf{F}}^2 \mathcal{F}(\mathbf{F};\bt)$.  In the trained network, some of these creases localize to  the learned decision boundaries (Figure \ref{SI_CreasedLandscape}F).  Zooming out reveals that training has not eliminated, but merely shifted, the creases found in the untrained network (Figure \ref{SI_CreasedLandscape}G).  The zoomed-out structure of the gradient $\nabla_{\mathbf{F}} \mathcal{F}(\mathbf{F})$ for both networks is similar; away from where the region of input space where the network was trained, the vector field $\nabla_{\mathbf{F}} \mathcal{F}(\mathbf{F})$ has the same behavior at limiting regions of the domain (\ref{SI_CreasedLandscape}H and I).  In particular, both networks contain a diagonal strip of mismatch between the vector fields approaching the top right quadrant of the domain.

To understand this structure, we compute the gradient of $\mathcal{F}$ using the multi-index notation of Equation \ref{eqmttzeta} in the main text
\begin{equation}
	-\nabla_{\mathbf{F}} \ln Z = -\frac{\nabla_{\mathbf{F}} Z}{Z} = - \frac{\sum_\mu \bar{\zeta}_\mu(\bt) \nabla_{\mathbf{F}}y^\mu(\mathbf{F})}{\sum_\mu \bar{\zeta}_\mu(\bt) y^\mu(\mathbf{F})}
\end{equation}
The gradient $\nabla_{\mathbf{F}}y^\mu(\mathbf{F})$ can be written as 
\begin{eqnarray}
	\nabla_{\mathbf{F}}y^\mu(\mathbf{F}) &=& \nabla_{\mathbf{F}} \prod_{a\in\mathcal{A}} e^{\mu_a F_a / 2} \\ &=& \frac{\boldsymbol{\mu}}{2}\prod_{a\in\mathcal{A}} e^{\mu_a F_a / 2} \\
	&=& \frac{\boldsymbol{\mu}}{2}y^\mu(\mathbf{F})
\end{eqnarray}
where $\boldsymbol{\mu}$ is a vector containing the the components of the multi-index $\mu$.  We have
\begin{eqnarray}
	-\nabla_{\mathbf{F}} \ln Z &=& -\frac{1}{2}\frac{\sum_\mu \boldsymbol{\mu} \bar{\zeta}_\mu(\bt) y^\mu(\mathbf{F})}{\sum_\mu \bar{\zeta}_\mu(\bt) y^\mu(\mathbf{F})} \\ 
	&=& -\frac{1}{2}\sum_k \frac{\sum_\mu \boldsymbol{\mu} \zeta^k_\mu(\bt) y^\mu(\mathbf{F})}{\sum_\mu \bar{\zeta}_\mu(\bt) y^\mu(\mathbf{F})} \\ 
	&\equiv& -\frac{1}{2} \sum_k \langle \boldsymbol{\mu}\rangle_k.
\end{eqnarray}
The quantity 
\begin{equation}
	\langle \boldsymbol{\mu}\rangle_k = \frac{\sum_\mu \boldsymbol{\mu} \zeta^k_\mu(\bt) y^\mu(\mathbf{F})}{\sum_\mu \bar{\zeta}_\mu(\bt) y^\mu(\mathbf{F})}
\end{equation}
records for node $k$ a direction in input space weighted by the relative contribution of the term $\zeta^k_\mu(\bt) y^\mu(\mathbf{F})$ to the sum $\bar{\zeta}_\mu(\bt) y^\mu(\mathbf{F})$.  The total gradient is proportional to a sum of $\langle \boldsymbol{\mu}\rangle_k$ over all nodes.  

In a region of input space for which $\pi_k(\mathbf{F};\bt) \approx 1$, then $-\nabla_{\mathbf{F}} \ln Z \approx - (1/2) \langle \boldsymbol{\mu}\rangle_k$.  At decision boundaries, where the steady-state occupation transitions to, say, $\pi_{k'}(\mathbf{F};\bt) \approx 1$, we will generally expect a region of high curvature in $\mathcal{F}$ if $\langle \boldsymbol{\mu}\rangle_k \neq \langle \boldsymbol{\mu}\rangle_{k'} $ because the gradient is transitioning from $(1/2) \langle \boldsymbol{\mu}\rangle_k$ to $(1/2) \langle \boldsymbol{\mu}\rangle_{k'}$.  Thus, a decision boundary will tend to co-localize to a strip of high curvature in $\mathcal{F}$.

While some creases can thus be created through training, by drawing new decision boundaries in input space, some creases are more fixed and result from the topological properties of the network.  These creases do not necessarily co-localize with decision boundaries.  For example, the horizontal and vertical creases parallel to the axes as well as the diagonal crease going off too the point $F_a\rightarrow \infty, \ F_b\rightarrow \infty$ appear in both the untrained and trained networks in Figures \ref{SI_CreasedLandscape}H and I.  These result from mismatched limiting behavior of $\nabla_{\mathbf{F}} \mathcal{F}$ at the boundaries of the domain.  For example, in Figure \ref{SI_CreasedLandscape}I, the mismatch between the limits at $\mathbf{F}=(-\infty, \infty)$ and $\mathbf{F} = (-\infty, -\infty)$ produces the crease separating the blue and magenta regions.  Additionally, the order of the limit matters.  In the top right quadrant, near $\mathbf{F} = (\infty, \infty)$, it can be shown that
\begin{equation}
	\lim_{F_a \rightarrow \infty} \lim_{F_b \rightarrow \infty} \nabla_\mathbf{F} \mathcal{F} \neq \lim_{F_b \rightarrow \infty} \lim_{F_a \rightarrow \infty} \nabla_\mathbf{F}\mathcal{F}.
\end{equation}
As a result, there is a different vector field when approaching $\mathbf{F} = (\infty, \infty)$ along $F_a$ and then $F_b$ than in the other order.  This produces a crease along the diagonal.  These creases which encode the behavior at the boundaries cannot be removed through training, because we assume training cannot set coefficients exactly to zero, it can only make them small.  For large enough $\mathbf{F}$, these terms will still eventually dominate and yield the same limiting behavior as in the untrained network.  

To summarize, the quantity $\mathcal{F}$, the analogous quantity to the energy in modern Hopfield networks, encodes both global, topological information about the Markov network through creases which result from mismatched behavior at the boundaries, as well as local information which results from training that produces new decision boundaries in the input space.  We leave to future work further exploration of possible connections between computations using Markov networks and other machine learning models.

\begin{figure*}[ht!]
	\begin{center}
		\includegraphics[width= \textwidth]{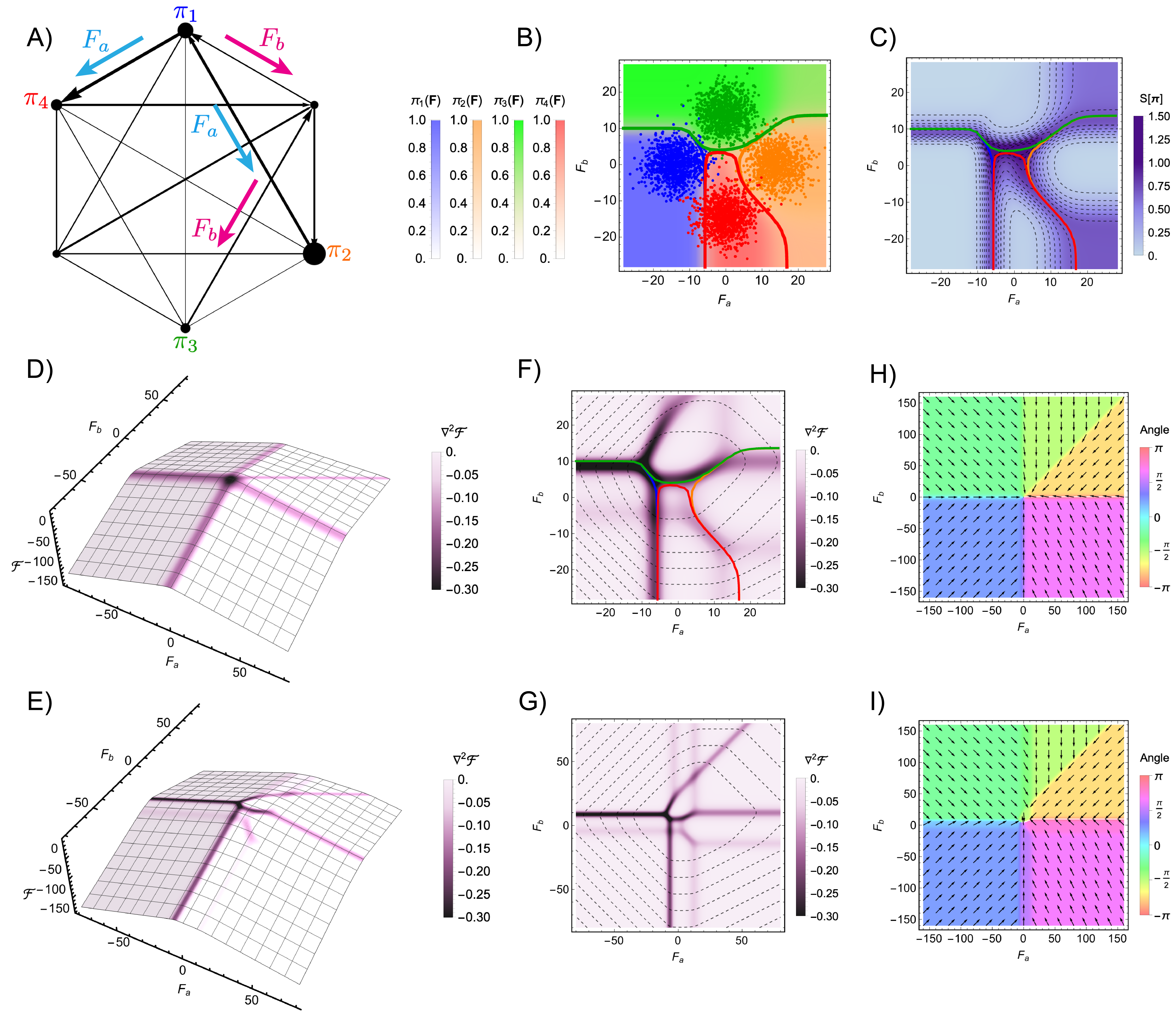}
		\caption{Creased landscape of $\mathcal{F}$.  A) Illustration of the problem definition and the learned parameters (cf. Figure \ref{SI_TrainedGraphsHexagon} above).  B)   Plot of the learned decision boundaries which solve the four-class classification problem.  C) Combination density and contour plot of the entropy $S[\boldsymbol{\pi}]$ in for the trained network.  D)  Plot of $\mathcal{F}$ for the untrained network (with $\bt = \mathbf{0}$).  E)  Plot of $\mathcal{F}$ for the trained network.  F)  Density plot of $\nabla_{\mathbf{F}}^2\mathcal{F}$ for the trained network.  Contours are drawn at fixed values of $\mathcal{F}$.  The decision of panel B boundaries are overlayed.  G)  Same as panel F, but zoomed out to show more of the large-scale structure.  H)  Vector plot of $\nabla_\mathbf{F} \mathcal{F}$ of the untrained graph.  Arrows are not drawn according to vector magnitude, only to show direction.  Coloring also indicates the direction of the vectors.  I)  Same as panel H, but for the trained graph.}
		\label{SI_CreasedLandscape}
	\end{center}
\end{figure*}

\section{How non-equilibrium driving allows expressivity}
\subsection{Expressivity from the parameters $E_j$, $B_{ij}$, and $F_{ij}$}

In the absence of any non-equilibrium driving, either through the learned parameters $F_{ij}$ or the input variables $F_a$, the steady-state distribution is a Boltzmann form $\pi_i \propto e^{-E_i}$ and does not depend on the $B_{ij}$ parameters.  Beating this restrictive functional form and achieving non-trivial classification expressivity thus requires non-equilibrium driving.  From the representation of the linear attention-like form of the matrix-tree expression in Equation \ref{eqmtt} of the main text, it is apparent that the key to successfully learning the classification task is the freedom to move the vectors $\bps(i;\bt)$ around, so that for the output nodes these vectors have large dot products with the input force vectors $\boldsymbol{\chi}(i,\mathbf{F})$ when $\mathbf{F}$ is in the corresponding region of input space.  We thus ask how each of the three types of parameters $E_j$, $B_{ij}$, and $F_{ij}$ affect this freedom.  

Let us first see how to transform a vector $\bps(i;\bt)$ rooted at node $i$ into a vector $\bps(j;\bt)$ rooted at node $j$.  To re-root the directed tree at node $j$ into that rooted at node $i$ will require flipping the edges $k\leftarrow l$ which connect nodes $i$ and $j$ in the tree $T^i_\alpha$.  This implies dividing the corresponding rates $W_{kl}$ from the product in the vector element $ \psi(T_i^\alpha; \bt)$ and multiplying by $W_{lk}$.  We define the product of these ratios for all flipped edges as $g^\alpha(i,j;\bt)$.  We then have
\begin{equation}
	\psi(T_j^\alpha; \bt) = g^\alpha(i,j;\bt) \psi(T_i^\alpha; \bt) 
\end{equation}
and note that $g^\alpha(j,i;\bt) = 1/g^\alpha(i,j;\bt)$. Now, the product of these ratios $W_{lk}/W_{kl}$ can only depend on $F_{kl}$ and the energies $E_i$ and $E_j$, and we can write
\begin{equation}
	g^\alpha(i,j;\bt) = g_E^\alpha(i,j;\bt_E) g_F^\alpha(i,j;\bt_F)
\end{equation}
where 
\begin{equation}
	g_E^\alpha(i,j;\bt_E) = e^{E_i - E_j}
\end{equation}
and 
\begin{equation}
	g_F^\alpha(i,j;\bt_F) = e^{-\sum'_{kl}F_{kl}}.
\end{equation}
The primed summation is only over edges flipped in the re-rooting process, and $\bt_E$ and $\bt_F$ contain only the $E_j$ and $F_{ij}$ parameters respectively.  Collecting the factors $g^\alpha(j,i;\bt)$ for each $\alpha$ along the diagonal of a $N_\alpha\times N_\alpha$ matrix $\mathbf{G}(i,j;\bt)$, we can write in matrix-vector notation 
\begin{equation}
	\boldsymbol{\psi}(i;\bt) = \mathbf{G}(i,j;\bt) * \boldsymbol{\psi}(j;\bt) = \mathbf{G}_E(i,j;\bt_E) * \mathbf{G}_F(i,j;\bt_F) * \boldsymbol{\psi}(j;\bt).
\end{equation}
The matrix $\mathbf{G}_E(i,j;\bt_E) = e^{E_i - E_j} \mathbf{I}$ is proportional to the identity matrix, but the matrix $\mathbf{G}_F(i,j;\bt_E)$ can have differing elements along the diagonal which allows transforming $\boldsymbol{\psi}(j;\bt)$ anisotropically.  

\begin{figure*}[ht!]
	\begin{center}
		\includegraphics[width= \textwidth]{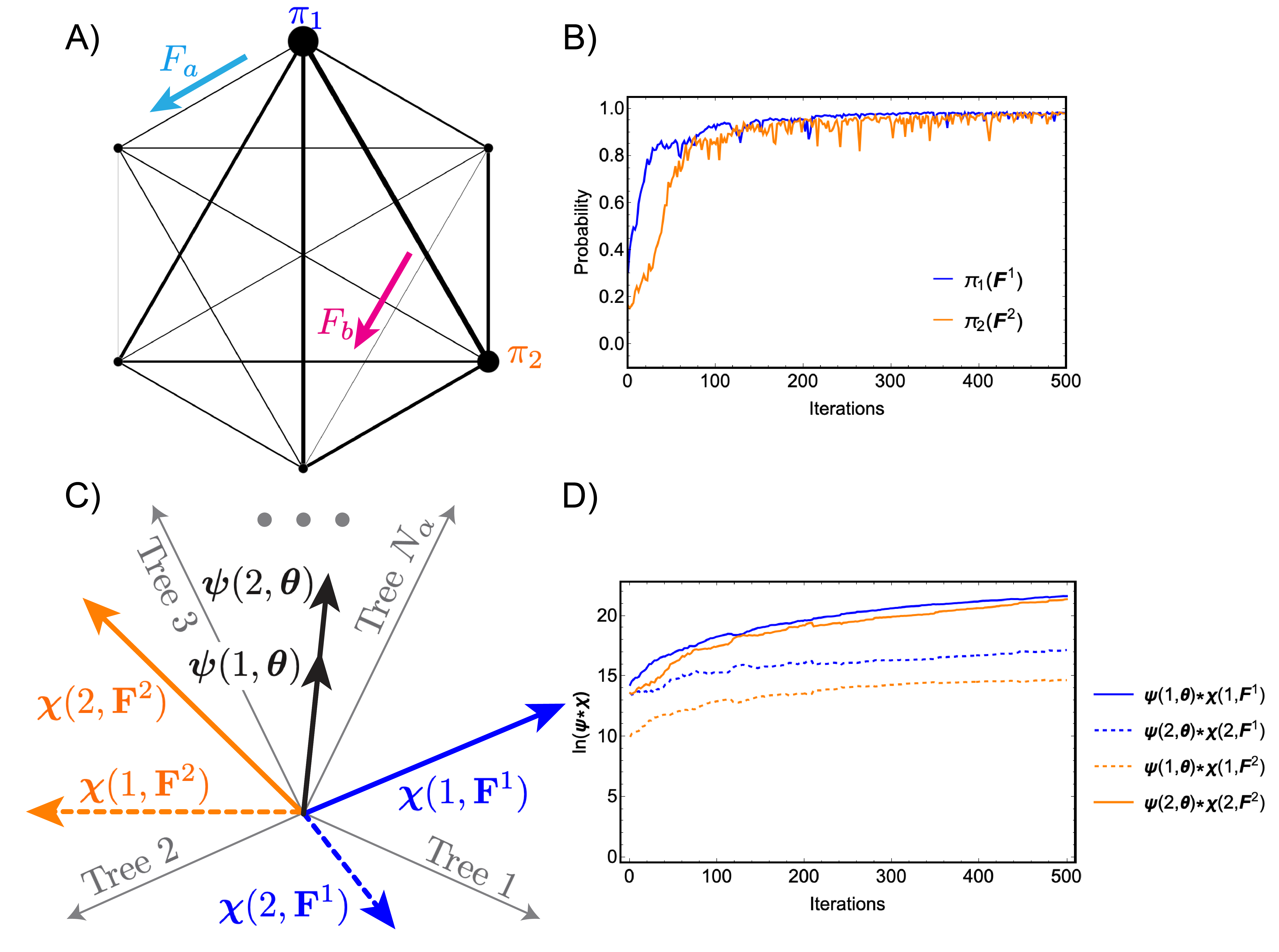}
		\caption{Successful learning in a fully connected 6 node graph where $F_{ij}$ parameters are not learned but the inputs are presented through $\mathbf{F}$.  A) Illustration of the problem definition and the learned parameters (cf. Figure \ref{SI_TrainedGraphsHexagon} above).  B)  Plot showing training convergence during training.  C)  Schematic illustration of the tree vector space for this problem.  D)  Plot showing how different pairs of vector dot products evolve during training.}
		\label{SI_VectorSpace}
	\end{center}
\end{figure*}

\begin{figure*}[ht!]
	\begin{center}
		\includegraphics[width= \textwidth]{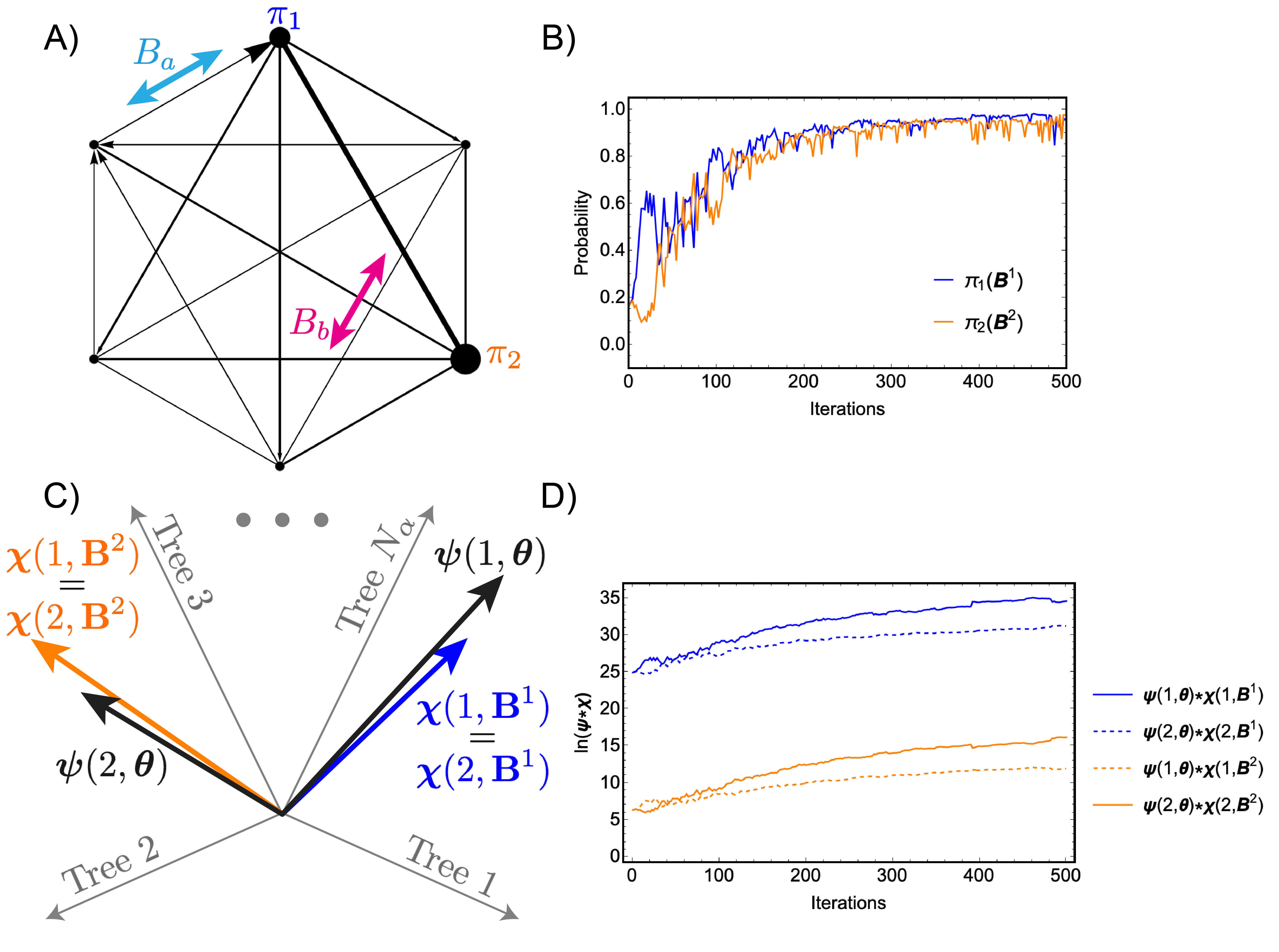}
		\caption{Successful learning in a fully connected 6 node graph where $F_{ij}$ parameters are learned and the inputs are presented through $\mathbf{B}$.  A) Illustration of the problem definition and the learned parameters (cf. Figure \ref{SI_TrainedGraphsHexagon} above).  B)  Plot showing training convergence during training.  C)  Schematic illustration of the tree vector space for this problem.  D)  Plot showing how different pairs of vector dot products evolve during training.}
		\label{SI_VectorSpaceBij}
	\end{center}
\end{figure*}

\begin{figure*}[ht!]
	\begin{center}
		\includegraphics[width= \textwidth]{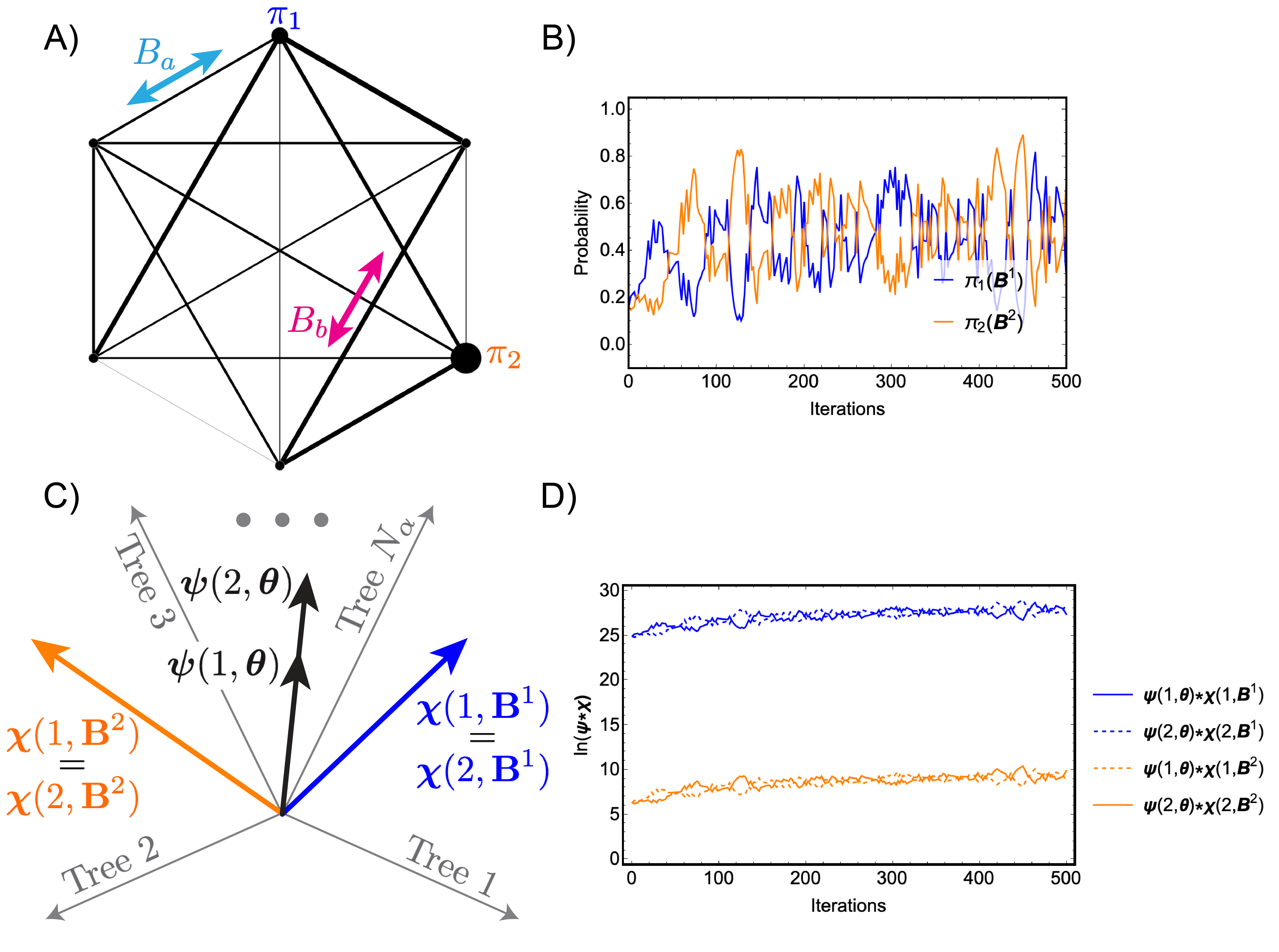}
		\caption{Unsuccessful learning in a fully connected 6 node graph where $F_{ij}$ parameters are not learned and the inputs are presented through $\mathbf{B}$.  A) Illustration of the problem definition and the learned parameters (cf. Figure \ref{SI_TrainedGraphsHexagon} above).  B)  Plot showing training convergence during training.  C)  Schematic illustration of the tree vector space for this problem.  D)  Plot showing how different pairs of vector dot products evolve during training.}
		\label{SI_VectorSpaceBijnoF}
	\end{center}
\end{figure*}

If we first assume that we only adjust $E_{j}$ values during training (i.e., $B_{ij} = F_{ij} = 0$), then this analysis implies that
\begin{equation}
	\boldsymbol{\psi}(i;\bt) = e^{E_\text{tot} - E_i}\mathbf{1}
\end{equation}
where $E_\text{tot} = \sum_k E_k$ and $\mathbf{1}$ is the vector of ones over tree space.  This severely limits the available space the vectors $\boldsymbol{\psi}(i;\bt)$ can occupy.  If we additionally learn $B_{ij}$ parameters, then 
\begin{equation}
	\boldsymbol{\psi}(i;\bt) = e^{E_\text{tot} - E_i}\tilde{\boldsymbol{\psi}}(\bt) 
\end{equation}
where $\tilde{\boldsymbol{\psi}}(\bt)$ is a single learnable vector that is shared among all nodes in the network.  If we learn $F_{ij}$'s, then the vectors $\boldsymbol{\psi}(i;\bt)$ can differ from each other in an element-by-element way, representing a large gain in expressivity. 

To illustrate the interpretation of the classification task in terms of the vectors $\boldsymbol{\psi}(i,\bt)$ and $\boldsymbol{\chi}(i,\mathbf{F})$, and how successful classification depends on non-equilibrium driving, we next consider three examples.  We consider here two ways of presenting inputs to the network: through contributions $\mathbf{F}$ to the anti-symmetric, non-equilibrium rate parameters, or through contributions $\mathbf{B}$ to the symmetric, equilibrium rate parameters.  If presenting inputs through $\mathbf{F}$, then for the reasons just outlined the vectors $\boldsymbol{\chi}(i,\mathbf{F})$ and $\boldsymbol{\chi}(j,\mathbf{F})$ can differ from each other for $i\neq j$.  If presenting inputs through $\mathbf{B}$, however, then $\boldsymbol{\chi}(i,\mathbf{F})= \boldsymbol{\chi}(j,\mathbf{F})$ for all $i,j$.

\subsection{Three examples}

\subsubsection{Inputs through $\mathbf{F}$, not learning $F_{ij}$}
First, we consider an example in which we present inputs through $\mathbf{F}$ but are not allowed to learn $F_{ij}$ parameters.  In this case learning is still generally possible, as illustrated in Figure \ref{SI_VectorSpace}.  Assuming the output nodes are labeled $1$ and $2$, the goal of learning in this case is to position the shared learnable vector $\tilde{\boldsymbol{\psi}}(\bt)$ so that 
\begin{equation}
	\boldsymbol{\psi}(1;\bt) * \boldsymbol{\chi}(1,\mathbf{F}^1) > \boldsymbol{\psi}(2;\bt) * \boldsymbol{\chi}(2,\mathbf{F}^1) \label{eqsmcp2}
\end{equation}
and 
\begin{equation}
	\boldsymbol{\psi}(2;\bt) * \boldsymbol{\chi}(2,\mathbf{F}^2) > \boldsymbol{\psi}(2;\bt) * \boldsymbol{\chi}(2,\mathbf{F}^1) \label{eqsmcp3}
\end{equation}
where $\mathbf{F}^1$ and $\mathbf{F}^2$ are typical inputs from classes 1 and 2.  To achieve this the vector $\tilde{\boldsymbol{\psi}}(\bt)$ should move near the subspace spanned by $\boldsymbol{\chi}(1,\mathbf{F}^1)$ and $\boldsymbol{\chi}(2,\mathbf{F}^2)$. It is apparent from studying how the dot products $\tilde{\boldsymbol{\psi}}(\bt) * \boldsymbol{\chi}(1,\mathbf{F}^1)$ and $\tilde{\boldsymbol{\psi}}(\bt) * \boldsymbol{\chi}(2,\mathbf{F}^2)$ change during training that something like this happens in Figure \ref{SI_VectorSpace}C.

\subsubsection{Inputs through $\mathbf{B}$, learning $F_{ij}$}
Second, we consider an example in which we present inputs through $\mathbf{B}$ but are allowed to learn $F_{ij}$ parameters (Figure \ref{SI_VectorSpaceBij}).  In this case learning is also possible, because although the input vectors $\boldsymbol{\chi}(i,\mathbf{B})$ are equal for every $i$ and $j$, they differ as $\mathbf{B}$ is varied.  This makes it possible to move the independent learnable vectors $\boldsymbol{\psi}(i;\bt)$ to satisfy Equations \ref{eqsmcp2} and \ref{eqsmcp3}.  

\subsubsection{Inputs through $\mathbf{B}$, not learning $F_{ij}$}
Finally, we consider an example in which we present inputs through $\mathbf{B}$ and cannot learn $F_{ij}$ parameters (Figure \ref{SI_VectorSpaceBijnoF}).  In this case learning is not possible, because the vectorial part of $\boldsymbol{\psi}(i,\bt)$ is shared for each $i$, and the input vectors $\boldsymbol{\chi}(i,\mathbf{B})$ are also equal for each $i$.  This prevents satisfying Equations \ref{eqsmcp2} and \ref{eqsmcp3}, which in this case simplify to the contradictory requirements
\begin{equation}
	e^{E_1}< e^{E_2} \ \text{and} \ e^{E_2}< e^{E_1}.
\end{equation}
If we were to gradually increase the threshold on the absolute values of the $F_{ij}$ parameters, then the vectorial parts of $\boldsymbol{\psi}(i,\bt)$ could start to split by from each other by increasing amounts, so that Figure \ref{SI_VectorSpaceBijnoF}C approaches Figure \ref{SI_VectorSpaceBij}C.  Figure \ref{NonEqDriving} in the main text illustrates how this continuously allows for more successful classification accuracy.
\end{onecolumngrid}

\bibliographystyle{unsrt}

\end{document}